\documentclass[aps,prd,superscriptaddress,nofootinbib,longbibliography]{revtex4-2}

\usepackage{amsmath,amssymb}
\usepackage{graphicx}
\usepackage{booktabs}
\usepackage{hyperref}
\usepackage{booktabs}
\usepackage{multirow}
\usepackage{color}
\usepackage{graphicx}
\def\be{\begin{equation}}
\def\ee{\end{equation}}
\def\bea{\begin{align}}
\def\eea{\end{align}}
\def\ben{\begin{equation*}}
\def\een{\end{equation*}}
\def\bean{\begin{align*}}
\def\eean{\end{align*}}

\usepackage{scalerel}
\usepackage{tikz}
\usetikzlibrary{svg.path}

\definecolor{orcidlogocol}{HTML}{A6CE39}
\tikzset{
  orcidlogo/.pic={
    \fill[orcidlogocol] svg{M256,128c0,70.7-57.3,128-128,128C57.3,256,0,198.7,0,128C0,57.3,57.3,0,128,0C198.7,0,256,57.3,256,128z};
    \fill[white] svg{M86.3,186.2H70.9V79.1h15.4v48.4V186.2z}
                 svg{M108.9,79.1h41.6c39.6,0,57,28.3,57,53.6c0,27.5-21.5,53.6-56.8,53.6h-41.8V79.1z M124.3,172.4h24.5c34.9,0,42.9-26.5,42.9-39.7c0-21.5-13.7-39.7-43.7-39.7h-23.7V172.4z}
                 svg{M88.7,56.8c0,5.5-4.5,10.1-10.1,10.1c-5.6,0-10.1-4.6-10.1-10.1c0-5.6,4.5-10.1,10.1-10.1C84.2,46.7,88.7,51.3,88.7,56.8z};
  }
}

\newcommand\orcidicon[1]{\href{https://orcid.org/#1}{\mbox{\scalerel*{
\begin{tikzpicture}[yscale=-1,transform shape]
\pic{orcidlogo};
\end{tikzpicture}
}{|}}}}

\begin{document}

\title{Geometrothermodynamics of Black Hole Shadows: A Review}

\author{Hernando Quevedo\textsuperscript{\orcidicon{0000-0003-4433-5550}}}
\email{quevedo@nucleares.unam.mx}
\affiliation{Instituto de Ciencias Nucleares, Universidad Nacional Aut\'onoma de M\'exico, AP 70543, Mexico City, Mexico}
\affiliation{Dipartimento di Fisica and ICRA, Universit\`a di Roma ``La Sapienza'', Roma, Italy}
\affiliation{Al-Farabi Kazakh National University, Al-Farabi av. 71, 050040 Almaty, Kazakhstan}

\author{Jose Miguel Ladino\textsuperscript{\orcidicon{0000-0001-9812-4949}}}
\email{miguel.ladino@correo.nucleares.unam.mx}
\affiliation{Instituto de Ciencias Nucleares, Universidad Nacional Aut\'onoma de M\'exico, AP 70543, Mexico City, Mexico}

\author{Carlos E. Romero-Figueroa\textsuperscript{\orcidicon{0000-0001-5548-7766}}}
\email{carlosed.romero@correo.nucleares.unam.mx}
\affiliation{Instituto de Ciencias Nucleares, Universidad Nacional Aut\'onoma de M\'exico, AP 70543, Mexico City, Mexico}

\author{Sasha A. Zaldivar\textsuperscript{\orcidicon{0009-0009-9350-055X}}}
\email{sasha@ciencias.unam.mx}
\affiliation{Instituto de Ciencias Nucleares, Universidad Nacional Aut\'onoma de M\'exico, AP 70543, Mexico City, Mexico}

\date{\today}
\begin{abstract}
Black-hole shadows and geometrothermodynamics (GTD) are defined in
different spaces but depend on the same macroscopic black-hole
parameters. We review their correspondence for Reissner--Nordstr\"om (RN),
Kerr, and Kerr--AdS black holes. The three GTD metrics reveal
complementary thermodynamic structures, including Davies instabilities,
extremal and degenerate limits, and regular curvature zeros, which can
be mapped to shadow configurations. Kerr--AdS extends this correspondence
to asymmetric shadows and multidimensional equilibrium states. We
formulate a local Jacobian-rank criterion for thermodynamic reconstruction
from shadow observables and discuss shadow--microstructure (SM) diagrams as
optical representations of GTD-classified equilibrium states. The
application to Sagittarius A* illustrates the inverse problem: shadow
observations constrain thermodynamic states and provide a basis for
identifying additional metric-dependent observables that could
distinguish among the GTD metric structures.
\end{abstract}

\maketitle
\tableofcontents

\section{Introduction}
\label{sec:introduction}

\begin{figure*}[h]
  \centering
  \begin{minipage}[t]{0.64\linewidth}
    \centering
    \includegraphics[width=\linewidth]{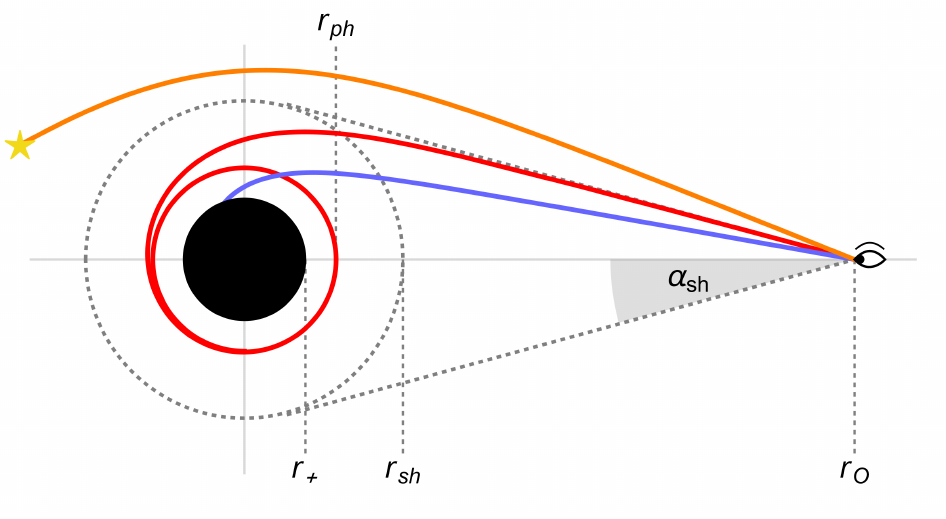}
    \\[2pt] \vspace{-0.6cm}\hspace{-9cm}\small{(a)}
  \end{minipage}\hfill
  \begin{minipage}[t]{0.36\linewidth}
    \centering
    \includegraphics[width=\linewidth]{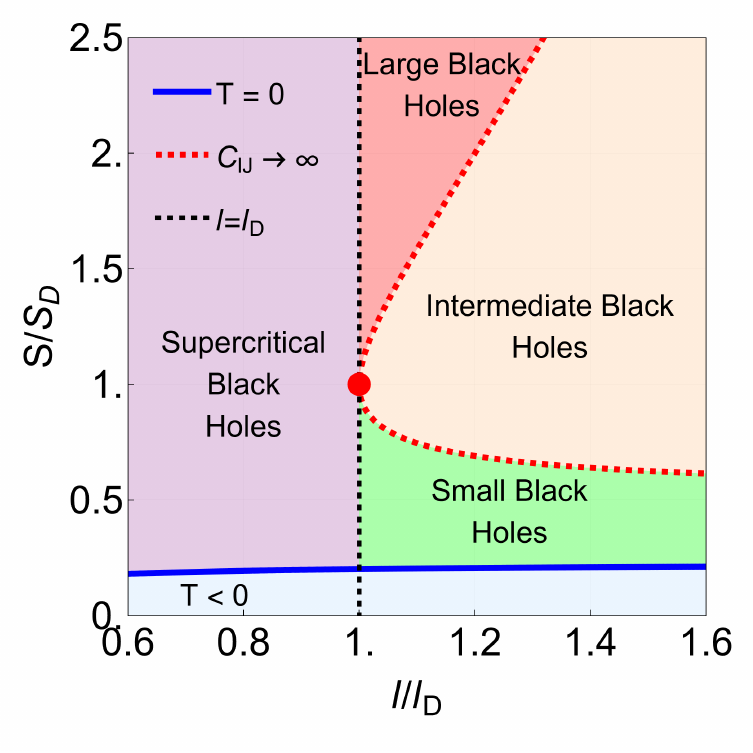}
    \\[2pt] \vspace{-0.6cm}\hspace{-3.5cm}\small{(b)}
  \end{minipage}
  \vspace{-0.7cm}
  \caption{(a) Shadow scheme of a spherically symmetric black hole, showing the radii of the event horizon $r_+$, photon sphere $r_{\text{ph}}$, critical shadow $r_{\text{sh}}$, and observer location $r_O$. Light trajectories illustrate a deflected ray from a distant star (orange), a critical ray defining the shadow's boundary and its angular size $\alpha_{sh}$ (red), and a ray captured by the black hole (blue). (b)~Thermodynamic phase diagram of Kerr-AdS black holes, showing normalized entropy $S/S_D$ vs AdS radius $l/l_D$. The red dot identifies the Davies point.}
  \label{fig:InicialFigures}
\end{figure*}

Black holes connect gravitational dynamics, quantum field theory, thermodynamics, and strong-field observations. The laws of black-hole mechanics acquire a thermodynamic interpretation when the horizon area is identified with entropy and the surface gravity with temperature \cite{Bekenstein1973,Hawking1975}. Independently, the propagation of null geodesics in black-hole spacetimes determines critical photon trajectories and, for a specified observer, the boundary of the black-hole shadow \cite{Synge1966,Luminet1979,Bardeen1973,PerlickTsupko2022,LadinoLarranaga2023} (see Fig.~\ref{fig:InicialFigures}(a)). The horizon-scale images of M87* and Sagittarius A* obtained by the Event Horizon Telescope have transformed black-hole shadows from a purely theoretical construction into an observational probe of the strong-field regime \cite{EHT2019,EHT2022,EHT2022IV,EHT2022VI}.

Thermodynamics and black-hole optics are nevertheless formulated in different spaces. Equilibrium thermodynamics is described on a manifold of thermodynamic states, whereas the shadow is constructed from null geodesics in spacetime and projected onto an observer's celestial screen. A thermodynamic curvature is therefore not a spacetime curvature, and a shadow observable is not, by itself, a thermodynamic variable. The connection between the two descriptions arises because they depend on the same macroscopic black-hole parameters. Mass, charge, angular momentum, and, when present, the anti--de Sitter (AdS) curvature radius determine both the spacetime geometry and the thermodynamic fundamental equation.

The possibility of relating optical and thermodynamic information was first explored through characteristic photon orbits. For AdS black holes, quantities associated with unstable null geodesics were shown to reproduce characteristic features of thermodynamic phase structure \cite{WeiLiuWang2019}. This idea was subsequently extended from photon-orbit variables to the shadow itself. In particular, the shadow size serves as a useful optical representation of thermodynamic branches and phase behavior, although different shadow observables do not necessarily encode the same thermodynamic information \cite{ZhangGuo2020,BelhajChakhchiElMoumniKhalloufiMasmar2020} (see Fig.~\ref{fig:InicialFigures}(b) for an example of the thermodynamic phases of Kerr-AdS black holes). These results established what is now commonly referred to as shadow thermodynamics.

A second geometric ingredient is provided by thermodynamic geometry. Hessian metrics constructed from thermodynamic potentials have been used to characterize equilibrium fluctuations, critical behavior, and effective thermodynamic interactions \cite{Ruppeiner1979,Ruppeiner1995}. In black-hole applications, however, the resulting curvature can depend on the thermodynamic potential, representation, and ensemble. Geometrothermodynamics (GTD) was introduced to formulate equilibrium thermodynamics in a Legendre-invariant geometric framework \cite{Quevedo2007}. In GTD, Legendre-invariant metrics are defined on the thermodynamic phase space and induce metrics on the equilibrium space. Their curvature singularities can then be compared with independently established thermodynamic response functions and stability boundaries. A wide array of physical systems have been successfully investigated using GTD, such as ideal and real gases \cite{QuevedoQuevedoSanchez2022,QuevedoSanchezTajVazquez2011,quevedo2023geometrothermodynamic,romero2026}, ideal quantum gases \cite{zaldivar2023ideal}, magnetic materials \cite{QuevedoQuevedoSanchez2026}, the Ising model \cite{BravettiLopezMonsalvoNettelQuevedo2014}, chemical reactions \cite{quevedo2014geometric}, econophysics \cite{2023IJGMM..2050057Q,e28070762}, and cosmological models \cite{romero2026quasi,e25060944,PhysRevD.86.063508,2019EPJC...79..577B,2014GrCo...20..208Q,Sanchezz2026}. Most notably for the present study, GTD has also served as a fundamental tool for analyzing black-hole thermodynamics within numerous frameworks of gravity \cite{QuevedoSanchez2008,LadinoRomeroQuevedo2024,LadinoRomeroQuevedo2025,LarranagaMojica2011,Gogoi,larranaga2012geometric,BravettiMomeniMyrzakulovQuevedo2013,AkbarQuevedoSaifullahSanchezTaj2011,TharanathSureshKuriakose2015,Zhang2018GTD,ChannuieMomeni2018,GhaffariLuciano2025,Kurihara2018,Sanchez2016,QuevedoQuevedoSanchez2020,HuChenWang2017,GanChenWang2016,2025IJGMM..2250049J,SekhmaniEtAl2025,TajQuevedo2012,MoLiu2014,QuevedoSanchez2009,LucianoSheykhi2023,LucianoSaridakis2023,2023EPJC...83..710R,Beissen2024,JankeJohnstonKenna2010,QuevedoQuevedoSanchez2016Phantom,HendiNaderi2015,HanChen2012,BravettiMomeniMyrzakulovAltaibayeva2013,Quevedo:2011fk,RizwanKumaraRajaniVaidAjith2019,QuevedoQuevedoSanchez2016EMD,ahmed2026shadow,LadinoRomeroQuevedo2026,quevedo2019quasi,Maya2026,Hernando2026,SHAH2026117662,Baghbani2026,QUEVEDO2026117321,Alessa2026,Biswas2026,Gogoi2026,BiswasSatyajit,QUEVEDO2026170601,ladino2026nakedsingularitiesblackhole}.

The combination of GTD with black-hole shadows has recently been developed for charged and rotating black holes. For the Reissner--Nordström--anti--de Sitter (RN--AdS) and Kerr--AdS black holes, the shadow was used to represent thermodynamic phase structure and GTD curvature in optical variables \cite{LadinoRomeroQuevedo2024,LadinoRomeroQuevedo2025}. For asymptotically flat RN and Kerr black holes, the Davies instabilities and the curvature structure of different GTD metrics were mapped to the shadow and used to construct shadow--microstructure (SM) diagrams, with an application to Sagittarius A* \cite{LadinoRomeroQuevedo2026}. More recently, GTD was applied to quantum Oppenheimer-Snyder spacetimes, successfully linking observable shadow properties, such as photon rings and their gap parameters, to the black hole's thermodynamic stability and microscopic interaction structure \cite{ladino2026nakedsingularitiesblackhole}. These developments motivate the central question of this review: to what extent can black-hole shadow observables encode the intrinsic thermodynamic and GTD structure of the equilibrium space?

A central aspect of the present analysis is the systematic construction of the three GTD metrics from the same fundamental equation. This common-potential formulation permits their geometric structures to be compared on equal footing. For the black holes considered here, the three metrics generally encode complementary information: Davies instabilities, extremal or degenerate limits, and additional regular curvature structures need not be represented by the same GTD metric. Their optical images therefore provide a natural basis for extending the SM construction.

A careful answer requires the thermodynamic, geometric, and optical ingredients to be distinguished before they are combined. Thermodynamic response functions and GTD structures provide complementary information and must be compared in order to establish the physical meaning of the geometric results. The order in which these calculations are performed is not essential. Once the relevant thermodynamic states and GTD structures have been identified, they can be represented by shadow observables. This comparison prevents a curvature singularity from being interpreted automatically as a phase transition and prevents an optical feature from acquiring thermodynamic meaning merely because it occurs at the same parameter value.

The review focuses on three representative four-dimensional black holes: the asymptotically flat RN and Kerr solutions, followed by the Kerr--AdS black hole. The first two isolate the charged and rotating cases without an additional cosmological scale. Kerr--AdS then introduces a higher-dimensional thermodynamic state space, ensemble dependence, and genuine AdS critical and global phase phenomena \cite{KubiznakMann2012,KubiznakMannTeo2017}. These three spacetimes also provide a natural progression on the optical side, from a circular shadow to a rotating asymmetric shadow and finally to a rotating shadow in a non-Minkowskian asymptotic geometry.

For consistency with the black-hole and shadow literature used throughout this review, we employ units $G=c=\hbar=k_{\rm B}=1$ unless stated otherwise. We use the metric signature $(-,+,+,+)$. 

The RN spacetime can be written in Schwarzschild-like coordinates as
\be
ds^2
=
-f_{\rm RN}(r)dt^2
+\frac{dr^2}{f_{\rm RN}(r)}
+r^2\left(d\theta^2+\sin^2\theta\,d\varphi^2\right),
\label{eq:RNmetricIntro}
\ee
where
\be
f_{\rm RN}(r)
=
1-\frac{2M}{r}+\frac{Q^2}{r^2}.
\label{eq:RNfIntro}
\ee
Here $M$ is the mass and $Q$ is the electric charge. The event horizons are located at
\be
r_{\pm}=M\pm\sqrt{M^2-Q^2},
\label{eq:RNhorizonsIntro}
\ee
so the black-hole domain is $|Q|\leq M$. The extremal configuration corresponds to $|Q|=M$. These same parameters determine the thermodynamic fundamental equation and the photon-sphere structure discussed later \cite{PerlickTsupko2022,LadinoRomeroQuevedo2026}.

The Kerr spacetime is written in Boyer--Lindquist coordinates as
\be
\begin{split}
ds^2={}&
-\left(1-\frac{2Mr}{\rho^2}\right)dt^2
-\frac{4Mar\sin^2\theta}{\rho^2}\,dt\,d\varphi
+\frac{\rho^2}{\Delta}\,dr^2
\\
&
+\rho^2 d\theta^2
+
\left(
r^2+a^2+\frac{2Ma^2r\sin^2\theta}{\rho^2}
\right)
\sin^2\theta\,d\varphi^2 ,
\end{split}
\label{eq:KerrmetricIntro}
\ee
with
\be
\rho^2=r^2+a^2\cos^2\theta,
\quad
\Delta=r^2-2Mr+a^2,
\quad
a=\frac{J}{M}.
\label{eq:KerrFunctionsIntro}
\ee
The horizons are determined by the condition $\Delta(r_\pm)=0$, which yields
\be
r_{\pm}=M\pm\sqrt{M^2-a^2}.
\label{eq:KerrHorizonsIntro}
\ee
Then, the black-hole condition is $|a|\leq M$, or equivalently $|J|\leq M^2$. Rotation breaks the spherical symmetry of the optical problem. The shadow boundary is generated by a photon region formed by unstable spherical null geodesics, and its apparent shape depends on both the rotation parameter and the observer inclination \cite{Bardeen1973,GrenzebachPerlickLammerzahl2014,PerlickTsupko2022}. For an AdS background, the curvature radius $l$ characterizes the geometry. The corresponding Kerr--AdS spacetime can be written in Boyer--Lindquist coordinates as
\be
\begin{split}
ds^2={}&
-\frac{\Delta_r}{\rho^2}
\left(
dt-\frac{a\sin^2\theta}{\Xi}\,d\varphi
\right)^2
+\frac{\rho^2}{\Delta_r}\,dr^2
+\frac{\rho^2}{\Delta_\theta}\,d\theta^2
\\
&+
\frac{\Delta_\theta\sin^2\theta}{\rho^2}
\left(
a\,dt-\frac{r^2+a^2}{\Xi}\,d\varphi
\right)^2 ,
\end{split}
\label{eq:KerrAdSmetricIntro}
\ee
where
\be
\rho^2=r^2+a^2\cos^2\theta,
\quad
\Delta_r=(r^2+a^2)\left(1+\frac{r^2}{l^2}\right)-2mr,
\label{eq:KerrAdSradialIntro}
\ee
and
\be
\Delta_\theta
=
1-\frac{a^2}{l^2}\cos^2\theta,
\quad
\Xi
=
1-\frac{a^2}{l^2}.
\label{eq:KerrAdSangularIntro}
\ee
The parameter $m$ appearing in the metric is related to the conserved mass and angular momentum by
\be
M=\frac{m}{\Xi^2},
\qquad
J=\frac{am}{\Xi^2}=aM.
\label{eq:KerrAdSChargesIntro}
\ee
This distinction is important because the thermodynamic fundamental equation used later is written in terms of the physical mass $M$, whereas the null-geodesic equations are conveniently expressed using the metric functions above. The event horizon is determined by the largest real root $r_+$ of the condition $\Delta_r=0$, i.e., 
\be
(r_+^2+a^2)\left(1+\frac{r_+^2}{l^2}\right)-2mr_+ = 0.
\label{horKAdS}
\ee

The condition $\Xi>0$ is required in the standard Kerr--AdS domain, implying $|a|<l$. The same functions $\rho^2$, $\Delta_r$, $\Delta_\theta$, and $\Xi$ will be used later in the analysis of the Kerr--AdS photon region and shadow \cite{GrenzebachPerlickLammerzahl2014,LadinoRomeroQuevedo2025}.

These metrics make explicit the common parametric origin of the thermodynamic and optical descriptions. In RN, the parameter $Q$ changes both the fundamental equation and the photon sphere. In Kerr the angular momentum enters thermodynamics through $J$ and the spacetime through $a=J/M$. In Kerr--AdS, the additional scale $l$ enters both the fundamental equation and the null-geodesic geometry, while its thermodynamic scaling properties are determined by the quasihomogeneous structure \cite{quevedo2019quasi,QuevedoQuevedo2023} analyzed in Sec.~\ref{sec:GTD}. The correspondence studied in this review is therefore not an identification of spacetime and equilibrium geometries. It is a relation between two constructions that depend on a common set of black-hole parameters.

A central issue is then whether the optical observables contain enough independent information to characterize the relevant thermodynamic states. For one-parameter branches, this reduces locally to the regularity and monotonicity of the map between a horizon or thermodynamic parameter and a shadow observable. For rotating systems with more than one relevant parameter, several independent shadow observables may be required. This observation motivates a local Jacobian-rank criterion developed later in the review. The criterion will be presented as a methodological synthesis of the existing shadow-thermodynamic results rather than as an established theorem for arbitrary black-hole families.

The observational interpretation requires an additional distinction. As depicted in Fig.~\ref{fig:InicialFigures}(a), the theoretical shadow is the critical curve generated by null geodesics. The bright ring reconstructed from interferometric data depends also on plasma emission, radiative transfer, instrumental response, and the imaging or model-fitting procedure. Consequently, observational constraints on a theoretical shadow observable are model dependent \cite{EHT2019,EHT2022,EHT2022IV,EHT2022VI}. This distinction becomes essential when the GTD-shadow correspondence is confronted with Sagittarius A*.

This review is organized as follows. In Sec.~\ref{sec:bh_thermodynamics}, we summarize the laws of black-hole thermodynamics and derive the fundamental equations required in the subsequent analysis. In Sec.~\ref{sec:GTD}, we analyze the quasihomogeneous structure of these fundamental equations, construct the three GTD metrics for RN, Kerr, and Kerr--AdS black holes, and compare their geometric structures with the corresponding thermodynamic response functions and physical domains. Section~\ref{sec:BHshadows} develops the black-hole shadow construction independently of thermodynamics and applies it to the same three spacetimes. In Sec.~\ref{sec:GTDshadows}, we combine the optical and thermodynamic descriptions and formulate the GTD--shadow correspondence, including the local conditions under which shadow observables can parametrize or reconstruct equilibrium states. Section~\ref{sec:SMdiagrams} discusses SM diagrams as optical representations of GTD-classified equilibrium states. In Sec.~\ref{sec:observations}, we examine the inverse observational problem, with particular attention to Sagittarius A*, the interpretation of AdS examples, and the prospects for observationally constraining the different GTD structures. Finally, Sec.~\ref{sec:conclusions} summarizes the main results and discusses the open problems and perspectives of black-hole shadow geometrothermodynamics.

\section{Black-hole thermodynamics}
\label{sec:bh_thermodynamics}

Black-hole thermodynamics originates from the formal analogy between the laws of black-hole mechanics and the laws of ordinary thermodynamics. The four laws of black-hole mechanics were formulated by Bardeen, Carter, and Hawking \cite{BardeenCarterHawking1973}, while the identification of the horizon area with entropy and the quantum-mechanical origin of the black-hole temperature were established by Bekenstein and Hawking \cite{Bekenstein1973,Hawking1975}. In the following, we use geometrized units, $G=c=\hbar=k_{\rm B}=1$, consistently with the black-hole solutions considered in this review.

For a stationary black hole, the zeroth law states that the surface gravity $\kappa$ is constant over the event horizon. This property is the gravitational analogue of thermal equilibrium: after the identification of $\kappa$ with temperature, a stationary horizon is characterized by a uniform temperature.

The first law relates infinitesimal variations of the mass to variations of the horizon area and of the other conserved quantities. The first law identifies the mass as the thermodynamic energy in terms of the independent variables that enter the metric, say $M=M(E^a)$, and shows how changes in $E^a$ 
contribute to the energy balance, i.e., 
\be
dM=I_a dE^a, \qquad I_a =\frac{\partial M}{\partial E^a} ,
\label{BHfirstlaw}
\ee
where $I_a$ are the dual variables. The second law states that the horizon area cannot decrease in a classical physical process,
\be
dA\geq 0 .
\ee
After identifying the area with entropy, this becomes the black-hole counterpart of the ordinary entropy-increase law and gives the horizon area its direct thermodynamic significance.

The third law states that a state with vanishing surface gravity cannot be reached by a finite sequence of physical processes \cite{BardeenCarterHawking1973}. Since $T$ is proportional to $\kappa$, this is the black-hole analogue of the unattainability formulation of the third law of thermodynamics. Extremal black holes therefore appear as limiting zero-temperature configurations rather than states obtainable through a finite thermodynamic process. The thermodynamic interpretation follows from the Bekenstein--Hawking entropy and the Hawking temperature,
\be
S=\frac{A}{4},
\qquad
T=\frac{\kappa}{2\pi}.
\label{BHentropytemperature}
\ee
With these identifications, Eq.~(\ref{BHfirstlaw}) takes the form
\be
dM=T\,dS+ J_i d F^i , \quad J_i = \frac{\partial M}{\partial F^i},
\label{BHthermofirstlaw}
\ee
where $F^i$ represents the remaining variables that enter the mass function and $J_i$ are the corresponding dual variables.

The entropy is therefore determined directly by the geometry of the event horizon. For each black-hole family, the horizon radius and its area express the entropy in terms of the parameters that characterize  the solution. We interpret the resulting relation as the fundamental equation in the entropy representation. Its inversion gives the mass as a function of the corresponding thermodynamic variables and hence the fundamental equation in the energy representation. 

As an example of the application of black-hole thermodynamics, let us consider the four-dimensional Kerr--Newman--AdS solution that describes a stationary, axially symmetric, and electrically charged black hole in an asymptotically anti-de Sitter spacetime \cite{CaldarelliCognolaKlemm2000}. We parameterize the negative cosmological constant in terms of the AdS curvature radius $l$ as
\be
\Lambda=-\frac{3}{l^2}.
\label{KNAdSLambda}
\ee
In Boyer--Lindquist-type coordinates, the line element can be written as
\be
\begin{split}
ds^2= &
-\frac{\Delta_r}{\rho^2}
\left(
dt-\frac{a\sin^2\theta}{\Xi}d\phi
\right)^2
+\frac{\rho^2}{\Delta_r}dr^2
+\frac{\rho^2}{\Delta_\theta}d\theta^2 \\ 
& +\frac{\Delta_\theta\sin^2\theta}{\rho^2}
\left(
a\,dt-\frac{r^2+a^2}{\Xi}d\phi
\right)^2 ,
\label{KNAdSmetric}
\end{split}
\ee
where
\be
\rho^2=r^2+a^2\cos^2\theta,
\quad
\Delta_r=
(r^2+a^2)
\left(
1+\frac{r^2}{l^2}
\right)
-2mr+q^2,
\label{KNAdSfunctions1}
\ee
and
\be
\Delta_\theta=
1-\frac{a^2}{l^2}\cos^2\theta,
\qquad
\Xi=
1-\frac{a^2}{l^2}.
\label{KNAdSfunctions2}
\ee
The parameters $m$, $a$, and $q$ appearing in the metric are related to the physical mass, angular momentum, and electric charge by
\be
M=\frac{m}{\Xi^2},
\qquad
J=aM,
\qquad
Q=\frac{q}{\Xi}.
\label{KNAdSphysicalparameters}
\ee
The uncharged limit $q\rightarrow0$ reduces the solution to Kerr--AdS, whereas the asymptotically flat limit $l\rightarrow\infty$ gives the Kerr--Newman spacetime. The horizons are determined by the zeros of the radial function $\Delta_r$. The outer black-hole horizon $r_+$ therefore satisfies
$ 
\Delta_r(r_+)=0,
$ 
or explicitly,
\be
(r_+^2+a^2)
\left(
1+\frac{r_+^2}{l^2}
\right)
-2mr_+
+q^2=0.
\label{KNAdShorizonequation}
\ee
Thus, $r_+$ is the outer positive root of the horizon equation associated with the black hole. The area of the horizon follows directly from the properties of the metric as
\be
A=
\frac{4\pi(r_+^2+a^2)}{\Xi}.
\label{KNAdSarea}
\ee
Then, the Bekenstein--Hawking entropy is given as 
\be
S=
\frac{\pi(r_+^2+a^2)}{\Xi}.
\label{KNAdSentropy}
\ee
Consequently, the entropy is determined by the horizon radius, the rotation parameter, and the AdS curvature radius through $\Xi$. Eqs.~(\ref{KNAdShorizonequation}) and (\ref{KNAdSentropy}), together with the relations Eq.~(\ref{KNAdSphysicalparameters}), provide the starting point for eliminating the metric parameters in favor of the thermodynamic variables and obtaining the corresponding fundamental equation that in the mass representation can be expressed as 
\be
\begin{split}
M(S,J,Q,l)
= & 
\Bigg[
\frac{S}{4\pi}
\left(
1+\frac{\pi Q^2}{S}
+\frac{S}{\pi l^2}
\right)^2 
+
\frac{\pi J^2}{S}
\left(
1+\frac{S}{\pi l^2}
\right)
\Bigg]^{1/2}.
\label{KNAdSfundamental}
\end{split}
\ee
It seems reasonable to consider the radius of curvature $l$ as a constant parameter. Then, for this fundamental equation, the first law reads
\be
dM = TdS + \Omega d J + \Psi_Q d Q ,
\ee
where $\Omega$ and $\Psi_Q$ denote the angular velocity at the horizon and the electric potential, respectively, which, together with the Hawking temperature $T$, characterize the thermodynamic behavior of the black hole. However, scaling arguments and consistency with the Smarr relation require the first law to include an additional work term associated with variations of the AdS radius $l$ \cite{romero2024extended1, QuevedoQuevedoSanchez2016Phantom}. Thus, $l$ should be treated as a fluctuating thermodynamic variable in the equilibrium space. This extended thermodynamic description, based on the fundamental relation Eq.~\eqref{KNAdSfundamental}, is developed in Sec.~\ref{Kerr-AdS sec}.

\section{Geometrothermodynamics of black holes}
\label{sec:GTD}

GTD provides a geometric formulation of equilibrium thermodynamics in which the thermodynamic phase space is endowed with a contact structure and with metrics that are invariant under Legendre transformations \cite{Quevedo2007,QuevedoSanchez2008,QuevedoSanchezTajVazquez2011}. The equilibrium space is then obtained as a submanifold of the phase space, and its curvature is used to characterize thermodynamic interaction and the singular structure associated with changes of stability and phase behavior. This formalism has been applied to a wide range of black holes and other thermodynamic systems \cite{QuevedoQuevedoSanchez2022,QuevedoSanchezTajVazquez2011,quevedo2023geometrothermodynamic,romero2026,zaldivar2023ideal,QuevedoQuevedoSanchez2026,BravettiLopezMonsalvoNettelQuevedo2014,quevedo2014geometric,2023IJGMM..2050057Q,e28070762,romero2026quasi,e25060944,PhysRevD.86.063508,2019EPJC...79..577B,2014GrCo...20..208Q,Sanchezz2026,QuevedoSanchez2008,LadinoRomeroQuevedo2024,LadinoRomeroQuevedo2025,LarranagaMojica2011,Gogoi,larranaga2012geometric,BravettiMomeniMyrzakulovQuevedo2013,AkbarQuevedoSaifullahSanchezTaj2011,TharanathSureshKuriakose2015,Zhang2018GTD,ChannuieMomeni2018,GhaffariLuciano2025,Kurihara2018,Sanchez2016,QuevedoQuevedoSanchez2020,HuChenWang2017,GanChenWang2016,2025IJGMM..2250049J,SekhmaniEtAl2025,TajQuevedo2012,MoLiu2014,QuevedoSanchez2009,LucianoSheykhi2023,LucianoSaridakis2023,2023EPJC...83..710R,Beissen2024,JankeJohnstonKenna2010,QuevedoQuevedoSanchez2016Phantom,HendiNaderi2015,HanChen2012,BravettiMomeniMyrzakulovAltaibayeva2013,Quevedo:2011fk,RizwanKumaraRajaniVaidAjith2019,QuevedoQuevedoSanchez2016EMD,ahmed2026shadow,LadinoRomeroQuevedo2026,quevedo2019quasi,Maya2026,Hernando2026,SHAH2026117662,Baghbani2026,QUEVEDO2026117321,Alessa2026,Biswas2026,Gogoi2026,BiswasSatyajit,QUEVEDO2026170601,ladino2026nakedsingularitiesblackhole}.

For the purposes of this review, two aspects are particularly important. First, the GTD construction must be consistent with the quasihomogeneous scaling properties of the fundamental equation. Second, a curvature singularity is not interpreted independently of ordinary thermodynamics. Its physical meaning must be established by comparing it with the response functions, stability conditions, extremal boundaries, or global thermodynamic transitions of the system. The examples below will illustrate this procedure for the RN, Kerr, and Kerr--AdS black holes. This order is useful because the first two cases isolate the basic GTD mechanism in asymptotically flat spacetimes, whereas Kerr--AdS introduces an additional thermodynamic scale, a richer response-function structure, and ensemble dependence \cite{quevedo2014ensemble,CaldarelliCognolaKlemm2000}.

A comment on ensemble terminology is necessary. In black-hole
thermodynamics, the specification of an ensemble should be understood
in terms of the thermodynamic variables that are allowed to fluctuate
and those that are held fixed, together with the corresponding
Legendre-transformed thermodynamic potential \cite{quevedo2014ensemble}. We therefore avoid
identifying an ensemble solely by the name of a thermodynamic potential.
For example, fixing an extensive variable and fixing its conjugate
intensive variable define different thermodynamic response problems.
The appropriate potential follows from the corresponding Legendre
transformation of the fundamental equation. In particular, the term
``grand canonical'' should not be identified generically with the Gibbs
free energy; the relevant potential depends on the variables exchanged
with the reservoir and on the thermodynamic formulation used.
Throughout this review, whenever ensemble dependence is discussed, we
will, therefore, state explicitly which variables are held fixed.

\subsection{General GTD framework}

The starting point of GTD is a $(2n+1)$-dimensional thermodynamic phase space ${\cal T}$ with coordinates
\be
Z^A=\{\Phi,E^a,I_a\},\quad A=0,1,...,2n,
\label{eq:GTDcoordinates}
\ee
where $\Phi$ is a thermodynamic potential \cite{callen2006thermodynamics}, $E^a$ are extensive variables, and $I_a$ are the corresponding intensive variables. The importance of introducing the phase space with this set of coordinates is that 
in this setting, a Legendre transformation can be expressed as a particular coordinate transformation on $\mathcal{T}$ \cite{Arnold}. Specifically, consider the change
\be
\{Z^A\}\longrightarrow\{\widetilde Z^A\}=\{\widetilde\Phi,\widetilde E^a,\widetilde I_a\},
\ee
defined by
\be
\Phi=\widetilde{\Phi}-\widetilde{E}^k\widetilde{I}_k,\qquad
E^i=-\widetilde{I}^i,\qquad E^j=\widetilde{E}^j,
\ee
and
\be
I^i=\widetilde{E}^i,\qquad I^j=\widetilde{I}^j.
\ee
Here, the index sets $i$ and $j$ form a disjoint partition of $\{1,\ldots,n\}$, and the summation in the first relation extends over the variables involved in the transformation. The choice $i=\emptyset$ leaves all thermodynamic variables unchanged and therefore corresponds to the identity transformation. At the opposite extreme, $i=\{1,\ldots,n\}$ transforms every conjugate pair and defines a total Legendre transformation. Intermediate choices of $i$ correspond to partial Legendre transformations. Furthermore, according to Darboux theorem \cite{Arnold}, on the phase space it is possible to introduce a contact structure by means of the Gibbs one-form
\be
\Theta=d\Phi-I_a dE^a,
\label{eq:GTDcontact}
\ee
which satisfies the nonintegrability condition $\Theta\wedge(d\Theta)^n\neq0$ \cite{Quevedo2007,QuevedoQuevedoSanchez2026}.
It is easy to see that the Gibbs one-form is invariant with respect to Legendre transformations as defined above. 
 In GTD, the metric structure of ${\cal T}$ is also required to be invariant under these transformations. A detailed analysis of the action of above  transformations on an arbitrary metric $G=G_{AB}dZ^A dZ^B$ shows  that Legendre-invariant metrics can be divided into three classes  \cite{Quevedo2007,QuevedoQuevedoSanchez2026}
\be
G^I=\Theta^2+
\left(\xi_{ab}E^a I^b\right)
\left(\delta_{cd}\,dE^c dI^d\right),
\label{eq:GIphase}
\ee
\be
G^{II}=\Theta^2+
\left(\xi_{ab}E^a I^b\right)
\left(\eta_{cd}\,dE^c dI^d\right),
\label{eq:GIIphase}
\ee
and
\be
G^{III}=\Theta^2+
\sum_a \xi_a(E^a I_a)^{2k+1}dE^a dI_a,
\label{eq:GIIIphase}
\ee
where $\eta_{ab}={\rm diag}(-1,1,\ldots,1)$, $\xi_{ab}$ is diagonal, $\xi_a$ are real constants, and $k$ is an integer. The $n$-dimensional equilibrium space ${\cal E}$ with coordinates $E^a$ is embedded in ${\cal T}$ by a map $\varphi:{\cal E}\rightarrow{\cal T}$ satisfying
\be
\varphi^*(\Theta)=0
\label{eq:equilibriumcondition}
\ee
a condition which is equivalent to the first law and the equilibrium equations
\be
d\Phi=I_a dE^a,
\qquad
I_a=\Phi_{,a}.
\label{eq:equilibriumequations}
\ee
In local coordinates, the embedding map $\varphi: {\cal E}\rightarrow {\cal T}$ implies that the coordinates of ${\cal T}$ become functions of the coordinates of ${\cal E}$, i.e.,  $Z^A =\{\Phi(E^a), E^a, I_a(E^b)\}$. The first function $\Phi=\Phi(E^a)$ represents the fundamental equation, whereas the dual variables $I_a$ become functions of $E^a$ by means of the definition $I_a = \Phi_a= \Phi_a(E^b)$. 

An important property of the embedding map $\varphi$ is that it induces a metric $g$ on ${\cal E}$ by means of the pullback, i.e. 
\be 
g=\varphi^*(G).
\ee 
 This means that for a particular thermodynamic system with fundamental equation $\Phi=\Phi(E^a)$
the pullback induces on ${\cal E}$ three different metrics $g^ I=\varphi^*(G^I)$, $g^{II}=\varphi^*(G^{II})$, and $g^{III}=\varphi^*(G^{III})$. It is then reasonable to demand that each metric represents a different aspect of the same thermodynamic system and that all metrics be  compatible with each other. It turns out that to reach this goal, we only need to fix the arbitrary constants of the metrics of the phase space by considering the scaling properties of the fundamental equation. 

Ordinary thermodynamic systems are homogeneous in the sense that, under a scaling $E^a\rightarrow \lambda E^a$, the function $\Phi(E^a)$ behaves as $\Phi(\lambda E^a)= \lambda^{\beta_\Phi} \Phi(E^a)$, where  $\lambda$ is a positive parameter and $\beta_\Phi$ is the degree of homogeneity of the function $\Phi(E^a)$. There are more general systems, such as black holes, described by quasihomegeneous functions \cite{quevedo2019quasi} that satisfy  the condition 
\be \Phi(\lambda^{\beta_1}E^1,\ldots, \lambda ^{\beta_n}E^n) = \lambda ^{\beta_\Phi} \Phi (E ^ 1,\ldots, E ^ n).
\ee 
In the limiting case $\lambda_a=1$ $(a=1,...,n)$, we return to the homogeneous case. Furthermore, if we choose 
 $\xi_a=\beta_a$, $\xi_{ab}={\rm diag}(\beta_1,\ldots,\beta_n)$, and $k=0$,  the three equilibrium metrics turn out to be compatible in the sense that, as we will see below, their curvature singularities represent the equilibrium conditions in a consistent way. Moreover, with this choice the metrics\footnote{Notation: $\Phi_{,ab} \equiv \partial^2 \Phi/\partial E^a \partial E^b$.} can be expressed as \cite{QuevedoQuevedoSanchez2026}
\be
g^I=
\Sigma\, \Phi_{,ab}\,dE^a dE^b,
\label{eq:gIgeneral}
\ee
\be
g^{II}=
\Sigma\, \sum_{a,b=1}^n \eta_{aa}\Phi_{,ab}\,dE^a dE^b,
\label{eq:gIIgeneral}
\ee
\be
g^{III}=
\sum_{a,b=1}^n \beta_a E^a\Phi_{,a}\Phi_{,ab}\,dE^a dE^b,
\label{eq:gIIIgeneral}
\ee
where the conformal factor appearing in the first two metrics is given by
\be
\Sigma=\sum_{a=1}^n \beta_a E^a\Phi_{,a}.
\label{eq:SigmaGTD}
\ee

We notice that in the case of quasihomogeneous systems, the Euler  and the Gibbs-Duhem relations turn out to depend explicitly on the constants $\beta_a$ as 
\be
\sum_{a=1}^n \beta_a I_a E^a = \beta_\Phi \Phi , \label{EulerID}
\ee
\be
\sum_{a=1}^n \left[ (\beta_a-\beta_\Phi) I_a dE^a + \beta_a E^a dI_a\right] =0.
\label{identities}
\ee
These relations are valid when all independent variables of the function $\Phi(E^a)$ are allowed to vary. If, instead, at least one of the variables $E^a$ is kept fixed, it can happen that the Euler relation is not satisfied. Each case must be verified separately. If the Euler relation is satisfied, the conformal factor $\Sigma$ can be written as $\Sigma = \beta_\Phi \Phi$, an equation that can be used to reduce the form of the metric functions and the curvature singularities.

In the case of two thermodynamic degrees of freedom ($n=2$) with  $E^a=(E^1,E^2)$,  the resulting line elements can be written as \cite{QuevedoQuevedo2023}
\be
\label{gI2D} 
g^{I}  =   \Sigma \left[\Phi_{,11} (d E^1)^2 + 2 \Phi_{,12} dE^1 dE^2 + \Phi_{,22}(dE^2)^2\right],\, 
\ee
\be 
g^{II}  =  \Sigma \left[-\Phi_{,11} (d E^1)^2  + \Phi_{,22} (dE^2)^2\right]\,,
\label{gII2D} 
\ee 
\be
\begin{split}
g^{III}  =  & \beta_1 E^1 \Phi_{,1} \Phi_{,11} (dE^1)^2 + \Sigma \Phi_{,12} dE^1 dE^2 \\
& + \beta_ 2 E^2 \Phi_{,2} \Phi_{,22} (dE^2)^2 \, ,
\label{gIII2D}
\end{split}
\ee
where $\Sigma  =  \beta_1 E^1\Phi_{,1}  + \beta_2 E^2\Phi_{,2} $. The corresponding curvature scalars can be represented as 
\be
{\cal R}^I = \frac{N^I}{D^I}, \quad
{\cal R}^{II} = \frac{N^{II}}{D^{II}}, \quad
{\cal R}^{III} = \frac{N^{III}}{D^{III}}, 
\ee
where $N^I$, $N^{II}$, and $N^{III}$ are large expressions given explicitly in terms of the derivatives of  $\Phi$, and 
\be
D^I=2\Sigma^3 (\Phi_{,11}\Phi_{,22}-\Phi_{,12}^2)^2\ ,
\label{eq:DI2D}
\ee
\be
D^{II}=2\Sigma^3 \Phi_{,11}^2\Phi_{,22}^2\ ,
\label{eq:DII2D}
\ee
\be
D^{III}=\left(  \Sigma ^2 \Phi_{,12}^2 - 4\beta_1\beta_2E^1  E^2 \Phi_{,1} \Phi_{,2} \Phi_{,11} \Phi_{,22}\right)^2\ .
\label{eq:DIII2D}
\ee
We see that if no cancellations occur with the zeros of the numerators, the zeros of $D^I$, $D^{II}$, and $D^{III}$  determine the curvature singularities of the equilibrium space, which should describe the geometric and thermodynamic properties of the one system with fundamental equation $\Phi=\Phi(E^1,E^2)$. Therefore, we demand that the zeros of the three denominators lead to compatible conditions. First, suppose that $D^{II}=0$, i.e, $\Phi_{,11}\Phi_{,22}=0$. Then, the zeros of $D^I$ and $D^{III}$ are compatible with the condition $\Phi_{,12}=0$. Secondly, suppose that $D^{I}=0$, i.e., $\Phi_{,11}\Phi_{,22} = \Phi_{,12}^2\neq 0$.  Then, $D^{II}\neq 0$, and the equation $D^{III}=0$ leads to \cite{QuevedoQuevedo2023} 
\be
\Phi_{,12}(\beta_1E^1 \Phi_{,1} - \beta_2 E^2 \Phi_{,2}) = 0 ,
\ee
a condition that cannot be satisfied in general because the equation $(\beta_1E^1 \Phi_{,1} - \beta_2 E^2 \Phi_{,2})=0$ fixes the value of the fundamental equation as  $\Phi(E^1, E^2) = f( E^2 
{E^1}^{\beta_2/\beta_1})$, where $f$ is an arbitrary function of its argument. Thirdly, suppose that $D^{III}=0$, i.e., $\Sigma ^2 \Phi_{,12}^2 =  4\beta_1\beta_2E^1  E^2 \Phi_{,1} \Phi_{,2} \Phi_{,11} \Phi_{,22}\neq 0$. Then, $D^{II}\neq 0$ and the equation $D^I=0$ is equivalent to
\be
\Phi_{,11} \Phi_{,22} (\beta_1 E^1 \Phi_{,1} - \beta_2 E^2 \Phi_{,2}) = 0 ,
\ee
which again constrains the generality of the fundamental equation. The compatibility analysis shows that the independent equilibrium conditions that can be extracted from the three GTD curvature denominators without imposing additional restrictions on the fundamental equation are 
\be
I: \quad \Phi_{,11}\Phi_{,22}-\Phi_{,12}^2=0\ , \label{condI}
\ee
\be
II: \quad \Phi_{,11}\Phi_{,22}=0 \label{condII} \ ,
\ee
\be
III: \quad \Phi_{,12}=0 \label{condIII}\ .
\ee

These conditions are precisely the equilibrium conditions discussed in standard thermodynamics \cite{callen2006thermodynamics}. GTD provides their geometric realization as the possible singularity conditions of the curvature of the equilibrium space. These conditions should be understood as the independent thermodynamic structures encoded by the three metric families; the actual curvature singularities of a particular system must still be checked after possible numerator cancellations and restrictions to the physical domain are taken into account. Their specific physical manifestation must then be determined for each system.

These conditions  show why the three GTD metrics need not have identical singularity structures. They probe different combinations of the second derivatives of the fundamental equation and can therefore characterize different thermodynamic structures, including singularities of distinct response functions. 

This point is also useful when comparing GTD with Hessian thermodynamic geometries. In the standard Hessian thermodynamic geometry based on entropy fluctuations, the metric is constructed from second derivatives of a chosen thermodynamic potential and its curvature depends on the thermodynamic representation and on the state space or ensemble used \cite{Ruppeiner1979,Ruppeiner1995,Ruppeiner2008,Sahay2017}. GTD adds a Legendre-invariant phase-space structure and, in the quasihomogeneous formulation used here, a systematic relation between the scaling properties of the fundamental equation and the equilibrium metrics. The physical significance of the GTD curvature singularities is established by comparing the resulting equilibrium conditions with the thermodynamic properties of the system, including its response functions, stability boundaries, extremal states, and other physically distinguished configurations.

From a statistical-mechanical viewpoint, different ensembles also
differ in the fluctuations that are permitted by the coupling to the
reservoir, with the corresponding thermodynamic potentials arising
naturally in the gravitational path-integral formulation
\cite{gibbons1977action}. Their stability conditions and response functions can
therefore be different. For ordinary additive systems, different
ensembles may become equivalent in the thermodynamic limit under
appropriate conditions. Black holes, however, are gravitational and
generally nonadditive systems, so such equivalence cannot be assumed
a priori. In the present review, we do not develop a fluctuation theory
of black-hole ensembles. Ensemble dependence is used in the more
restricted thermodynamic sense of specifying the variables held fixed,
the corresponding thermodynamic potential, and the response functions
relevant to stability.

\subsection{Reissner--Nordstr\"om black hole}
\label{subsec RNGTD}
The asymptotically flat RN black hole provides the simplest charged example. In the mass representation \cite{LadinoRomeroQuevedo2026}, the fundamental equation is
\be
M(S,Q)=\frac{S+\pi Q^2}{2\sqrt{\pi S}}.
\label{eq:MRN}
\ee
We first determine its quasihomogeneous structure before constructing the GTD metrics. Requiring $M(\lambda^{\beta_S}S,\lambda^{\beta_Q}Q)=\lambda^{\beta_M}M(S,Q)$ gives the quasihomogeneous coefficients $\beta_S=2\beta_M$ and $\beta_Q=\beta_M$. Substituting these relations directly into Euler's identity, Eq.~\eqref{EulerID}, gives $2\beta_M S M_{,S}+\beta_M Q M_{,Q}=\beta_M M$, or equivalently $2S M_{,S}+Q M_{,Q}=M$. Consequently, the conformal factor of $g^I$ and $g^{II}$ is $\Sigma=\beta_M M$. Moreover, the derivatives required for the three metrics are
\be
M_{,S}
=
\frac{S-\pi Q^2}{4\sqrt{\pi}S^{3/2}},
\quad
M_{,Q}
=
\frac{\sqrt{\pi}Q}{\sqrt S},
\label{eq:RNfirstderivatives}
\ee
and
\be
M_{,SS}
=
\frac{3\pi Q^2-S}{8\sqrt{\pi}S^{5/2}},
\quad
M_{,SQ}
=
-\frac{\sqrt{\pi}Q}{2S^{3/2}},
\quad
M_{,QQ}
=
\frac{\sqrt{\pi}}{\sqrt S}.
\label{eq:RNsecondderivatives}
\ee
Substitution into Eqs.~(\ref{gI2D})--(\ref{gIII2D}) yields
\be
g^I= \frac{\beta_M (S+\pi Q^2)}{2S}\left[ 
\frac{(3\pi Q^2-S)}
{8\pi S^2}\,dS^2
-
\frac{ Q}
{S}\,dS\,dQ
+\,dQ^2\right] ,
\label{eq:RNgI}
\ee
\be
g^{II}
=
\frac{\beta_M(S+\pi Q^2)(S-3\pi Q^2)}
{16\pi S^3}\,dS^2
+
\frac{\beta_M(S+\pi Q^2)}
{2S}\,dQ^2 ,
\label{eq:RNgII}
\ee
\be
\begin{split}
g^{III}={}&
\frac{\beta_M(S-\pi Q^2)(3\pi Q^2-S)}
{16\pi S^3}\,dS^2
-
\frac{\beta_M Q(S+\pi Q^2)}
{4S^2}\,dS\,dQ
\\
&+
\frac{\beta_M\pi Q^2}{S}\,dQ^2 .
\end{split}
\label{eq:RNgIII}
\ee

We now evaluate the three equilibrium conditions Eqs.~(\ref{condI})--(\ref{condIII}). For the first metric,
\be
I: \quad M_{,SS}M_{,QQ}-M_{,SQ}^2
=
\frac{\pi Q^2-S}{8S^3} = 0 \ \Rightarrow \  
 S=\pi Q^2.
\label{eq:RNI}
\ee
For the second metric,
\be
II: \quad M_{,SS}M_{,QQ}
=
\frac{3\pi Q^2-S}{8S^3}\ \Rightarrow\ 
 S=3\pi Q^2.
\label{eq:RNII}
\ee
Finally, for the third metric 
\be
III: \quad M_{,SQ}
=
-\frac{\sqrt{\pi}Q}{2S^{3/2}} \ \Rightarrow \ 
 Q=0.
\label{eq:RNIII}
\ee
Thus, GTD produces three different distinguished conditions before any thermodynamic or response functions are invoked. To find out the thermodynamic meaning of the curvature singularities, we consider the  first law
\be
dM=T\,dS+\Psi_Q\,dQ,
\label{eq:RNfirstlaw}
\ee
with dual variables
\be
T=M_{,S}
=
\frac{S-\pi Q^2}{4\sqrt{\pi}S^{3/2}},
\qquad
\Psi_Q=M_{,Q}
=
\frac{\sqrt{\pi}Q}{\sqrt S}.
\label{eq:RNduals}
\ee
Consequently, condition $I$, Eq.~(\ref{eq:RNI}), corresponds to the extremal limit $T\rightarrow 0$. 
To analyze the response functions systematically, we consider the response matrix
\be
\begin{pmatrix}
dT\\
d\Psi_Q
\end{pmatrix}
=
\begin{pmatrix}
M_{,SS}&M_{,SQ}\\
M_{,SQ}&M_{,QQ}
\end{pmatrix}
\begin{pmatrix}
dS\\
dQ
\end{pmatrix}.
\label{eq:RNresponsematrix}
\ee
Then, the heat capacities at fixed charge and fixed electric potential are \cite{LadinoRomeroQuevedo2026}
\be
C_Q
=
T\left(\frac{\partial S}{\partial T}\right)_Q
=
\frac{M_{,S}}{M_{,SS}}
=
\frac{2S(S-\pi Q^2)}{3\pi Q^2-S},
\label{eq:RNCQ}
\ee
and
\be
C_{\Psi_Q}
=
T\left(\frac{\partial S}{\partial T}\right)_{\Psi_Q}
=
T\,
\frac{M_{,QQ}}
{M_{,SS}M_{,QQ}-M_{,SQ}^2}
=
-2S,
\label{eq:RNCphi}
\ee
respectively. The electric susceptibilities at fixed entropy and fixed temperature read
\be
\left(\frac{\partial Q}{\partial\Psi_Q}\right)_S
=
\frac{1}{M_{,QQ}}
=
\sqrt{\frac{S}{\pi}},
\label{eq:RNchiS}
\ee
and
\be
\left(\frac{\partial Q}{\partial\Psi_Q}\right)_T
=
\frac{M_{,SS}}
{M_{,SS}M_{,QQ}-M_{,SQ}^2}
=
\sqrt{\frac{S}{\pi}}\,
\frac{3\pi Q^2-S}{\pi Q^2-S}.
\label{eq:RNchiT}
\ee
The inverse mixed response is
\be
\left(\frac{\partial Q}{\partial T}\right)_{\Psi_Q}
=
\left(\frac{\partial S}{\partial\Psi_Q}\right)_T
=
-\frac{M_{,SQ}}
{M_{,SS}M_{,QQ}-M_{,SQ}^2}
=
\frac{4\sqrt{\pi}Q S^{3/2}}
{\pi Q^2-S}.
\label{eq:RNcrossinverse}
\ee
In addition, the direct thermo-electric coupling is
\be
\left(\frac{\partial T}{\partial Q}\right)_S
=
\left(\frac{\partial\Psi_Q}{\partial S}\right)_Q
=
M_{,SQ} 
=
-\frac{\sqrt{\pi}Q}{2S^{3/2}},
\label{eq:RNcrossdirect}
\ee

The physical interpretation of the three GTD conditions is now immediate. Condition $I$, Eq.~\eqref{eq:RNI}, coincides with extremality and with the divergences of the isothermal electric susceptibility, Eq.~\eqref{eq:RNchiT}, and the inverse mixed response, Eq.~\eqref{eq:RNcrossinverse}. Condition~$II$, Eq.~\eqref{eq:RNII}, corresponds precisely to the divergence of $C_Q$, Eq.~\eqref{eq:RNCQ}
\be
S_D=3\pi Q^2,
\label{eq:RNDavies}
\ee
which corresponds to the RN Davies spinodal instability \cite{Davies1978,LadinoRomeroQuevedo2026}. More generally, for asymptotically flat RN and Kerr black holes, the Davies point is associated with a divergence of the heat capacity and separates locally stable from unstable configurations, rather than signaling a phase transition with phase coexistence \cite{LadinoRomeroQuevedo2026,Romero-Figueroa:2026fan}. A first-order small/large black-hole transition, by contrast, is determined globally by the equality of the appropriate free energies \cite{CaldarelliCognolaKlemm2000}. Moreover, condition~$III$, Eq.~\eqref{eq:RNIII} is the Schwarzschild limit and simultaneously the state at which the direct thermo-electric coupling vanishes, Eq.~\eqref{eq:RNcrossdirect}. It remains to verify that the candidate GTD conditions are reflected in the actual curvature or degeneracy structure. A direct calculation gives \cite{LadinoRomeroQuevedo2026}
\be
{\cal R}_{\rm RN}^{I}
=
-\frac{8\pi^2Q^2S^2(\pi Q^2-3S)}
{\beta_M(\pi Q^2-S)^2(\pi Q^2+S)^3},
\label{eq:RNRcurvI}
\ee
\be
{\cal R}_{\rm RN}^{II}
=
-\frac{8\pi S^2(4\pi^2Q^4-5\pi Q^2S-S^2)}
{\beta_M(\pi Q^2+S)^3(3\pi Q^2-S)^2},
\label{eq:RNRcurvII}
\ee
and
\be
{\cal R}_{\rm RN}^{III}
=
-\frac{64\pi S^2(\pi Q^2+S)}
{\beta_M(13\pi^2Q^4-14\pi Q^2S+5S^2)^2}.
\label{eq:RNRcurvIII}
\ee
The numerator of ${\cal R}_{\rm RN}^{I}$ does not cancel the extremal factor, so $g^I$ is singular at $S=\pi Q^2$. Likewise, $g^{II}$ is singular precisely at the Davies locus. The denominator of ${\cal R}_{\rm RN}^{III}$ has no real zero for $S>0$, whereas
\be
\det g^{III}
=
-\frac{\beta_M^2Q^2
(13\pi^2Q^4-14\pi Q^2S+5S^2)}
{64S^4}
\ee
vanishes at $Q=0$. The third metric therefore identifies the disappearance of the thermo-electric coupling, Eq.~\eqref{eq:RNcrossdirect} rather as a curvature singularity.

The RN example illustrates the complementary information carried by the three GTD metrics. The second metric reproduces the known Davies instability, while the first metric geometrically identifies extremality together with a singular electric-response sector. The third metric distinguishes the Schwarzschild limit through the vanishing of the mixed thermo-electric response. These structures are thermodynamically different and should not be grouped under a single notion of phase transition.

\subsection{Kerr black hole}

For the asymptotically flat Kerr black hole the fundamental equation is \cite{LadinoRomeroQuevedo2024}
\be
M(S,J)
=
\sqrt{\frac{S^2+4\pi^2J^2}{4\pi S}}.
\label{eq:MKerr}
\ee
Its quasihomogeneity condition gives $\beta_S=2\beta_M$ and $\beta_J=2\beta_M$. Substituting these coefficients directly into Euler's identity, Eq.~\eqref{EulerID}, yields $2\beta_M S M_{,S}+2\beta_M J M_{,J}=\beta_M M$, or equivalently $2S M_{,S}+2J M_{,J}=M$. The conformal factor is therefore $\Sigma=\beta_M M$. The first derivatives are
\be
M_{,S}
=
\frac{S^2-4\pi^2J^2}
{4\sqrt{\pi}S^{3/2}\sqrt{S^2+4\pi^2J^2}},
\qquad
M_{,J}
=
\frac{2\pi^{3/2}J}
{\sqrt S\sqrt{S^2+4\pi^2J^2}},
\label{eq:Kerrfirstderivatives}
\ee
and the second derivatives are
\be
M_{,SS}
=
\frac{
48\pi^4J^4+24\pi^2J^2S^2-S^4
}{
8\sqrt{\pi}S^{5/2}(S^2+4\pi^2J^2)^{3/2}
},
\label{eq:KerrMSS}
\ee
\be
M_{,SJ}
=
-\frac{
\pi^{3/2}J(4\pi^2J^2+3S^2)
}{
S^{3/2}(S^2+4\pi^2J^2)^{3/2}
},
\qquad
M_{,JJ}
=
\frac{
2\pi^{3/2}S^{3/2}
}{
(S^2+4\pi^2J^2)^{3/2}
}.
\label{eq:Kerrsecondderivatives}
\ee
Then, the GTD metrics Eqs.~(\ref{gI2D})--(\ref{gIII2D}) can be written as
\be
\begin{split}
g^I={}&
\frac{\beta_M
(48\pi^4J^4+24\pi^2J^2S^2-S^4)}
{16\pi S^3(S^2+4\pi^2J^2)}
\,dS^2
\\
&-
\frac{\pi\beta_MJ(4\pi^2J^2+3S^2)}
{S^2(S^2+4\pi^2J^2)}
\,dS\,dJ
+
\frac{\pi\beta_M S}
{S^2+4\pi^2J^2}
\,dJ^2 ,
\end{split}
\label{eq:KerrgI}
\ee
\be
g^{II}
=
-\frac{\beta_M
(48\pi^4J^4+24\pi^2J^2S^2-S^4)}
{16\pi S^3(S^2+4\pi^2J^2)}
\,dS^2
+
\frac{\pi\beta_M S}
{S^2+4\pi^2J^2}
\,dJ^2 ,
\label{eq:KerrgII}
\ee
\be
\begin{split}
g^{III}={}&
\frac{\beta_M(S^2-4\pi^2J^2)
(48\pi^4J^4+24\pi^2J^2S^2-S^4)}
{16\pi S^3(S^2+4\pi^2J^2)^2}
\,dS^2
\\
&-
\frac{\pi\beta_MJ(4\pi^2J^2+3S^2)}
{2S^2(S^2+4\pi^2J^2)}
\,dS\,dJ
+
\frac{8\pi^3\beta_MJ^2S}
{(S^2+4\pi^2J^2)^2}
\,dJ^2 .
\end{split}
\label{eq:KerrgIII}
\ee
The three equilibrium conditions, Eqs.~(\ref{condI})--(\ref{condIII}), can now be evaluated
\be
I: \quad M_{,SS}M_{,JJ}-M_{,SJ}^2
=
-\frac{\pi}{4S^3} = 0 ,
\label{eq:KerrIeval}
\ee
\be
II:\quad M_{,SS}M_{,JJ}
=
\frac{
\pi\left(
48\pi^4J^4+24\pi^2J^2S^2-S^4
\right)
}{
4S(S^2+4\pi^2J^2)^3 
}=0,
\label{eq:KerrII}
\ee
\be
III: \quad M_{,SJ}
=
-\frac{
\pi^{3/2}J(4\pi^2J^2+3S^2)
}{
S^{3/2}(S^2+4\pi^2J^2)^{3/2}
} =0. \label{eq:KerrIII}
\ee
It follows that condition $I$, Eq.~(\ref{eq:KerrIeval}), cannot be satisfied in general, and condition $III$, Eq.~(\ref{eq:KerrIII}), implies the vanishing of the angular momentum, $J=0$. We now consider the first law 
\be
dM=T\,dS+\Omega\,dJ,
\label{eq:Kerrfirstlaw}
\ee
with dual variables
\be
T = M_{,S}
=
\frac{S^2-4\pi^2J^2}
{4\sqrt{\pi}S^{3/2}\sqrt{S^2+4\pi^2J^2}},
\qquad
\Omega = M_{,J}
=
\frac{2\pi^{3/2}J}
{\sqrt S\sqrt{S^2+4\pi^2J^2}}.
\label{eq:Kerrduals}
\ee
Extremality is given by $S^2=4\pi^2J^2$ and is distinct from all three conditions above. The response matrix is
\be
\begin{pmatrix}
dT\\
d\Omega
\end{pmatrix}
=
\begin{pmatrix}
M_{,SS}&M_{,SJ}\\
M_{,SJ}&M_{,JJ}
\end{pmatrix}
\begin{pmatrix}
dS\\
dJ
\end{pmatrix},
\label{eq:Kerrresponsematrix}
\ee
and can be used to calculate the heat capacities at fixed angular momentum and angular velocity, which read \cite{LadinoRomeroQuevedo2026}
\be
C_J = T\left(\frac{\partial S}{\partial T}\right)_J
=
\frac{M_{,S}}{M_{,SS}}
=
\frac{
2S(S^2-4\pi^2J^2)(S^2+4\pi^2J^2)
}{
48\pi^4J^4+24\pi^2J^2S^2-S^4
},
\label{eq:KerrCJ}
\ee
and
\be
C_\Omega
=
T\left(\frac{\partial S}{\partial T}\right)_\Omega
=
\frac{2S^3(4\pi^2J^2-S^2)}
{(S^2+4\pi^2J^2)^2},
\label{eq:KerrCOmega}
\ee
respectively. 
The rotational susceptibilities at fixed entropy and fixed temperature are
\be
\left(\frac{\partial J}{\partial\Omega}\right)_S
=
\frac{1}{M_{,JJ}}
=
\frac{(S^2+4\pi^2J^2)^{3/2}}
{2\pi^{3/2}S^{3/2}},
\label{eq:KerrchiS}
\ee
and
\be
\left(\frac{\partial J}{\partial\Omega}\right)_T
=
\frac{M_{,SS}}
{M_{,SS}M_{,JJ}-M_{,SJ}^2}
=
-\frac{
\sqrt S
(48\pi^4J^4+24\pi^2J^2S^2-S^4)
}{
2\pi^{3/2}(S^2+4\pi^2J^2)^{3/2}
}.
\label{eq:KerrchiT}
\ee
Furthermore, the direct thermo-rotational coupling is
\be
\left(\frac{\partial T}{\partial J}\right)_S
=
\left(\frac{\partial\Omega}{\partial S}\right)_J
=
M_{,SJ}=
-\frac{
\pi^{3/2}J(4\pi^2J^2+3S^2)
}{
S^{3/2}(S^2+4\pi^2J^2)^{3/2}
} .
\label{eq:Kerrcrossdirect}
\ee
We see that condition $II$, Eq.~(\ref{eq:KerrII}), is precisely the divergence of $C_J$ and hence determines the Kerr Davies instability \cite{Davies1978,LadinoRomeroQuevedo2026}. Solving Eq.~(\ref{eq:KerrII}) for the physical branch gives
\be
S_D = 2\pi J \sqrt{3 + 2\sqrt{3}}
\label{eq:KerrDavies}
\ee
Furthermore, condition~$III$, Eq.~\eqref{eq:KerrIII}, as in the RN case, corresponds to the Schwarzschild limit, where the thermo-rotational coupling, Eq.~\eqref{eq:Kerrcrossdirect}, vanishes. Condition $I$, Eq.~\eqref{eq:KerrIeval} has no physical zero and therefore does not define a response-function singularity in the Kerr domain. The actual curvature calculation confirms this interpretation. For the first metric Eq.~\eqref{eq:KerrgI}
\be
{\cal R}_{\rm Kerr}^{I}=0.
\label{eq:KerrRI}
\ee
Thus, $g^I$ is flat and has no critical points. For the second metric Eq.~\eqref{eq:KerrgII}
\be
{\cal R}_{\rm Kerr}^{II}
=
-\frac{
8\pi(S^2+4\pi^2J^2)(S^2+12\pi^2J^2)
(48\pi^4J^4+24\pi^2J^2S^2-5S^4)
}{
\beta_M S
(48\pi^4J^4+24\pi^2J^2S^2-S^4)^2
},
\label{eq:KerrRII}
\ee
so its curvature diverges exactly at the Davies point, Eq.~\eqref{eq:KerrDavies}. Additionally, the scalar curvature vanishes at the regular physical zero-curvature condition
\be
48\pi^4J^4+24\pi^2J^2S^2-5S^4=0,
\label{eq:KerrRIIzero}
\ee
which does not correspond to a phase transition, but rather separates regions with opposite signs of the curvature, as discussed in more detail in Sec.~\ref{sec:zero_temperature}. Finally, the third scalar curvature is
\be
{\cal R}_{\rm Kerr}^{III}
=
-\frac{
768\pi S^3(S^2+4\pi^2J^2)^4
(80\pi^4J^4-32\pi^2J^2S^2+11S^4)
}{
\beta_M
\left(
256\pi^8J^8+2048\pi^6J^6S^2
+736\pi^4J^4S^4
-128\pi^2J^2S^6
+17S^8
\right)^2
}.
\label{eq:KerrRIII}
\ee
It has no curvature singularity in the physical Kerr domain. Nevertheless,
\be
\det g^{III}\propto J^2,
\ee
so the metric becomes degenerate at $J=0$, consistently with the disappearance of the mixed thermo-rotational response, Eq.~\eqref{eq:Kerrcrossdirect}. Thus, the Kerr case exhibits a more selective geometric response. The second GTD metric reproduces the known Davies instability and also reveals an additional zero-curvature state. The first metric is flat and therefore does not capture thermodynamic instabilities, while the third metric is curved for rotating states and distinguishes the Schwarzschild limit through the disappearance of the thermo-rotational coupling.

\subsection{Kerr--AdS black hole}
\label{Kerr-AdS sec}
The Kerr--AdS black hole introduces the AdS curvature radius $l$ and therefore requires a three-dimensional equilibrium space. We use the fundamental equation \cite{LadinoRomeroQuevedo2025}
\be
M(S,J,l)=
\frac{
\sqrt{S+\pi l^2}\,
\sqrt{\pi l^2S^2+S^3+4\pi^3J^2l^2}
}{
2\pi^{3/2}l^2\sqrt S
}.
\label{eq:MKerrAdS}
\ee
By applying the condition of quasihomogeneity, it can be seen that the above function does not satisfy this condition unless the radius of curvature $l$ is considered thermodynamic variable \cite{romero2024extended1,QuevedoQuevedoSanchez2016Phantom}. This fact was previously  established by considering the cosmological constant $\Lambda$ thermodynamic variable associated with the pressure, giving rise to what is now known as extended thermodynamics and black hole chemistry \cite{KubiznakMann2012}. Consequently, to apply quasihomogeneous GTD to the Kerr-AdS black hole we will treat the radius of curvature as an additional independent variable.

It is easy to see that the quasihomogeneous coefficients satisfy $\beta_S=2\beta_M$, $\beta_J=2\beta_M$, and $\beta_l=\beta_M$. Substituting these relations into Euler's identity, Eq.~\eqref{EulerID}, gives $2\beta_M S M_{,S}+2\beta_M J M_{,J}+\beta_M l M_{,l}=\beta_M M$, or equivalently $2S M_{,S}+2J M_{,J}+lM_{,l}=M$, which is satisfied identically. Hence the conformal factor is $\Sigma=\beta_M M$.

Choosing the coordinates of the equilibrium space as $E^a= (S,J,l)$, the GTD metrics take the form \cite{LadinoRomeroQuevedo2025}
\be
\begin{split}
g^I=\beta_M M\big[
&M_{,SS}dS^2
+2M_{,SJ}dS\,dJ
+2M_{,Sl}dS\,dl
+M_{,JJ}dJ^2
\\
&+2M_{,Jl}dJ\,dl
+M_{,ll}dl^2
\big],
\end{split}
\label{eq:KerrAdSgI}
\ee
\be
g^{II}
=
\beta_M M\big[
-M_{,SS}dS^2
+M_{,JJ}dJ^2
+2M_{,Jl}dJ\,dl
+M_{,ll}dl^2
\big],
\label{eq:KerrAdSgII}
\ee
\be
\begin{split}
g^{III}={}&
2\beta_M S M_{,S}M_{,SS}\,dS^2
+
2\beta_M J M_{,J}M_{,JJ}\,dJ^2
+
\beta_M l M_{,l}M_{,ll}\,dl^2
\\
&+
2\beta_M(SM_{,S}+JM_{,J})M_{,SJ}\,dS\,dJ
\\
&+
\beta_M(2SM_{,S}+lM_{,l})M_{,Sl}\,dS\,dl
\\
&+
\beta_M(2JM_{,J}+lM_{,l})M_{,Jl}\,dJ\,dl .
\end{split}
\label{eq:KerrAdSgIII}
\ee
All derivatives are obtained directly from Eq.~\eqref{eq:MKerrAdS}; the six second derivatives are omitted here for brevity, as their explicit forms obscure the factorized equilibrium structure. They are given in the Appendix of \cite{LadinoRomeroQuevedo2025}. For $g^I$, the characteristic condition is the degeneracy of the complete Hessian
\be
\begin{split}
I:\quad
0={}&
M_{,SS}\left(M_{,Jl}^2-M_{,JJ}M_{,ll}\right)
+
M_{,JJ}M_{,Sl}^2
+
M_{,ll}M_{,SJ}^2
\\
&-
2M_{,SJ}M_{,Sl}M_{,Jl}.
\end{split}
\label{eq:KerrAdSI}
\ee
Direct substitution of the fundamental relation, Eq.~(\ref{eq:MKerrAdS}), shows that condition $I$, Eq.~(\ref{eq:KerrAdSI}), cannot be satisfied in the black-hole regime \cite{LadinoRomeroQuevedo2025}. In fact, after extracting the positive denominator, its numerator is a sum of positive terms proportional to nonnegative even powers of $J$ and positive powers of $S$ and $l$. Furthermore, the second metric, Eq.~(\ref{eq:KerrAdSgII}), gives
\be
II:\qquad
M_{,SS}
\left(
M_{,Jl}^2-M_{,JJ}M_{,ll}
\right)
=0.
\label{eq:KerrAdSII}
\ee
It therefore contains two branches,
\be
M_{,SS}=0,
\label{eq:KerrAdSIIa}
\ee
and
\be
M_{,Jl}^2-M_{,JJ}M_{,ll}=0.
\label{eq:KerrAdSIIb}
\ee
The first branch can be written explicitly as
\be
\begin{split}
0={}&
64\pi^7J^4Sl^6
+48\pi^8J^4l^8
+48\pi^3J^2S^5l^2
+120\pi^4J^2S^4l^4
\\
&+
96\pi^5J^2S^3l^6
+24\pi^6J^2S^2l^8
+3S^8
+8\pi S^7l^2
+6\pi^2S^6l^4
-\pi^4S^4l^8 .
\end{split}
\label{eq:KerrAdSMSSzero}
\ee
The second branch is equivalent to
\be
\begin{split}
0={}&
-8\pi^6J^4l^4
+6\pi^3J^2S^3l^2
+12\pi^4J^2S^2l^4
+6\pi^5J^2Sl^6
\\
&+
3S^6
+9\pi S^5l^2
+9\pi^2S^4l^4
+3\pi^3S^3l^6 .
\end{split}
\label{eq:KerrAdSsusczero}
\ee
For the third metric, Eq.~(\ref{eq:KerrAdSgIII}), the independent three-dimensional equilibrium conditions require one of the following simultaneous sets:
\be
M_{,SS}=M_{,SJ}=M_{,Sl}=0,
\label{eq:KerrAdSIIIa}
\ee
\be
M_{,SJ}=M_{,JJ}=0,
\label{eq:KerrAdSIIIb}
\ee
\be
M_{,Sl}=M_{,ll}=0,
\label{eq:KerrAdSIIIc}
\ee
or
\be
M_{,JJ}=M_{,ll}=0.
\label{eq:KerrAdSIIId}
\ee
For the fundamental relation, Eq.~(\ref{eq:MKerrAdS}), $M_{,JJ}>0$ and $M_{,ll}>0$ in the regular domain, while $M_{,Sl}$ does not vanish there. Hence none of the sets Eqs.~(\ref{eq:KerrAdSIIIa})--(\ref{eq:KerrAdSIIId}) generates an additional physical condition \cite{LadinoRomeroQuevedo2025}. We next determine the thermodynamic meaning of these GTD conditions. For the Kerr--AdS black hole, the extended thermodynamic first law takes the form
\be
dM=T\,dS+\Omega\,dJ+\Psi_l\,dl,
\label{eq:KerrAdSfirstlaw}
\ee
where
\be
T=M_{,S}
=
\frac{
3S^4+4\pi S^3l^2+\pi^2S^2l^4-4\pi^4J^2l^4
}{
4\pi^{3/2}S^{3/2}l^2
\sqrt{S+\pi l^2}
\sqrt{S^3+\pi S^2l^2+4\pi^3J^2l^2}
},
\label{eq:KerrAdST}
\ee
\be
\Omega=M_{,J}
=
\frac{
2\pi^{3/2}J\sqrt{S+\pi l^2}
}{
\sqrt S
\sqrt{S^3+\pi S^2l^2+4\pi^3J^2l^2}
},
\label{eq:KerrAdSOmega}
\ee
and
\be
\Psi_l=M_{,l}
=
-\frac{
\sqrt S
\left(
2\pi^3J^2l^2+S^3+\pi S^2l^2
\right)
}{
\pi^{3/2}l^3
\sqrt{S+\pi l^2}
\sqrt{S^3+\pi S^2l^2+4\pi^3J^2l^2}
}.
\label{eq:KerrAdSPsi}
\ee

The complete linear response matrix is
\be
\begin{pmatrix}
dT\\
d\Omega\\
d\Psi_l
\end{pmatrix}
=
\begin{pmatrix}
M_{,SS}&M_{,SJ}&M_{,Sl}\\
M_{,SJ}&M_{,JJ}&M_{,Jl}\\
M_{,Sl}&M_{,Jl}&M_{,ll}
\end{pmatrix}
\begin{pmatrix}
dS\\
dJ\\
dl
\end{pmatrix}.
\label{eq:KerrAdSresponsematrix}
\ee
Condition~$I$, Eq.~\eqref{eq:KerrAdSI}, is therefore the loss of invertibility of the complete response matrix. Since it has no generic zero in the physical domain, the generalized susceptibilities obtained from the full inverse Hessian do not develop a corresponding interior singularity. At fixed $J$ and $l$, the heat capacity is
\be
C_{lJ}
=
T\left(\frac{\partial S}{\partial T}\right)_{l,J}
=
\frac{M_{,S}}{M_{,SS}},
\label{eq:KerrAdSClJ}
\ee
and hence diverges precisely on the branch Eq.~(\ref{eq:KerrAdSIIa}). At fixed $S$ and $l$, and at fixed $S$ and $J$, respectively, the elementary susceptibilities are
\be
\left(\frac{\partial J}{\partial\Omega}\right)_{S,l}
=
\frac{1}{M_{,JJ}},
\qquad
\left(\frac{\partial l}{\partial\Psi_l}\right)_{S,J}
=
\frac{1}{M_{,ll}}.
\label{eq:KerrAdSelementaryresp}
\ee
At fixed $S$ and $\Omega$, the coupled rotational--AdS susceptibility is
\be
\left(\frac{\partial l}{\partial\Psi_l}\right)_{S,\Omega}
=
\frac{M_{,JJ}}
{M_{,JJ}M_{,ll}-M_{,Jl}^2},
\label{eq:KerrAdSkappa}
\ee
and it diverges on the second branch, Eq.~(\ref{eq:KerrAdSIIb}). The remaining responses in this $(J,l)$ sector have the same denominator
\be
\left(\frac{\partial J}{\partial\Omega}\right)_{S,\Psi_l}
=
\frac{M_{,ll}}
{M_{,JJ}M_{,ll}-M_{,Jl}^2},
\qquad
\left(\frac{\partial J}{\partial\Psi_l}\right)_{S,\Omega}
=
-\frac{M_{,Jl}}
{M_{,JJ}M_{,ll}-M_{,Jl}^2}.
\label{eq:KerrAdScoupledresponses}
\ee
Thus, the two factors of condition~$II$, Eq.~(\ref{eq:KerrAdSII}), have an immediate response-function interpretation. The first branch, defined by $M_{,SS}=0$, corresponds to the canonical Davies-type spinodal points, as shown in the thermodynamic phase diagram for Kerr-AdS black holes in Fig.~\ref{fig:InicialFigures}(b). At fixed $J$ and $l$, two such points can occur, corresponding to a local maximum $(S_m,T_m)$ and a local minimum $(S_{min},T_{min})$ in the $T$--$S$ plane. At the critical condition, given by Eq.~\eqref{critical point KerrAdS}, these two points merge into a single inflection point $(S_c,T_c)$, characterized by
\be
M_{,SS}=0,
\qquad
M_{,SSS}=0.
\label{eq:KerrAdScritical}
\ee
By solving these conditions, we can obtain the numerical critical values
\be
S_c \approx 28.718873\, J, \qquad l_c \approx 6.464017\, \sqrt{J}, \qquad T_c \approx 0.041749\, J^{-1/2}, \qquad M_c \approx 1.878321\, \sqrt{J}.
\label{eq:KerrAdScriticalcantidades}
\ee
The second branch of condition~$II$, Eq.~(\ref{eq:KerrAdSIIb}), corresponds to the coupled rotational--AdS susceptibility, but the solutions found in Ref.~\cite{LadinoRomeroQuevedo2025} lie in the region $\Omega l\geq1$ and are excluded by the physical restriction $\Omega l<1$. In summary, Kerr--AdS exhibits a clear GTD selection: condition~$I$ does not generate a generic physical singularity, while condition~$III$ has no physical solution in the regular domain. Condition~$II$ contains both the canonical thermal spinodals and a coupled susceptibility sector. The physically relevant canonical critical structure is encoded in the factor $M_{,SS}$, whereas the first-order transition remains a global thermodynamic phenomenon \cite{LadinoRomeroQuevedo2026,CaldarelliCognolaKlemm2000}. GTD thus organizes the possible equilibrium conditions within a single geometric framework, distinguishing the sectors compatible with the physical domain from those excluded by its restrictions.

\subsection{Comparison with thermodynamic geometry}

The preceding calculations modify the comparison with thermodynamic geometry in an important way. The three GTD metrics should be compared as complementary geometries constructed from the same fundamental equation, rather than selected retrospectively so that each reproduces a predetermined response function.

For the RN black hole, the standard entropy-based thermodynamic geometry is flat in the usual two-dimensional state space and does not reproduce the fixed-charge Davies instability \cite{Ruppeiner2008}. The GTD description separates three structures. The first metric is curved and becomes singular at extremality, where an electric-response sector also diverges. The second metric reproduces the Davies singularity of $C_Q$. The third metric has no curvature singulatity in the charged physical domain but becomes degenerate in the Schwarzschild limit, precisely where the direct thermo-electric coupling vanishes. GTD therefore supplies a broader geometric classification than one based only on the fixed-charge heat capacity.

The Kerr case is complementary. Thermodynamic geometry gives a nontrivial curvature whose singularity structure depends on the thermodynamic state space and ensemble \cite{Ruppeiner2008}. In GTD, the first metric constructed directly from $M(S,J)$ is flat and has no physical equilibrium condition of type~$I$. The second metric reproduces the fixed-$J$ Davies instability through a curvature singularity. In addition, its curvature vanishes at a regular thermodynamic state, which in the effective microscopic interpretation corresponds to a non-interacting configuration \cite{Ruppeiner1995}. The third metric is curved for rotating states, has no Davies point, and degenerates when $J=0$, where the thermo-rotational coupling disappears. Thus, the three GTD metrics distinguish curvature, response singularities, and the vanishing of mixed thermodynamic coupling as different pieces of information.

For Kerr--AdS, thermodynamic geometry can reproduce ensemble-dependent instabilities, critical behavior, and effective interaction sectors once the appropriate fluctuation space is specified \cite{Sahay2017,WeiLiu2021,DehyadegariSheykhi2021}. The main GTD distinction is structural. Quasihomogeneity requires the complete thermodynamic set $(S,J,l)$, Euler's identity fixes the conformal factor, and the three GTD conditions separate the determinant of the full response matrix, the canonical thermal and coupled $(J,l)$ response sectors, and the possible simultaneous degeneracies associated with the third metric. In the physical canonical domain analyzed in Ref.~\cite{LadinoRomeroQuevedo2025}, only the thermal branch of condition~$II$ supplies the relevant local critical structure. Another important distinction, illustrated by Schwarzschild--AdS, is that the thermodynamic geometry is flat in the $(S,l)$ space \cite{xu2020ruppeiner}, while a nontrivial curvature requires reinterpreting the cosmological constant as the thermodynamic pressure \cite{KubiznakMann2012}. In GTD, by contrast, the AdS scale $l$ can be treated directly as a thermodynamic variable, as in the present Kerr--AdS analysis. Moreover, the GTD curvature remains nonzero in the asymptotically flat limit $l \to \infty$ of AdS black holes, consistently describing these systems as interacting without introducing an alternative thermodynamic representation \cite{dehyadegari2017critical}.

These results, together with the analyses of RN, Kerr, and Kerr--AdS black holes, show that the three GTD metrics need not identify the same thermodynamic singularities, but instead probe complementary equilibrium structures. The second metric reproduces the Davies or Davies-type thermal response singularity in all three examples considered here. The first metric probes the determinant of the full response matrix and, depending on the system, can identify extremality or remain regular. The third metric is sensitive to mixed thermodynamic couplings and dimensional reductions, or can have no admissible singularity in the physical domain. This complementary structure is the information that will later be represented in terms of black-hole shadows. We now turn to the corresponding zero-curvature loci, which provide another perspective on the thermodynamic structure.



\subsection{Non-interacting thermodynamic states}
\label{sec:zero_temperature}

The geometric description of thermodynamic interactions is closely
related to the microscopic correlations encoded in thermodynamic
curvature. In thermodynamic geometry, the scalar curvature scales with
the correlation volume of the underlying microscopic degrees of
freedom, $R\sim\xi^d$, where $\xi$ is the correlation length and $d$
is the spatial dimensionality \cite{Ruppeiner1995}. Thus, a vanishing
thermodynamic curvature can be associated with the vanishing of the
corresponding effective microscopic correlations. A similar
geometric interpretation can be pursued within the framework of GTD. In our previous
analysis of real-fluid models, we investigated the temperature at
which the GTD scalar curvature $\mathcal{R}$ vanishes, and found a near-universal
description when this zero-curvature temperature is expressed in
terms of the Boyle temperature scale for different fluid models
\cite{romero2026}. Motivated by this connection, we explore here the corresponding
non-interacting states (N-states) in black-hole thermodynamics, defined by $\mathcal{R}=0$. 

\begin{figure}[htbp]
\centering
\includegraphics[
    width=0.92\textwidth
]{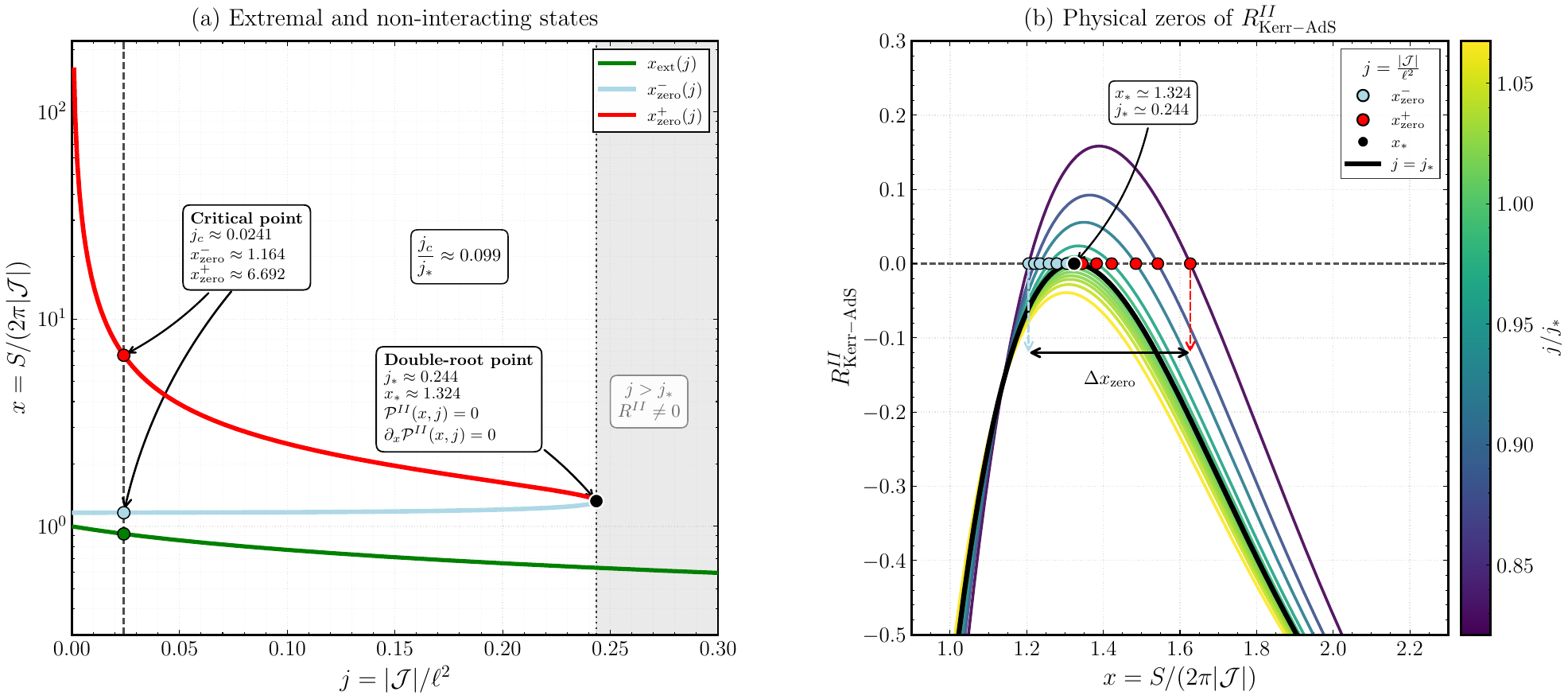}
\caption{
(a) Extremal and zero-curvature loci for the Kerr--AdS black hole in the
$(x,j)$ plane. The two zero-curvature branches merge at the
double-root point $(x_*,j_*)\simeq(1.324,0.244)$. The red and blue dots
indicate the critical points evaluated along the corresponding
zero-curvature branches, while the black dot marks the branch merger.
(b) The scalar curvature $R^{II}_{\rm Kerr-AdS}$ as a function of $x$
for representative values of $j$, illustrating the disappearance of
physical N-states for $j>j_*$.}
\label{fig:xext_xzero_KerrAdS}
\end{figure}

\begin{figure}[htbp]
\centering
\includegraphics[
    width=0.92\textwidth
]{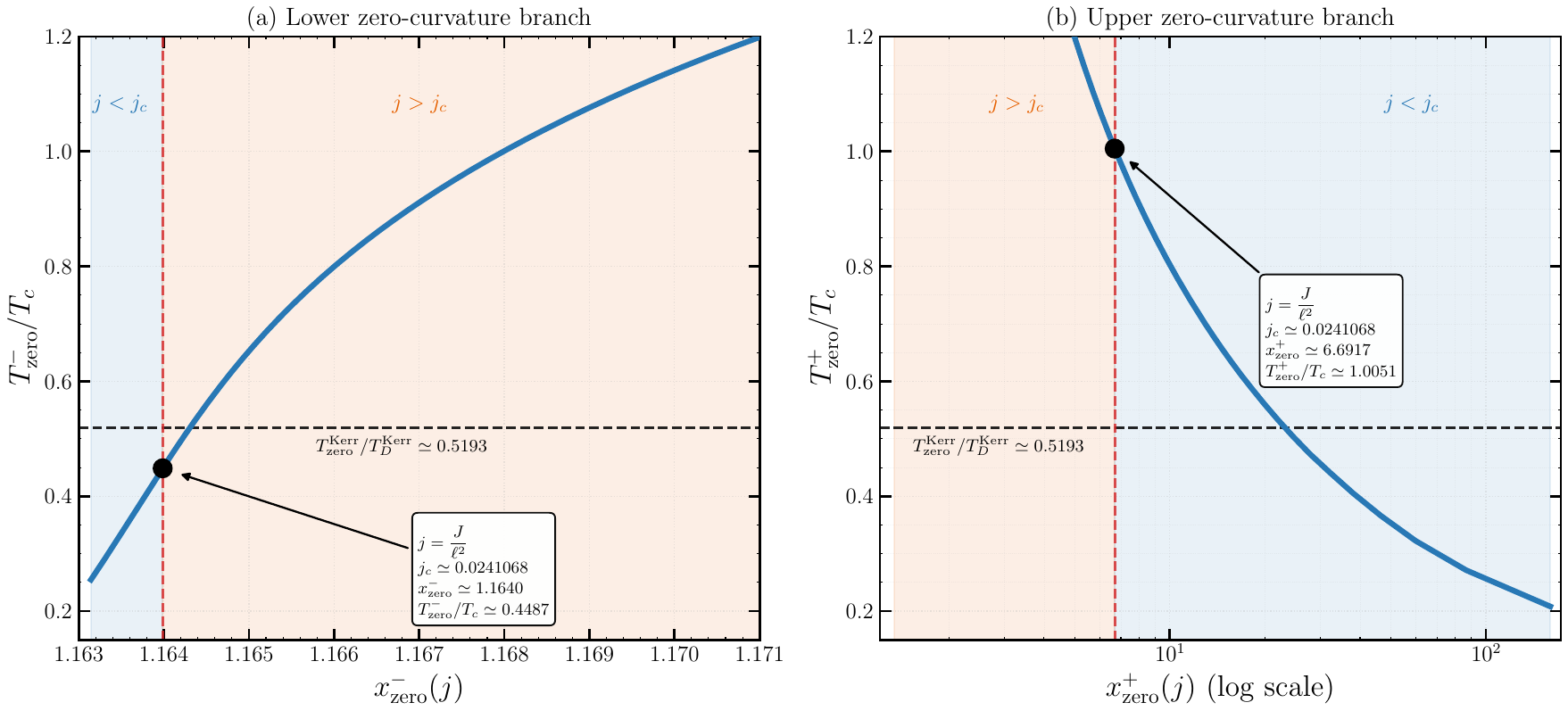}
\caption{
Dimensionless zero-curvature temperatures for the Kerr--AdS black hole.
The two panels show (a) $T_{\rm zero}^{-}/T_c$ and (b)
$T_{\rm zero}^{+}/T_c$ along the two physical branches of
$\mathcal{R}^{II}=0$, parametrized by $j=|J|/l^2$. The critical
configuration $j_c\simeq0.0241068$ is indicated by the corresponding
point on each branch, while
the horizontal black dashed line provides the asymptotically flat Kerr black hole
reference value given in Eq.~\eqref{Kerr ratio}.
}
\label{fig:Tzero_ratio_KerrAdS}
\end{figure}

Next, for simplicity, we focus on $\mathcal{R}^{II}$. For the asymptotically flat RN black hole, the scalar curvature given in
Eq.~\eqref{eq:RNRcurvII} has no physical zeros. Thus, N-states do not exist in this case. For the Kerr
black hole family, we introduce the dimensionless parameter $x=S/(2\pi |J|)$, with physical domain $x>1$, where $x=1$ corresponds to the extremal Kerr solution, for which the Hawking
temperature given in Eq.~\eqref{eq:Kerrduals} vanishes. Using the condition in Eq.~\eqref{eq:KerrRIIzero}, which in terms of
the dimensionless parameter $x$ reduces to
\begin{equation}
3+6x^2-5x^4=0
\quad\Longrightarrow\quad
x_{\rm zero}^{Kerr}
=
\sqrt{\frac{3+2\sqrt{6}}{5}}
\simeq1.257. \label{xzerokerr}
\end{equation}
Evaluating the Hawking temperature Eq.~\eqref{eq:Kerrduals} at
this root gives
$T_{\rm zero}^{\rm Kerr}\simeq0.0144147\,|J|^{-1/2}$.
On the other hand, the zero of the denominator of
Eq.~\eqref{eq:KerrRII} determines the Davies temperature,
$T_D^{\rm Kerr}\simeq0.0277603\,|J|^{-1/2}$. Interestingly, both
$x_{\rm zero}$ and $x_D$ are constant for the Kerr black hole,
leading to a universal ratio of the corresponding temperatures
\begin{equation}
\frac{T_{\rm zero}^{\rm Kerr}}
{T_D^{\rm Kerr}} \label{Kerr ratio}
\simeq0.5193.
\end{equation}
Next, to analyze the Kerr--AdS solution, it is useful to introduce the additional dimensionless parameter
$j=|J|/l^2$, in terms of which the Hawking temperature Eq.~\eqref{eq:KerrAdST}, can be written as
\begin{equation}
T^{Kerr-AdS}=
\frac{1}{8\pi\sqrt{|J|}}
\frac{
1-x^{-2}+8jx+12j^2x^2
}{
\left[
\dfrac{x^2+1}{2x}
+j+2jx^2+2j^2x^3
\right]^{1/2}
}.
\label{eq:KerrAdS_temperature}
\end{equation}
The physical black hole region in which $T\geq0$ is determined by $12j^2x^4+8jx^3+x^2-1\geq0$, with equality corresponding to the extremal black hole limit. Furthermore, using the results presented in \cite{LadinoRomeroQuevedo2025}, it can be shown that the thermodynamic critical point cannot be obtained analytically in closed form. However, in the small-rotation regime, the critical quantities can be obtained perturbatively. In terms of the dimensionless variables introduced above, the thermodynamic critical point is located at
\begin{equation}
x^{Kerr-AdS}_c \simeq 4.5059,
\qquad
j^{Kerr-AdS}_c \simeq 0.0241068, \qquad T_c^{\rm Kerr-AdS}
\simeq0.0418736\,|J|^{-1/2}. \label{critical point KerrAdS}
\end{equation}
Using the complete expression for $\mathcal{R}^{II}$ presented in
\cite{LadinoRomeroQuevedo2025}, the zero-curvature condition, which is
too lengthy to be written explicitly here, will be denoted by
$\mathcal{P}^{II}(x,j)$ and yields two physical solutions\footnote{A third mathematical root also appears;
however, it corresponds to states with $T<0$ and is therefore excluded
from the present work.} namely, $x_{\rm zero}^{-}$ and
$x_{\rm zero}^{+}$, as illustrated in
Fig.~\ref{fig:xext_xzero_KerrAdS}(a). The superscripts $-$ and $+$ correspond to the small black hole (SBH)
and large black hole (LBH) regimes, respectively; see
Fig.~\ref{fig:InicialFigures}(b), which displays the different
thermodynamic phases of the Kerr--AdS black hole. Moreover, Fig.~\ref{fig:xext_xzero_KerrAdS}(b) illustrates the behavior
of the zero-curvature points as a function of $j$. As $j$ increases,
the separation between the two physical branches decreases, and they
eventually merge at $j=j_*$. This behavior is particularly relevant
for the subsequent analysis of black hole shadows. Since
$j=|J|/l^2$, increasing $l$ at fixed $J$ corresponds to moving
towards smaller values of $j$. In this regime, the LBH
zero-curvature branch moves into the photon-region domain and is
therefore not observable in the shadow analysis, whereas the SBH
zero-curvature branch remains outside the photon region. 
In particular, Fig.~\ref{fig:xext_xzero_KerrAdS}(a) shows the evolution of the extremal black hole curve and the two physical zero-curvature branches. The lower and upper branches, remain separated over a finite range of $j$ and eventually merge at a double-root point, where the conditions
\begin{equation}
 \mathcal{P}^{II}(x_*,j_*)=0, \qquad
\partial_x\mathcal{P}^{II}(x_*,j_*)=0
\end{equation}
are simultaneously satisfied. Solving these equations numerically yields the merger point  $j_*\simeq0.244, x_*\simeq1.324$. For $j>j_*$, the Kerr--AdS spacetime exhibits a regime without non-interacting thermodynamic states, as indicated by the gray region Fig.~\ref{fig:xext_xzero_KerrAdS}(a). Thus, the finite merger point $j_*$ is a distinctive feature of Kerr--AdS and has no counterpart in asymptotically flat Kerr, RN, or RN--AdS black holes.\footnote{For RN--AdS, it can be shown that only one physical zero-curvature branch exists; see \cite{LadinoRomeroQuevedo2024} for details.}. It would be interesting to investigate whether this merger phenomenon persists in other black hole solutions, and whether the ratio $j_c/j_*$ exhibits a similar behavior. For Kerr--AdS, this ratio is found to be $j_c/j_* \simeq 0.099$. Furthermore, the corresponding zero-curvature temperatures are obtained by
evaluating the Hawking temperature Eq.~\eqref{eq:KerrAdS_temperature}
along each branch. Fig.~\ref{fig:Tzero_ratio_KerrAdS} shows the
dimensionless ratio $T_{\rm zero}/T_c$, where $T_c$ is the critical
temperature of the Kerr--AdS solution given in
Eq.~\eqref{critical point KerrAdS}.
At criticality, the temperatures satisfy
\begin{equation}
\left.
\frac{T_{\rm zero}^{-}}{T_c}
\right|_{j_c}
\simeq0.4487,
\qquad
\left.
\frac{T_{\rm zero}^{+}}{T_c}
\right|_{j_c}
\simeq1.0051.
\end{equation}

\begin{table}[htbp]
\centering
\renewcommand{\arraystretch}{1.4}
\caption{
Summary of the physical non-interacting states. For the asymptotically flat Kerr black hole, $T_c$ denotes the temperature at the Davies point, whereas for Kerr--AdS it denotes the critical temperature. The Kerr--AdS values are evaluated at the critical configuration $j=j_c$.
}
\label{tab:zero_curvature_summary}

\begin{tabular}{l@{\hspace{0.5cm}}c@{\hspace{0.5cm}}c@{\hspace{0.5cm}}c@{\hspace{1.0cm}}c}
\toprule
Black hole 
& N-states ($\mathcal{R}^{II}=0$) 
& $T_{\rm zero}/T_c$ 
& $j_*$ 
& Region of physical N-states\\
\midrule
RN 
& None
& -- 
& -- 
& None \\

Kerr 
& $x_{\rm zero}=1.257$ 
& $0.5193$ 
& -- 
& All allowed parameter space \\

\multirow{2}{*}{Kerr--AdS ($j=j_c$)}
& $x_{\rm zero}^{-}\simeq1.164$ 
& $0.4487$ 
& \multirow{2}{*}{$0.244$}
& \multirow{2}{*}{$j\leq j_* $ \qquad ($|J|\lesssim 0.244 l^2$)} \\

& $x_{\rm zero}^{+}\simeq6.692$ 
& $1.0051$ 
& 
& \\

\bottomrule
\end{tabular}
\end{table}

Finally, Table~\ref{tab:zero_curvature_summary} summarizes the main results of the numerical analysis. These results show that the existence and structure of the zero-curvature states depend on the black hole geometry. No physical zero-curvature state is found for asymptotically flat RN, whereas Kerr exhibits a single zero-curvature state with a constant $x_{\rm zero}$. In particular, this solution persists for every $J\neq0$, with no finite parameter threshold at which it disappears. In Kerr--AdS, by contrast, the zero-curvature solutions form two branches governed by the dimensionless parameter $j$, which eventually merge at the finite double-root point $j_*\simeq0.244$. A natural next step is to confront the zero-curvature states with independent astrophysical observables, such as black-hole shadows. Since shadow observables depend on the black-hole parameters through the null-geodesic structure, they can provide independent constraints on the thermodynamic parameter space. For Kerr--AdS, combining the condition $\mathcal{R}^{II}=0$ with shadow observables could therefore constrain the allowed zero-curvature branches and establish a connection between the microscopic structure encoded by GTD and observable black-hole signatures. This will be discussed in the following sections.

\section{Black-hole shadows}
\label{sec:BHshadows}

Black-hole shadows provide the optical component of the correspondence developed in this review. We use throughout the spacetime conventions and black-hole notation introduced in Sec.~\ref{sec:introduction}. They are determined by the propagation of null geodesics in the exterior spacetime and therefore belong conceptually to a different geometric setting from thermodynamics and geometrothermodynamics. The connection with the previous sections arises only because the same macroscopic parameters that specify the thermodynamic fundamental equation also determine the spacetime metric and its null geodesics.

The theory of black-hole shadows has a long history. The escape and capture of photons in the Schwarzschild geometry were studied by Synge \cite{Synge1966}, while Luminet gave an early analysis of the optical appearance of a black hole surrounded by emitting matter \cite{Luminet1979}. The rotating case was developed in the Kerr geometry by Bardeen \cite{Bardeen1973}. Subsequent work clarified the geometrical meaning of photon surfaces and photon regions, the apparent shape of rotating black holes, and observer-dependent shadow constructions \cite{ClaudelVirbhadraEllis2001,DeVries2000,GrenzebachPerlickLammerzahl2014}. Modern analytical developments and conventions are reviewed comprehensively in Ref.~\cite{PerlickTsupko2022}.

For the purposes of the present review, only the optical structures needed later for the thermodynamic correspondence will be developed. We first summarize the general construction of the critical shadow curve. We then specialize to the RN, Kerr, and Kerr--AdS black holes, in the same order used in Secs.~\ref{sec:bh_thermodynamics}--{\ref{sec:GTD}}. No thermodynamic meaning is assigned to the shadow observables in this section. That relation will be introduced only in the following section.
Throughout this section, we use geometrized units, $G=c=1$, for the null-geodesic and shadow calculations, independently of the thermodynamic scaling analysis in Sec.~\ref{sec:GTD}.

\subsection{General construction}

The theoretical shadow is defined by null geodesics that separate photon trajectories escaping to an observer from trajectories captured by the black hole. In a stationary spacetime, the null trajectories can be obtained from the Hamilton--Jacobi equation
\be
g^{\mu\nu}{\cal S}_{,\mu}{\cal S}_{,\nu}=0,
\label{eq:HJshadow}
\ee
where ${\cal S}$ is Hamilton's principal function. Killing symmetries generate conserved quantities. For a stationary and axisymmetric spacetime these include the photon energy $E$ and the axial angular momentum $L$. If the Hamilton--Jacobi equation is separable, an additional Carter-type constant ${\cal Q}$ can be introduced. It is useful to define the impact parameters
\be
\xi=\frac{L}{E},
\qquad
\eta=\frac{{\cal Q}}{E^2}.
\label{eq:impactparameters}
\ee

For a generic static and spherically symmetric spacetime written as
\be
ds^2=-f(r)dt^2+\frac{dr^2}{f(r)}
+r^2\left(d\theta^2+\sin^2\theta\,d\varphi^2\right),
\label{eq:sphericalmetricshadow}
\ee
the unstable circular photon orbit is determined by the extremum of the null effective potential. The corresponding photon-sphere radius $r_{\rm ph}$ satisfies
\be
r_{\rm ph} f_{,r}(r_{\rm ph})-2f(r_{\rm ph})=0.
\label{eq:photonspheregeneral}
\ee
The critical impact parameter is then~\cite{LadinoLarranaga2023,PerlickTsupko2022,ahmed2026shadow}
\be
b_{\rm c}^2=
\frac{r_{\rm ph}^2}{f(r_{\rm ph})}.
\label{eq:criticalimpactgeneral}
\ee
For a static observer at infinity in an asymptotically flat spacetime, $b_{\rm c}$ coincides with the linear shadow radius in the  screen of the observer. For a static observer at a finite radius $r_O$, the angular radius $\alpha_{\rm sh}$ obeys
\be
\sin^2\alpha_{\rm sh}
=
\frac{b_{\rm c}^2 f(r_O)}{r_O^2}
=
\frac{r_{\rm ph}^2}{f(r_{\rm ph})}
\frac{f(r_O)}{r_O^2}.
\label{eq:finiteshadowangle}
\ee
This expression already shows explicitly that a shadow is an observer-dependent quantity even though the critical photon orbit is determined by the spacetime geometry \cite{PerlickTsupko2022}. Rotating geometries are qualitatively different. The unstable null trajectories that generate the shadow boundary form a photon region rather than a single photon sphere \cite{GrenzebachPerlickLammerzahl2014}. For separable metrics, the radial photon potential $R(r)$ determines the spherical photon orbits through
\be
 R(r_{\rm sph})=0,
\qquad
R_{,r}(r_{\rm sph})=0,
\label{eq:sphericalphotonconditions}
\ee
with instability selected by the appropriate second-derivative condition. Solving these equations yields the critical impact parameters $\xi(r_{\rm sph})$ and $\eta(r_{\rm sph})$. Their projection onto an orthonormal tetrad at the observer determines the shadow curve. For a distant observer in an asymptotically flat rotating spacetime, with inclination angle $\theta_O$ relative to the rotation axis, the standard celestial coordinates can be written as \cite{Bardeen1973,HiokiMaeda2009,PerlickTsupko2022}
\be
\alpha=-\xi\csc\theta_O,
\qquad
\beta=\pm\sqrt{\eta+a^2\cos^2\theta_O-\xi^2\cot^2\theta_O},
\label{eq:Kerrcelestialgeneral}
\ee
where $a$ is the rotation parameter. The critical shadow curve is obtained parametrically by varying $r_{\rm sph}$ over the unstable spherical photon orbits.

A spherical shadow can often be characterized adequately by one radius. A rotating shadow generally cannot. Its horizontal and vertical diameters, displacement, area, effective radius, asymmetry, and distortion may contain independent information. The size and distortion observables introduced for Kerr-like shadows by Hioki and Maeda provide one commonly used characterization \cite{HiokiMaeda2009}. More general combinations of size and shape observables can improve the determination of black-hole parameters \cite{KumarGhosh2020}. This distinction will become important later because a multidimensional thermodynamic state space cannot in general be reconstructed from one optical scalar.

It is also important to distinguish the theoretical critical shadow curve from the photon ring and from the bright emission ring. The critical curve is a geometrical separatrix determined by null geodesics. The photon ring is produced by strongly lensed trajectories that spend a long time near the critical photon region, while the observed emission ring depends additionally on the emitting plasma, radiative transfer, and instrumental reconstruction. Horizon-scale imaging therefore constrains the theoretical shadow only through an astrophysical inference model \cite{EHT2019,EHT2022,PerlickTsupko2022}. This distinction will be retained when observational implications are discussed later.

\subsection{Reissner--Nordstr\"om shadow}
\label{subsec:RNshadow}

\begin{figure*}[h]
  \centering
  \begin{minipage}[t]{0.327\linewidth}
    \centering
    \includegraphics[width=\linewidth]{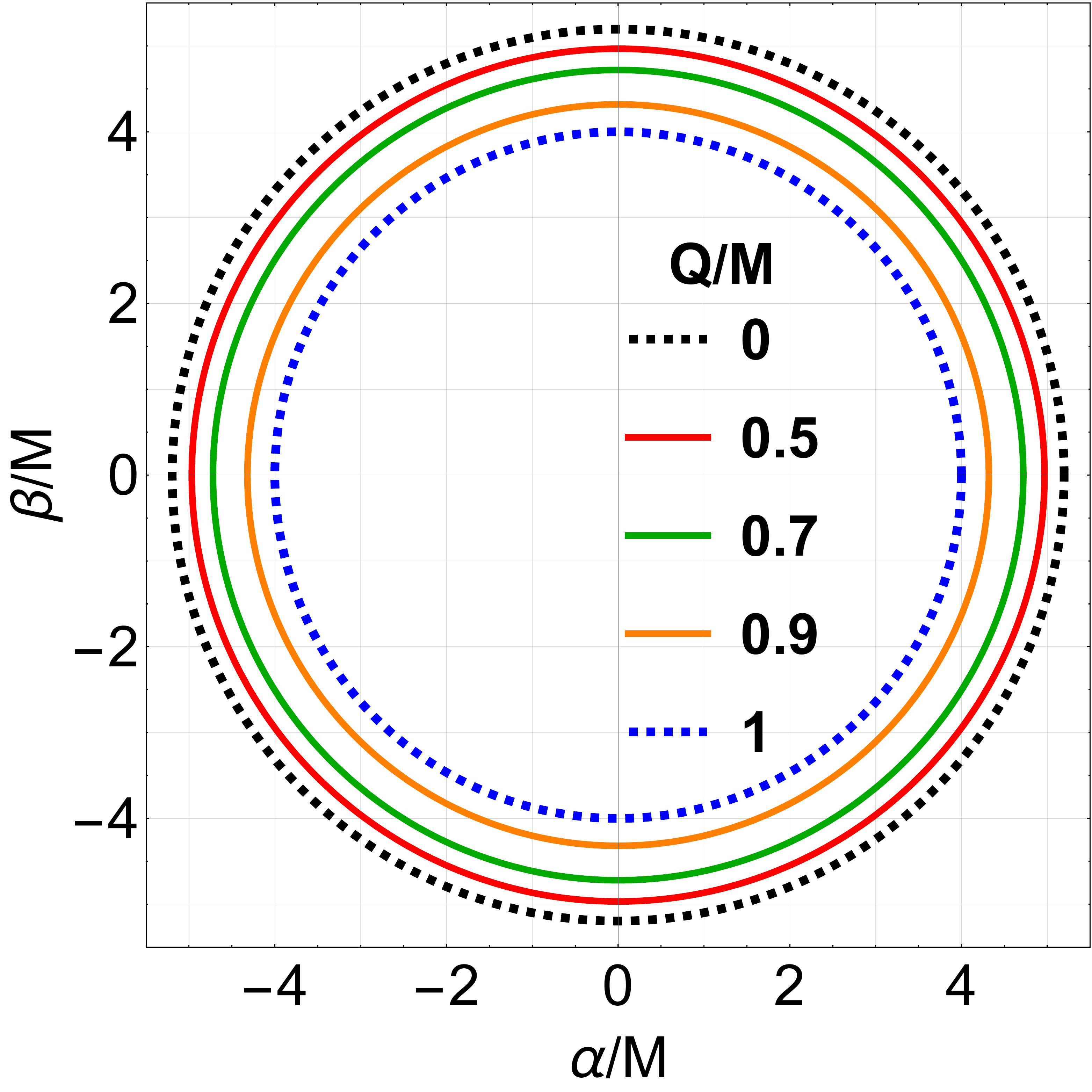}
    \\[2pt] \vspace{-0.6cm}\hspace{-3.5cm}\small{(a)}
  \end{minipage}\hfill
  \begin{minipage}[t]{0.327\linewidth}
    \centering
    \includegraphics[width=\linewidth]{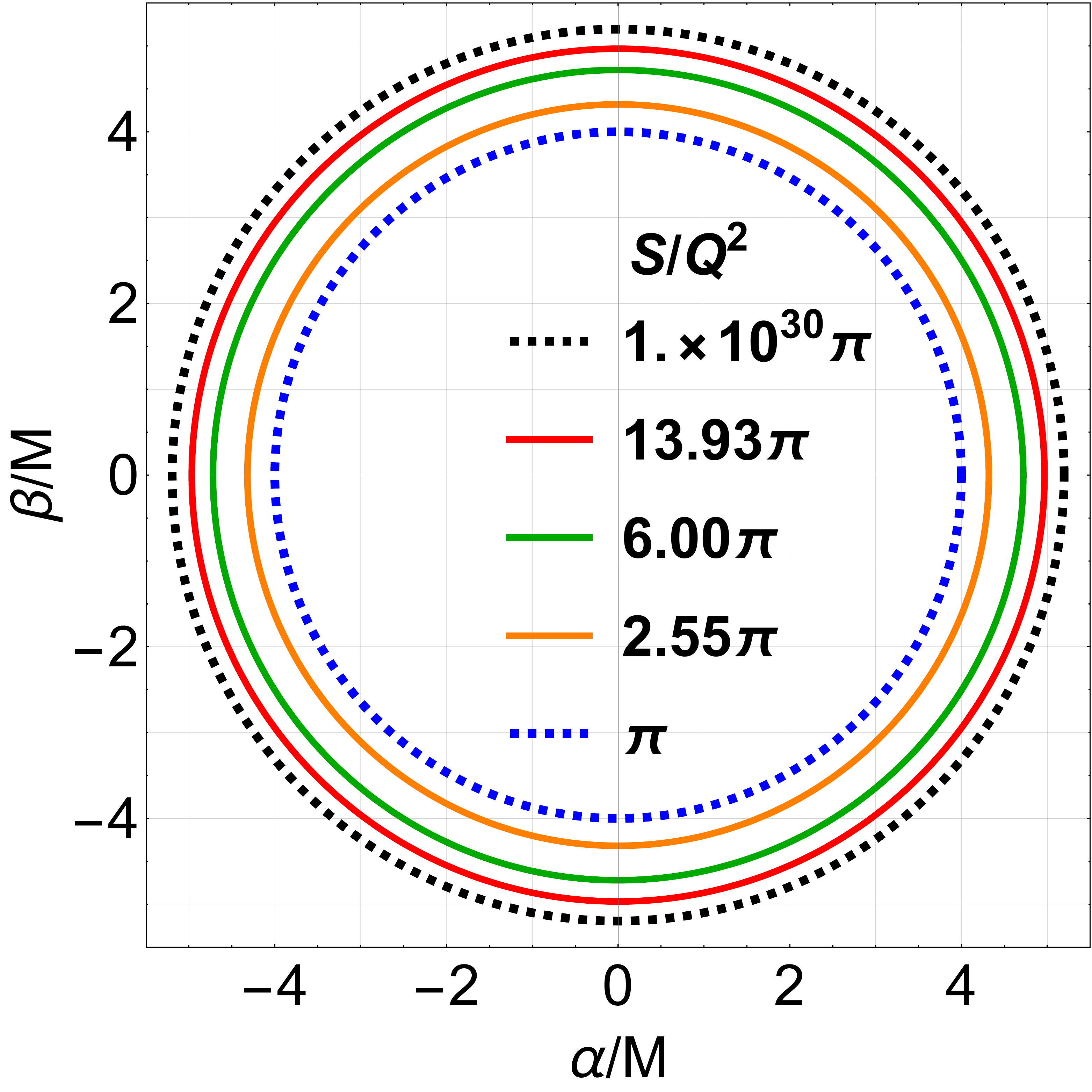}
    \\[2pt] \vspace{-0.6cm}\hspace{-3.5cm}\small{(b)}
  \end{minipage}\hfill
  \begin{minipage}[t]{0.346\linewidth}
    \centering
    \includegraphics[width=\linewidth]{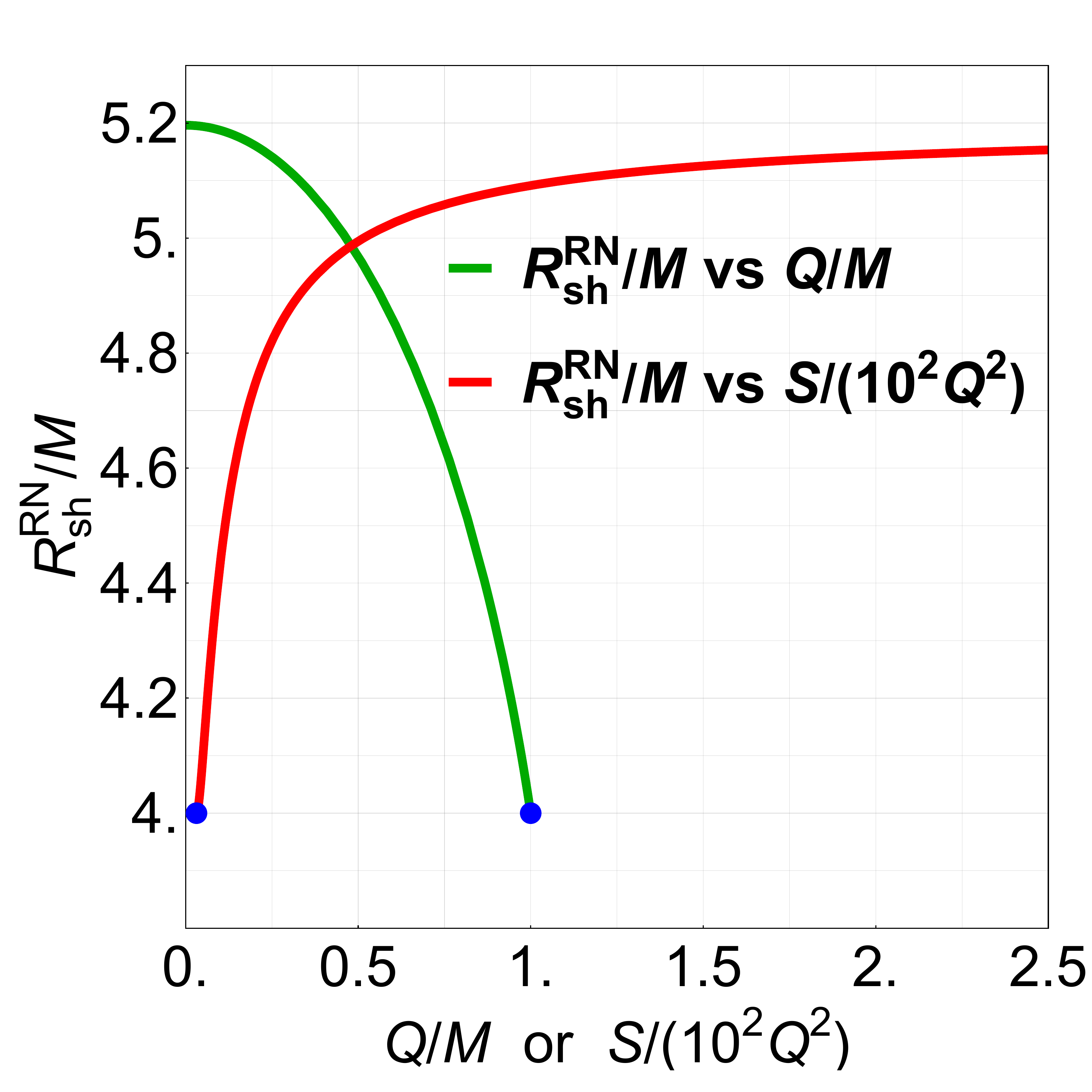}
    \\[2pt] \vspace{-0.6cm}\hspace{-3.5cm}\small{(c)}
  \end{minipage}
  \vspace{-0.8cm}
  \caption{Critical shadow boundaries for RN black holes ($r_0 \to \infty$). (a) Shadow contours in the parameter space for various values of $Q/M$. (b) Shadow boundaries mapped into the entropy space for distinct values of $S/Q^2$. (c) Critical shadow radius $R_{\text{sh}}^{\text{RN}}$ as a function of both $Q/M$ and $S/Q^2$. The dashed blue curves and blue dots denote the extremal RN limit, while the dashed black curves represent the Schwarzschild limit.}
  \label{fig:RNShadows}
\end{figure*}

The RN spacetime introduced in Eqs.~\eqref{eq:RNmetricIntro}--\eqref{eq:RNhorizonsIntro} is the simplest example in which the shadow depends on a second black-hole parameter. In the notation of the Introduction, the metric function is $f_{\rm RN}(r)$ and the physical black-hole domain is $|Q|\leq M$. Substituting $f_{\rm RN}(r)$ from Eq.~\eqref{eq:RNfIntro} into the photon-sphere condition Eq.~\eqref{eq:photonspheregeneral} gives two formal circular null-orbit radii. The outer unstable orbit that determines the shadow is
\be
r_{\rm ph}
=
\frac{1}{2}
\left(
3M+\sqrt{9M^2-8Q^2}
\right),
\label{eq:RNphotonradius}
\ee
with the corresponding asymptotic shadow radius given by the critical impact parameter
\be
R_{\rm sh}^{\rm RN}
=
\frac{r_{\rm ph}}{\sqrt{f_{\rm RN}(r_{\rm ph})}}.
\label{eq:RNshadowradius}
\ee
Equivalently, using the photon-sphere condition, Eq.~\eqref{eq:photonspheregeneral}, it can be written entirely in terms of $M$ and $Q$. The important feature for the present discussion is its smooth dependence on the charge throughout the black-hole domain. In the Schwarzschild limit, $Q=0$, one obtains $r_{\rm ph}=3M$ and
\be
R_{\rm sh}^{\rm RN}=3\sqrt{3}\,M.
\label{eq:RNshadowSchwlimit}
\ee
At extremality, $|Q|=M$, the unstable photon orbit is located at $r_{\rm ph}=2M$ and
\be
R_{\rm sh}^{\rm RN}=4M.
\label{eq:RNshadowextreme}
\ee
Fig. \ref{fig:RNShadows} illustrates the critical shadow boundaries of RN black holes for an observer at infinity in both parameter and entropy spaces. In the parameter space, the shadow contour monotonically shrinks as the charge $Q$ increases from the uncharged Schwarzschild limit ($Q=0$) to the extremal RN limit ($Q=M$); thus, the asymptotic RN shadow decreases from $3\sqrt{3}\,M$ to $4M$ as the charge increases from zero to its extremal value \cite{PerlickTsupko2022,LadinoRomeroQuevedo2026}. Analogously, in the entropy space, the shadow size increases from the extremal lower bound of $S=\pi Q^2$ toward its maximum extent in the Schwarzschild limit as $S \to \infty$.

Because the RN shadow is circular for a static observer, a single size observable is sufficient to characterize its critical curve once $M$ and the observer convention have been specified. At finite distance, however, the observable angular radius is given by Eq.~\eqref{eq:finiteshadowangle}, and the observer radius must be kept explicitly. This finite-distance formulation is particularly important in asymptotically non-flat geometries and will reappear below.

\subsection{Kerr shadow}
\label{subsec:Kerrshadow}

\begin{figure*}[h]
  \centering
  \begin{minipage}[t]{0.327\linewidth}
    \centering
    \includegraphics[width=\linewidth]{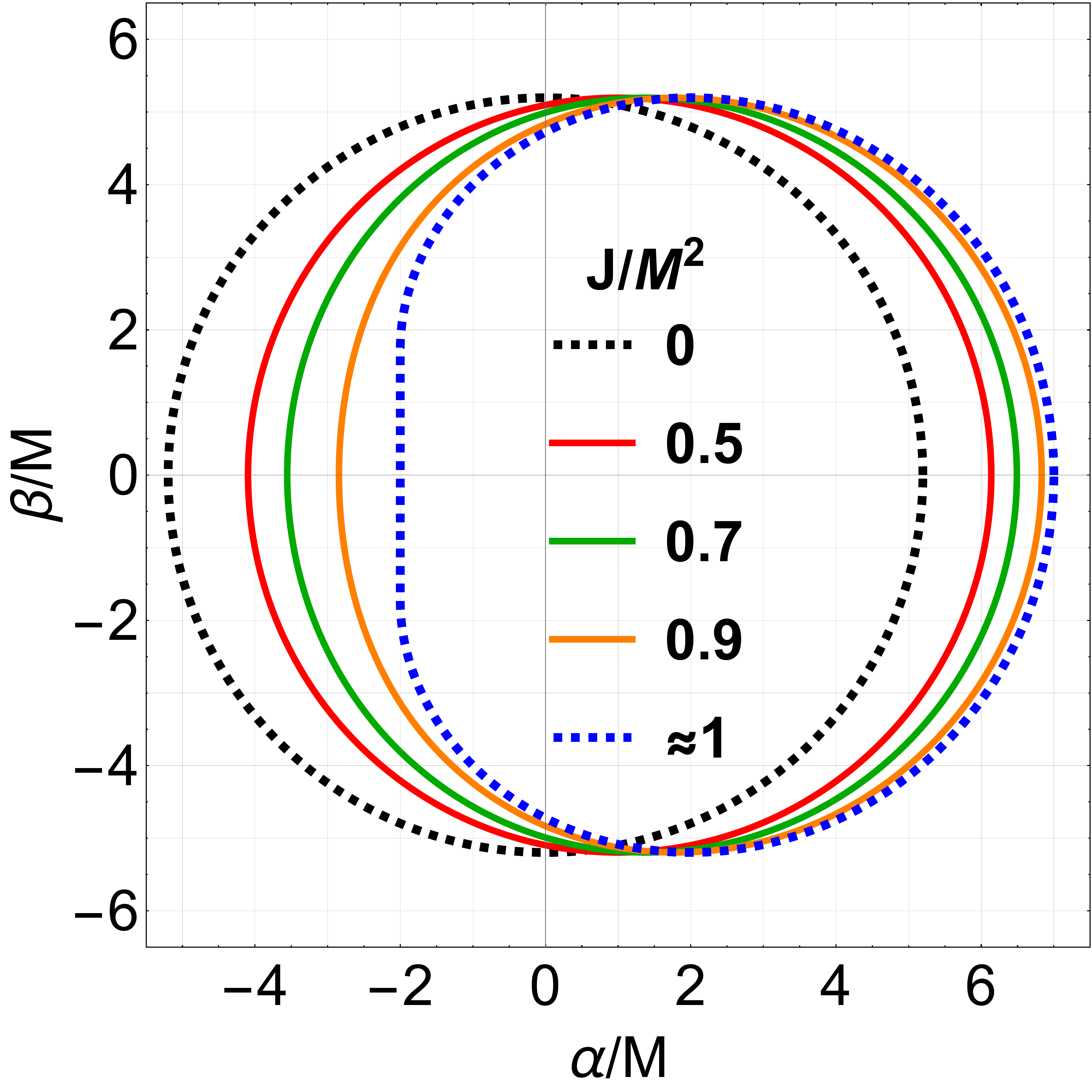}
    \\[2pt] \vspace{-0.6cm}\hspace{-3.5cm}\small{(a)}
  \end{minipage}\hfill
  \begin{minipage}[t]{0.327\linewidth}
    \centering
    \includegraphics[width=\linewidth]{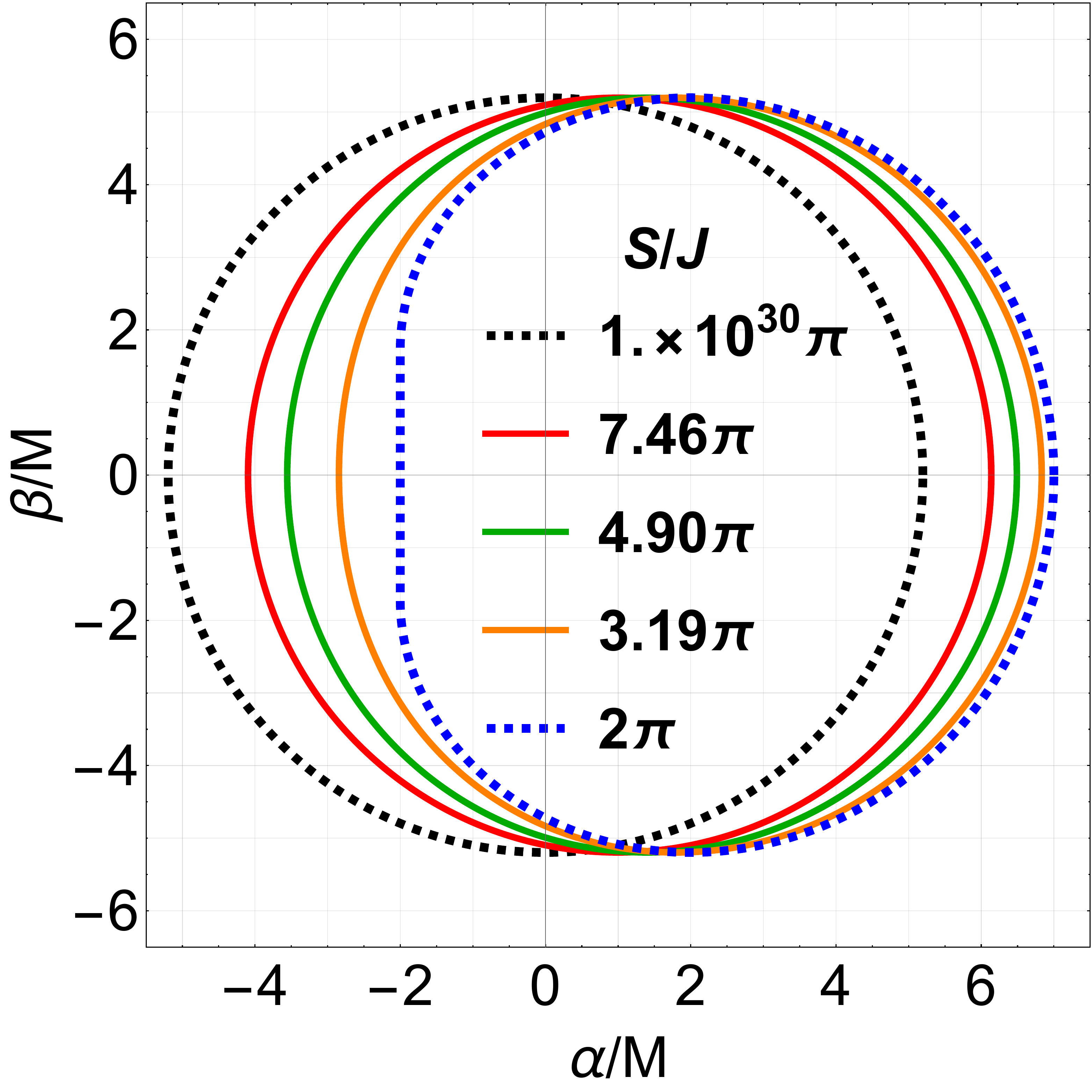}
    \\[2pt] \vspace{-0.6cm}\hspace{-3.5cm}\small{(b)}
  \end{minipage}\hfill
  \begin{minipage}[t]{0.346\linewidth}
    \centering
    \includegraphics[width=\linewidth]{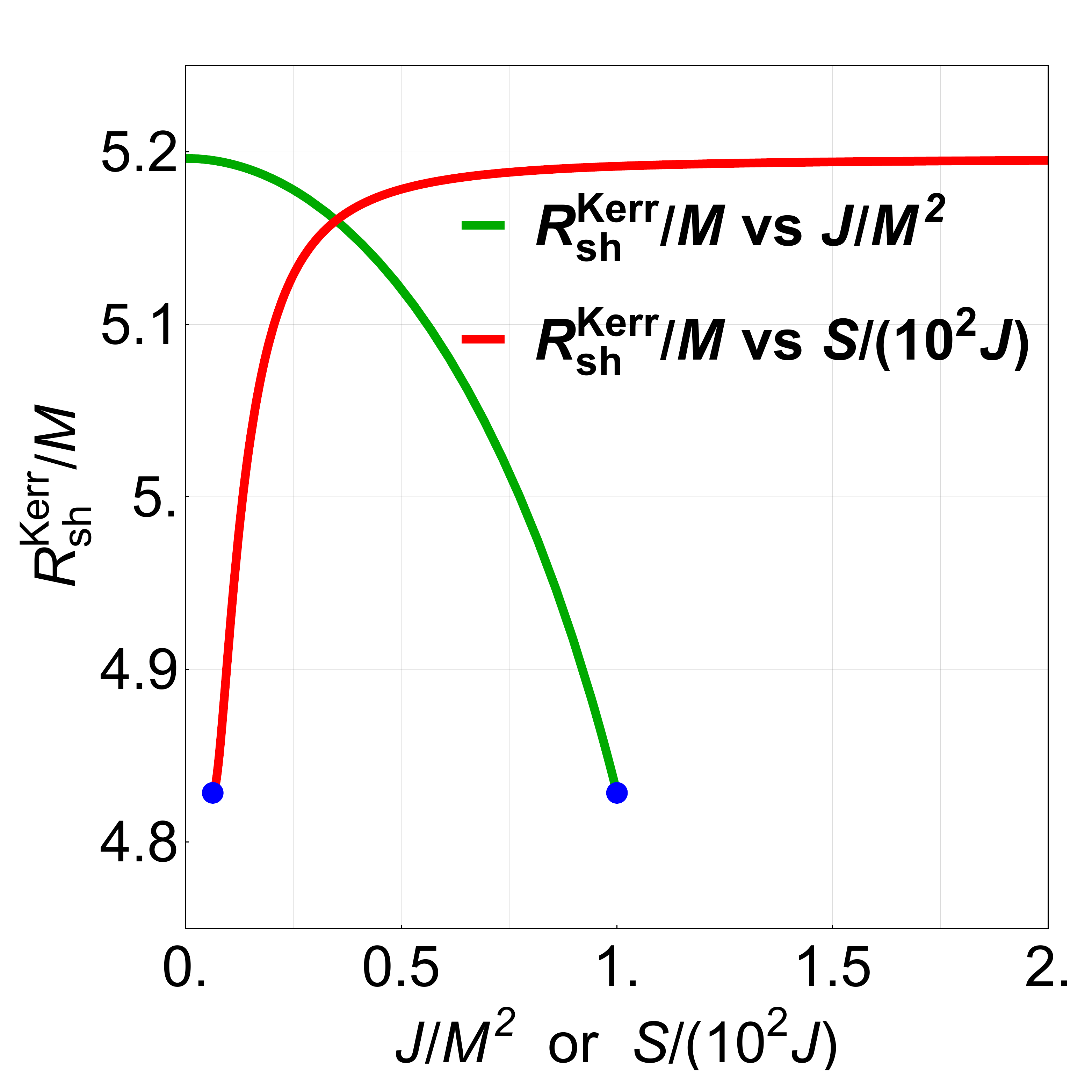}
    \\[2pt] \vspace{-0.6cm}\hspace{-3.5cm}\small{(c)}
  \end{minipage}
  \vspace{-0.8cm}
  \caption{Critical shadow boundaries for the Kerr black hole. (a) Shadow contours in the parameter space for various values of $J/M^2$ (  $\theta_0=90^\circ$, $r_0 \to \infty$). (b) Shadow boundaries mapped into the entropy space for distinct $S/J$ (  $\theta_0=90^\circ$, $r_0 \to \infty$). (c) Critical shadow radius $R_{\text{sh}}^{\text{Kerr}}$ as a function of both $J/M^2$ and $S/J$ ($\theta_0=0^o$). The dashed blue curves and blue dots denote the extremal Kerr limit, while the dashed black curves represent the Schwarzschild limit.}
  \label{fig:KerrShadows}
\end{figure*}

Rotation replaces the spherical photon orbit by a family of unstable spherical null geodesics and produces a shadow that is displaced and asymmetric for generic observer inclination. We use the Kerr metric and notation introduced in Eqs.~\eqref{eq:KerrmetricIntro}--\eqref{eq:KerrHorizonsIntro}, in particular $\rho^2=r^2+a^2\cos^2\theta$, $\Delta=r^2-2Mr+a^2$, and $a=J/M$. The null Hamilton--Jacobi equation is separable, and its radial potential can be written as \cite{Bardeen1973,PerlickTsupko2022}
\be
R(r)
=
\left[(r^2+a^2)E-aL\right]^2
-
\Delta
\left[
{\cal Q}+(L-aE)^2
\right].
\label{eq:Kerrradialpotential}
\ee
Thus, applying Eq.~\eqref{eq:sphericalphotonconditions} gives the critical impact parameters associated with a spherical photon orbit of radius $r$,
\be
\xi(r)
=
\frac{r^2(r-3M)+a^2(r+M)}
{a(M-r)},
\label{eq:Kerrxi}
\ee
and
\be
\eta(r)
=
\frac{r^3\left[4a^2M-r(r-3M)^2\right]}
{a^2(M-r)^2}.
\label{eq:KerrEta}
\ee
Substitution into Eq.~\eqref{eq:Kerrcelestialgeneral} yields the parametric shadow boundary for a distant observer, recovering the Schwarzschild result smoothly in the limit $J \to 0$ (or $a \to 0$), although Eqs.~\eqref{eq:Kerrxi} and \eqref{eq:KerrEta} themselves are not the most convenient expressions for taking that limit directly. While the Kerr equilibrium space is described by thermodynamic coordinates $(S,J)$ (Sec.~\ref{sec:GTD}) and the optical problem conventionally by spacetime parameters $(M,a)$, with $a=J/M$, the two descriptions refer to the same black-hole family, even though no thermodynamic interpretation is assigned to the optical variables at this stage. Nevertheless, Fig. \ref{fig:KerrShadows} establishes the shadow boundaries in both domains. The geometry of the Kerr shadow depends heavily on both the spin parameters $a/M$ (or angular momentum $J/M^2$) and the observer inclination $\theta_0$. For an observer on the rotation axis ($\theta_0=0^\circ$), the critical curve remains purely circular. Away from the axis, such as for an equatorial observer ($\theta_0=90^\circ$), frame dragging displaces the curve, generating an increasing left-right asymmetry and a reduction in overall size as the spin and inclination increase toward the extremal limit ($J \to M^2$). In entropy space, the shadow expands and shifts back toward perfect symmetry as the entropy $S$ increases from the extremal bound of $2\pi J$ toward the Schwarzschild limit ($S \to \infty$). These intertwined geometric and thermodynamic effects motivated several shadow observables designed to separate the overall size from the deformation of the critical curve \cite{HiokiMaeda2009,DeVries2000,KumarGhosh2020}.

For the later thermodynamic analysis it is useful not to reduce the Kerr shadow prematurely to one effective radius. A scalar size observable can be convenient when a one-parameter sequence is studied at fixed inclination, but the complete critical curve contains additional information. Horizontal and vertical diameters, displacement, asymmetry, and other shape descriptors may respond differently to changes in spin. The optical description should therefore retain enough independent observables to distinguish the relevant black-hole parameters whenever possible.

The asymptotically flat Kerr shadow is also the natural starting point for comparison with Sagittarius A*. The specific observer inclination and the shadow-size convention used in that application will be introduced only in the observational section. Here the essential point is that the Kerr shadow is generated by a photon region and is intrinsically a multi-observable object \cite{LadinoRomeroQuevedo2026}. 

\subsection{Kerr--AdS shadow}
\label{subsec:KerrAdSshadow}

\begin{figure*}[h]
  \centering
  \begin{minipage}[t]{0.330\linewidth}
    \centering
    \includegraphics[width=\linewidth]{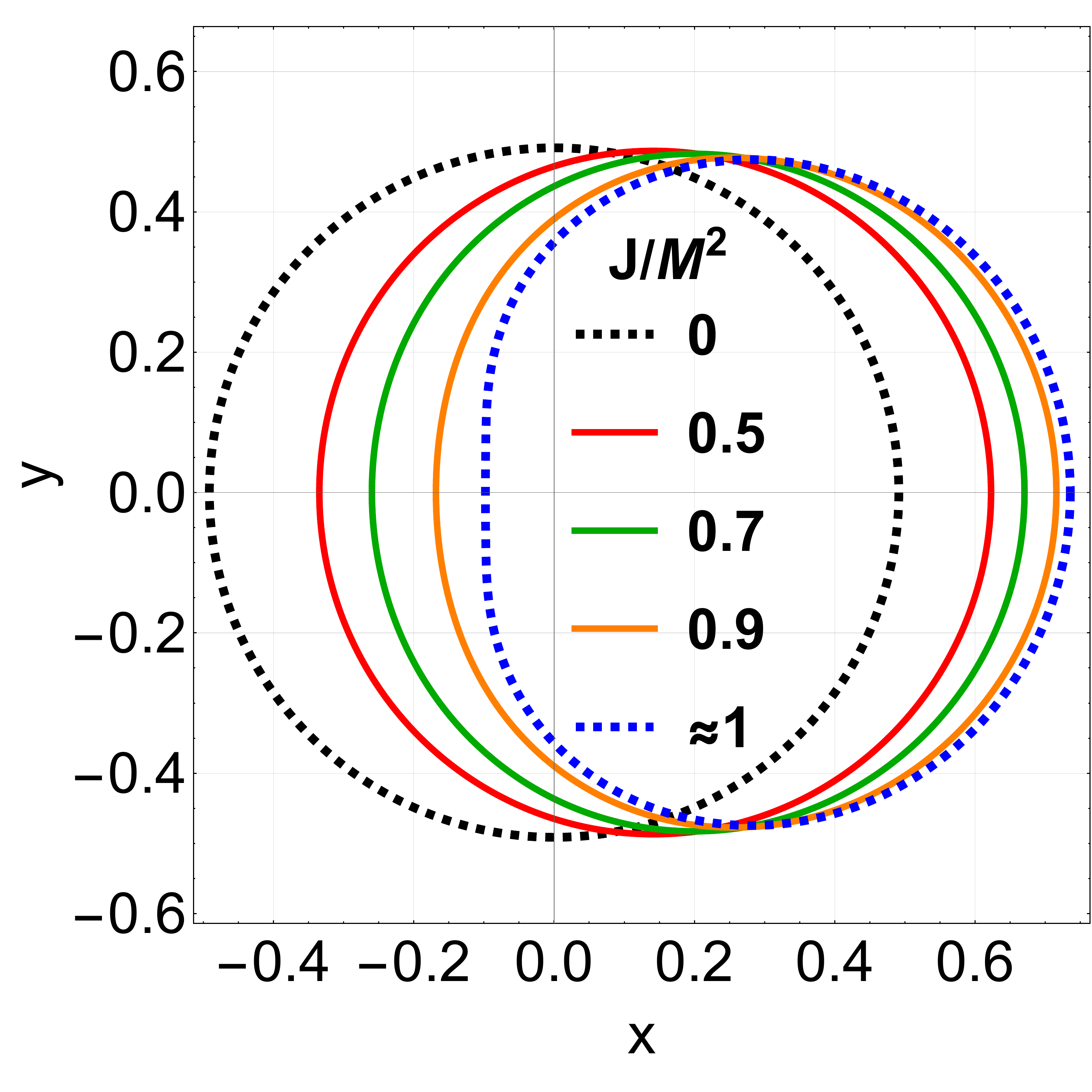}
    \\[2pt] \vspace{-0.6cm}\hspace{-3.5cm}\small{(a)}
  \end{minipage}\hfill
  \begin{minipage}[t]{0.330\linewidth}
    \centering
    \includegraphics[width=\linewidth]{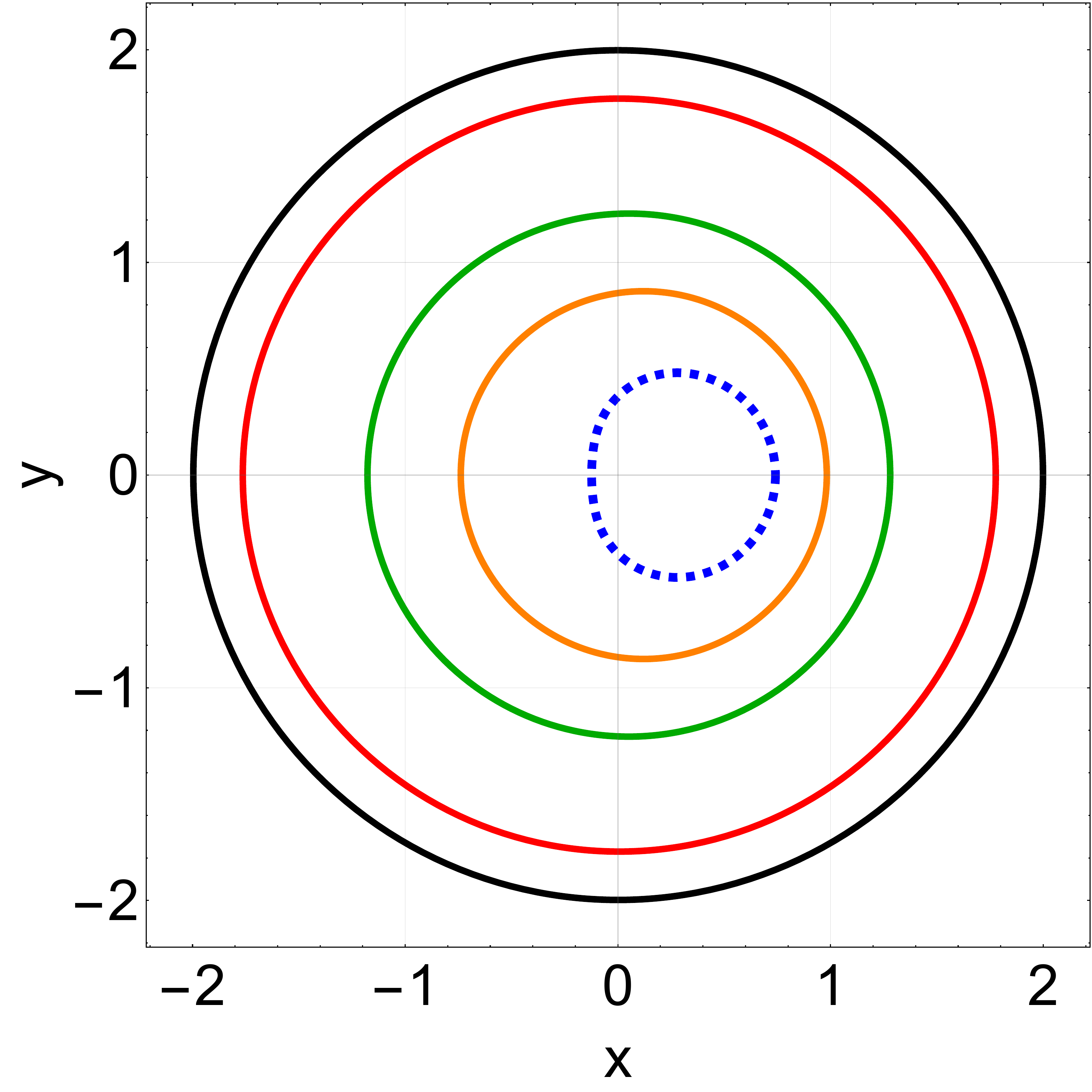}
    \\[2pt] \vspace{-0.6cm}\hspace{-3.5cm}\small{(b)}
  \end{minipage}\hfill
  \begin{minipage}[t]{0.34\linewidth}
    \centering
    \includegraphics[width=\linewidth]{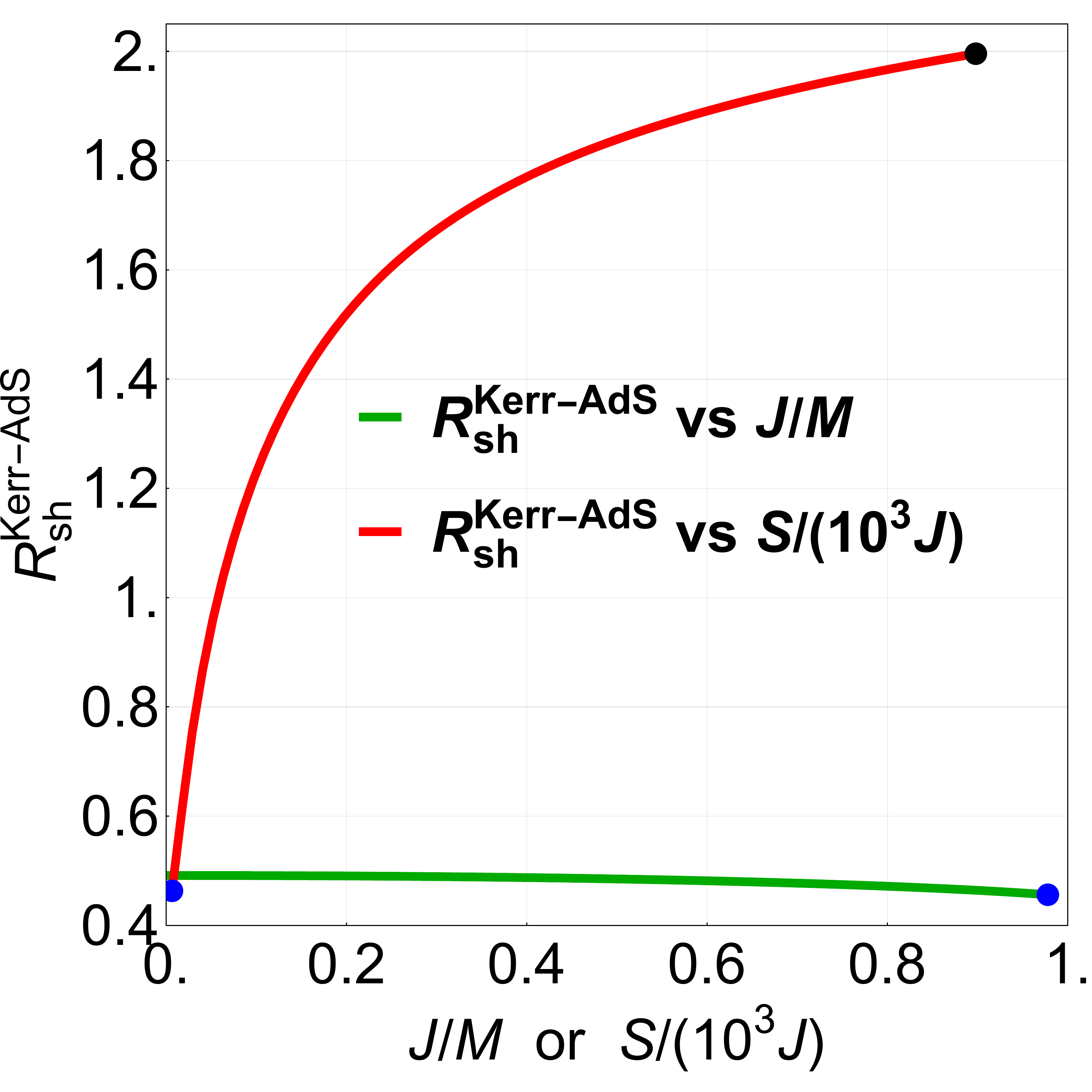}
    \\[2pt] \vspace{-0.6cm}\hspace{-3.5cm}\small{(c)}
  \end{minipage}
  \vspace{0.5cm} 
  \begin{minipage}[t]{0.330\linewidth}
    \centering
    \includegraphics[width=\linewidth]{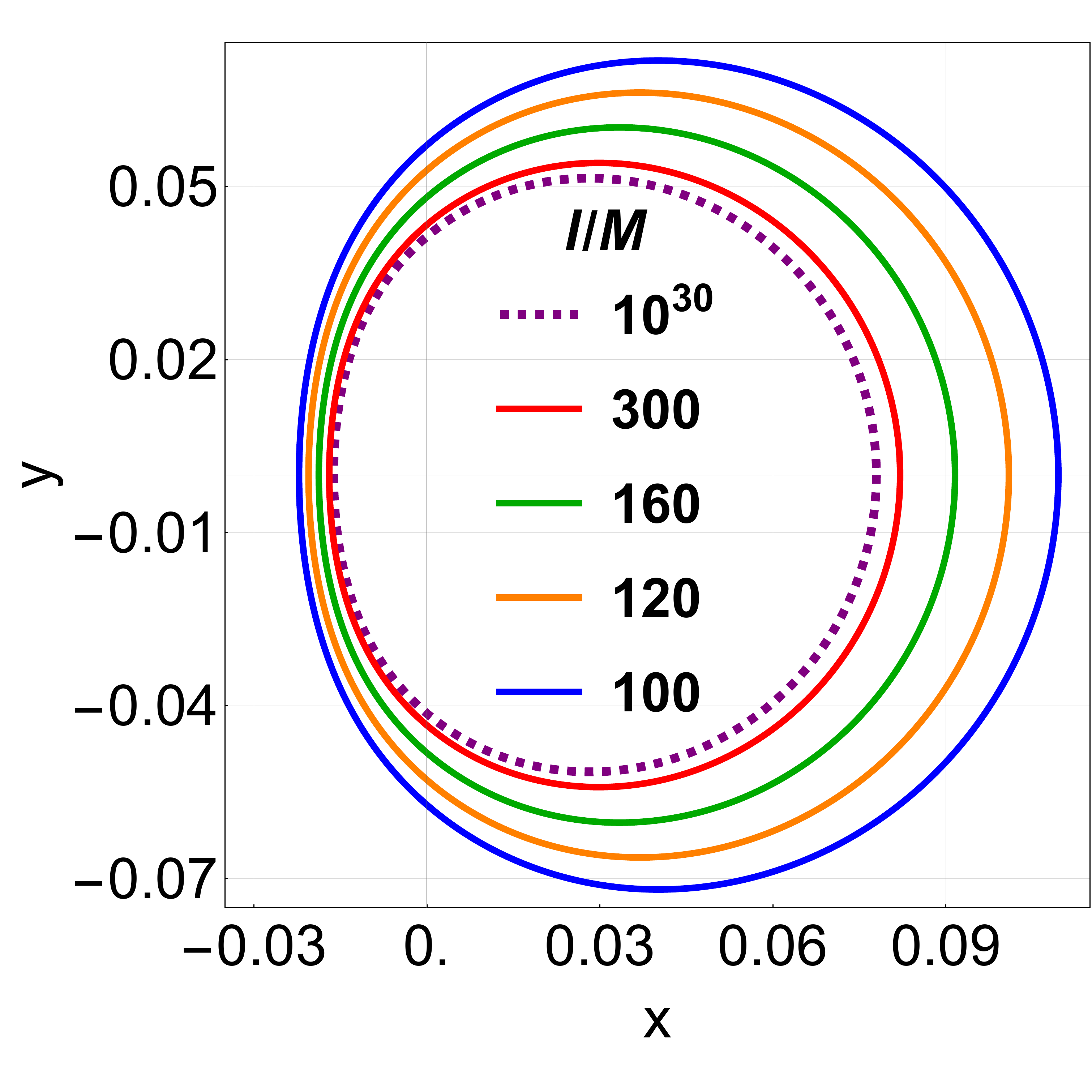}
    \\[2pt] \vspace{-0.6cm}\hspace{-3.5cm}\small{(d)}
  \end{minipage}\hfill
  \begin{minipage}[t]{0.330\linewidth}
    \centering
    \includegraphics[width=\linewidth]{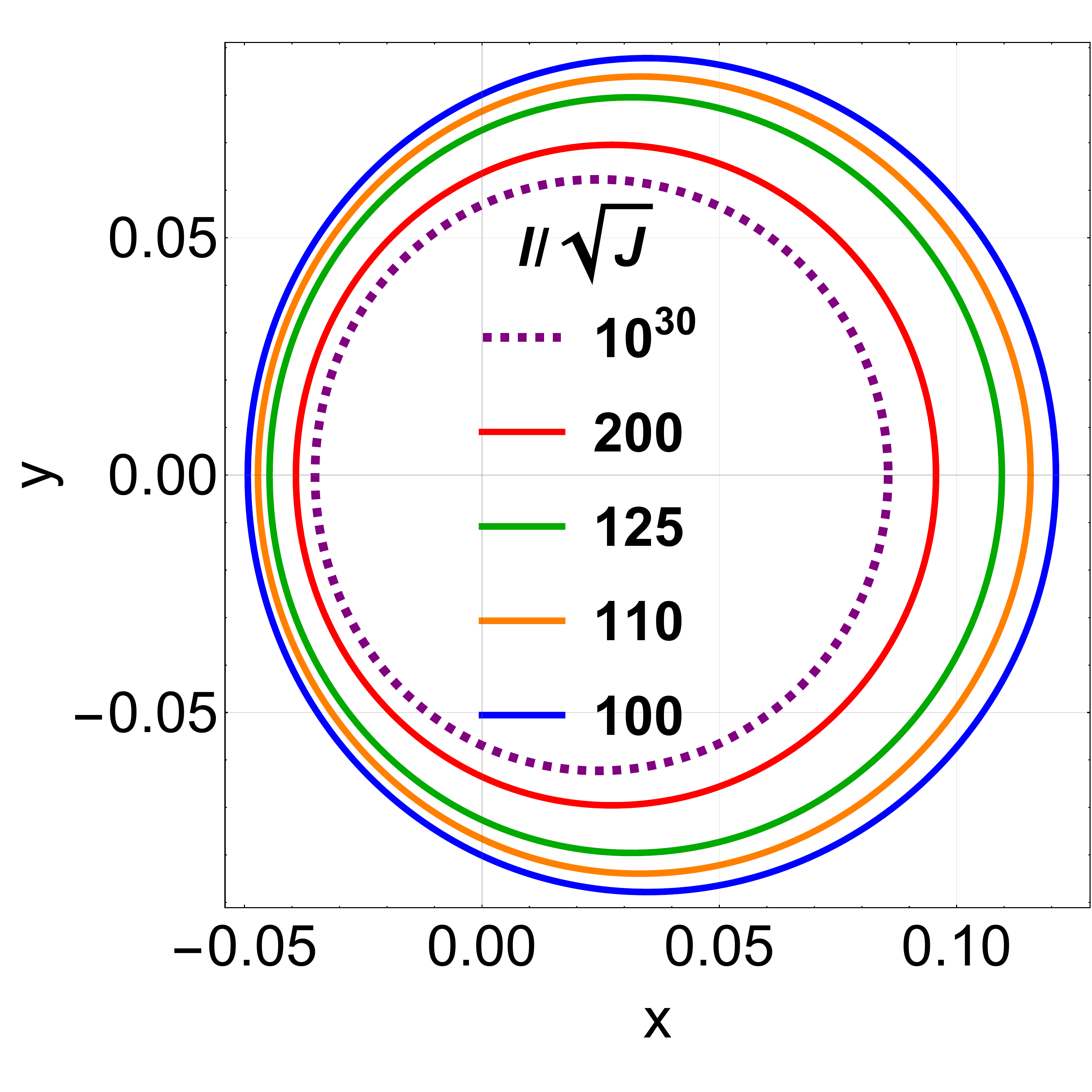}
    \\[2pt] \vspace{-0.6cm}\hspace{-3.5cm}\small{(e)}
  \end{minipage}\hfill
  \begin{minipage}[t]{0.34\linewidth}
    \centering
    \includegraphics[width=\linewidth]{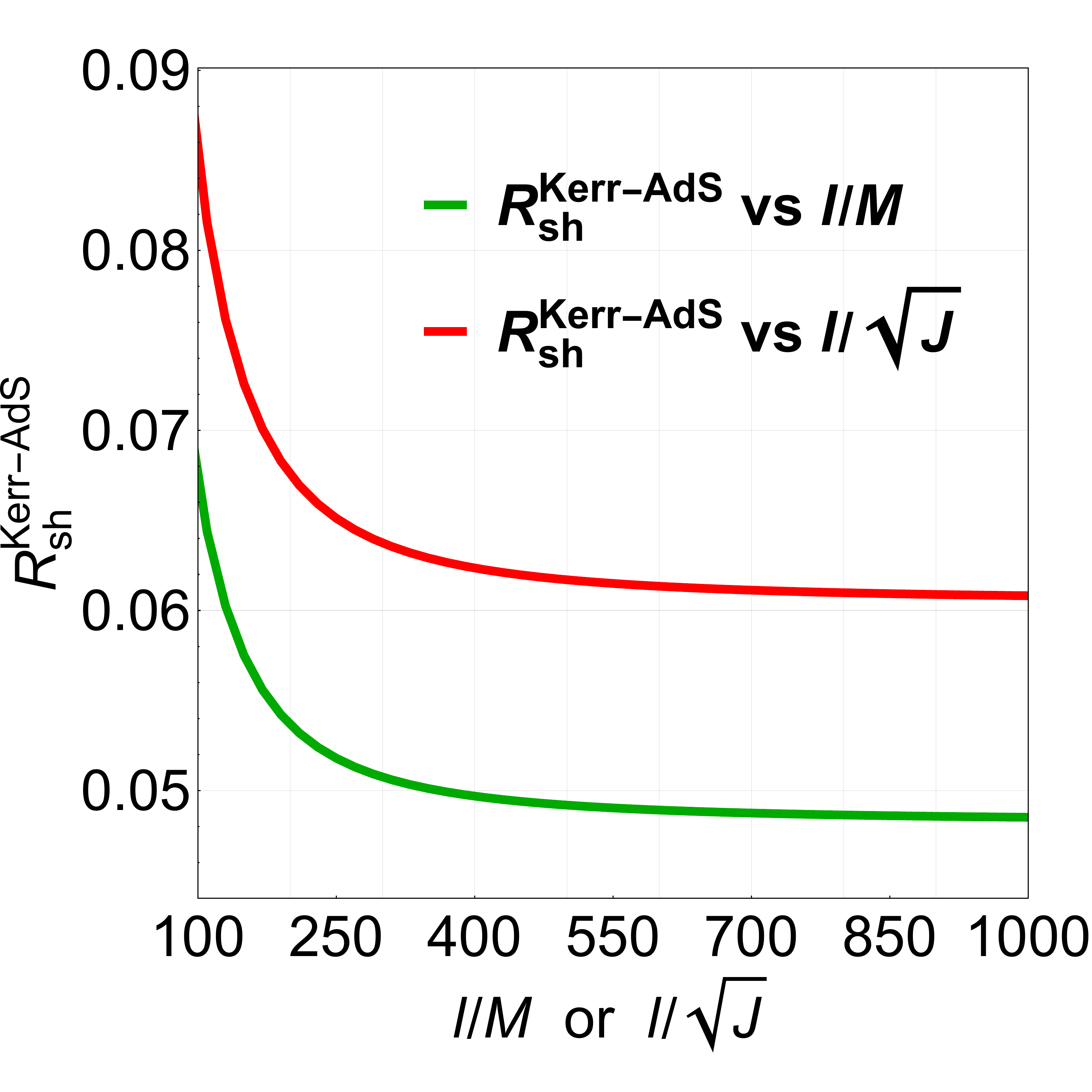}
    \\[2pt] \vspace{-0.6cm}\hspace{-3.5cm}\small{(f)}
  \end{minipage}
  
  \vspace{-0.5cm}
  \caption{Critical shadow boundaries for Kerr-AdS black holes. (a) Shadow contours in parameter space for distinct values of $J/M^2$ ($\theta_0=90^\circ$, $l = 10M$, $r_0 = 100M$). (b) Boundaries in entropy space for distinct $S/J \approx \{2\pi, \; 12.73\pi, \; 31.83\pi, \; 127.32\pi , \; 2.88 \pi \times 10^2  \}$ ($\theta_0=90^\circ$, $l = 100\sqrt{J}$, $r_0 = 100\sqrt{J}$). (c)~Critical shadow radius $R_{\text{sh}}^{\text{Kerr-AdS}}$ vs. $J/M$ and $S/J$ using parameters from (a) and (b) ($\theta_0=0^\circ$).  (d) Contours in parameter space for distinct $l/M$ ($\theta_0=90^\circ$, $J = 0.95M^2$, $r_0 = 1000M$). (e)~Boundaries in entropy space for distinct $l/\sqrt{J}$ ($\theta_0=90^\circ$, $S = 16J$, $r_0 = 1000\sqrt{J}$). (f)~$R_{\text{sh}}^{\text{Kerr-AdS}}$ vs. $l/M$ and $l/\sqrt{J}$ using parameters from (d) and (e) ($\theta_0=0^\circ$). Dashed curves denote: extremal Kerr-AdS (blue, with dots), Schwarzschild-AdS limit (black), and Kerr limit (purple).}
  \label{fig:KerrAdSShadows}
\end{figure*}
The Kerr--AdS geometry introduces a further optical scale through the AdS curvature radius $l$. We use the metric and notation introduced in Eqs.~\eqref{eq:KerrAdSmetricIntro}--\eqref{eq:KerrAdSChargesIntro}. The parameter $m$ is the metric parameter, whereas the conserved mass and angular momentum are $M=m/\Xi^2$ and $J=aM$. This distinction is important because the null-geodesic equations are naturally expressed in terms of $m$, $a$, and $l$, while the thermodynamic description of Sec.~\ref{sec:GTD} uses the conserved quantities. The Hamilton--Jacobi equation remains separable. Following Ref.~\cite{LadinoRomeroQuevedo2025}, the radial and polar functions for null geodesics can be written as
\be
R(r)
=
\Xi^2
\left[
(r^2+a^2)E-aL
\right]^2
-
{\cal Q}\Delta_r,
\label{eq:KerrAdSradialpotentialShadow}
\ee
and
\be
\Theta(\theta)
=
-\frac{\Xi^2}{\sin^2\theta}
\left(
aE\sin^2\theta-L
\right)^2
+
{\cal Q}\Delta_\theta.
\label{eq:KerrAdSpolarpotentialShadow}
\ee
We retain the impact parameters introduced in Eq.~\eqref{eq:impactparameters},
$\xi=L/E$ and $\eta={\cal Q}/E^2$, without introducing an additional rescaling.

The constant-radius photon trajectories that generate the critical curve satisfy
\be
R(r_{\rm sph})=0,
\qquad
R_{,r}(r_{\rm sph})=0,
\label{eq:KerrAdSsphericalconditions}
\ee
together with the condition $\Theta(\theta)\geq0$ along the allowed polar motion. Solving the two radial conditions for the critical impact parameters gives the analytic expressions obtained in Ref.~\cite{LadinoRomeroQuevedo2025},
\be
\xi(r_{\rm sph})
=
\frac{r_{\rm sph}^2+a^2}{a}
-
\frac{4r_{\rm sph}\Delta_r(r_{\rm sph})}
{a\Delta_{r,r}(r_{\rm sph})},
\label{eq:KerrAdSxicriticalShadow}
\ee
and
\be
\eta(r_{\rm sph})
=
\frac{
16r_{\rm sph}^2\Xi^2\Delta_r(r_{\rm sph})
}{
\left[\Delta_{r,r}(r_{\rm sph})\right]^2
}.
\label{eq:KerrAdSetacriticalShadow}
\ee
These quantities depend on $r_{\rm sph}$, which is itself determined by the spacetime parameters $(m,a,l)$. The allowed interval of constant-radius photon trajectories determines the critical curve of the shadow. In the asymptotically flat limit $l\rightarrow\infty$, one has $\Xi\rightarrow1$, and the Kerr critical-impact-parameter expressions are recovered.

The observer prescription is essential in Kerr--AdS because the spacetime is not asymptotically flat. Following the finite-distance construction used in Ref.~\cite{LadinoRomeroQuevedo2025}, the observer is placed at $(r_0,\theta_0)$ and equipped with an orthonormal tetrad. The direction of an incoming photon is then specified by the local celestial angles $\rho$ and $\sigma$. These angles are defined in the  tetrad of the observer and should be distinguished from the Cartesian screen coordinates $(x,y)$ introduced below through stereographic projection. The critical curve is obtained by varying $r_{\rm sph}$ over the allowed photon region and projecting the corresponding light directions onto the  celestial sphere of the observer.

For completeness, the celestial coordinates derived in Ref.~\cite{LadinoRomeroQuevedo2025} can be written as
\be
\sin\rho
=
\pm
\sqrt{
1-\frac{1}{\Xi^2}
+
\frac{
16r_{\rm sph}^2\Delta_r(r_0)\Delta_r(r_{\rm sph})
}{
\left[
4r_{\rm sph}\Xi\Delta_r(r_{\rm sph})
+
(r_0^2-r_{\rm sph}^2)\Xi\Delta_{r,r}(r_{\rm sph})
\right]^2
}
},
\label{eq:KerrAdSrhoShadow}
\ee
and
\be
\begin{split}
\sin\sigma
={}&
\frac{
\Xi\csc\theta_0\sqrt{\Delta_r(r_0)}
}{
2a\sqrt{\Delta_\theta(\theta_0)}
}
\frac{
\left[
a^2+2r_{\rm sph}^2+a^2\cos(2\theta_0)
\right]\Delta_{r,r}(r_{\rm sph})
-
8r_{\rm sph}\Delta_r(r_{\rm sph})
}{
\sqrt{
16r_{\rm sph}^2\Delta_r(r_0)\Delta_r(r_{\rm sph})
+
(\Xi^2-1)
\left[
4r_{\rm sph}\Delta_r(r_{\rm sph})
+
(r_0^2-r_{\rm sph}^2)\Delta_{r,r}(r_{\rm sph})
\right]^2
}
}.
\end{split}
\label{eq:KerrAdSsigmaShadow}
\ee
A stereographic projection onto a Cartesian image plane is then given by
\be
x=-2\tan\left(\frac{\rho}{2}\right)\sin\sigma,
\qquad
y=-2\tan\left(\frac{\rho}{2}\right)\cos\sigma.
\label{eq:KerrAdSscreenShadow}
\ee
Thus, the Kerr--AdS shadow depends on the spacetime parameters $(m,a,l)$ or $(M,J,l)$, and on the observer position $(r_0,\theta_0)$. In contrast with the RN case, a single radius does not generally characterize the complete optical information.

Fig. \ref{fig:KerrAdSShadows} presents the critical shadow boundaries for Kerr-AdS black holes, mapping the influence of both the angular momentum $J$ and the AdS radius $l$ across the parameter and entropy spaces. For an equatorial observer, increasing $J$ toward the extremal limit (or equivalently, reducing the entropy $S$ toward its lower bound) enhances the left-right asymmetry and monotonically shrinks the overall shadow contour. Additionally, a smaller AdS radius $l$ expands the shadow boundary relative to the asymptotically flat Kerr limit, confirming that the apparent shadow radius strictly decreases as $l$ grows and the spacetime approaches asymptotic flatness. Furthermore, any thermodynamic mapping of shadows for AdS spacetimes in the entropy space requires determining the physically valid domain for the entropy $S$. We restrict $S$ to lie between a minimum value (the extremal black hole limit) and a maximum value, which occurs when the outer photon orbit reaches the observer's radial coordinate $r_O$. Thus, while for asymptotically flat spacetimes the interval is simply $S \in [S_{\rm ext}, \infty)$, for Kerr-AdS black holes we consider $S \in [S_{\rm ext}, S_{\rm max}]$, where $S_{\rm max}$ corresponds to the photon orbit reaching the observer. For simplicity, we shall refer to this upper bound as the Schwarzschild-AdS limit of the shadow in the entropy space.

\subsection{Shadow observables and observer dependence}
\label{subsec:shadowobservables}

For an asymmetric shadow, two scalar observables can be used to characterize its size and distortion: the radius $R_{\rm sh}$ of a reference circle and the distortion parameter $\delta_{\rm sh}$ \cite{HiokiMaeda2009,LadinoRomeroQuevedo2025}.
Denoting by $(x_t,y_t)$ the top point of the shadow, by $(x_r,0)$ and $(x_l,0)$ its right and left intersections with the horizontal axis, and by $(x_l',0)$ the left intersection of the reference circle, these observables are
\be
R_{\rm sh}
=
\frac{(x_t-x_r)^2+y_t^2}
{2|x_r-x_t|},
\qquad
\delta_{\rm sh}
=
\frac{|x_l-x_l'|}
{R_{\rm sh}}.
\label{eq:KerrAdSshadowobservables}
\ee
These observables compress different aspects of the full critical curve and will be useful later when thermodynamic states are represented in optical variables. Note that for all rotating black hole shadow analyses in this work, the shadow radius is measured using the four-point reference circle defined in Eq.~\eqref{eq:KerrAdSshadowobservables}.

\begin{figure*}[h]
  \centering
  \begin{minipage}[t]{0.33\linewidth}
    \centering
    \includegraphics[width=\linewidth]{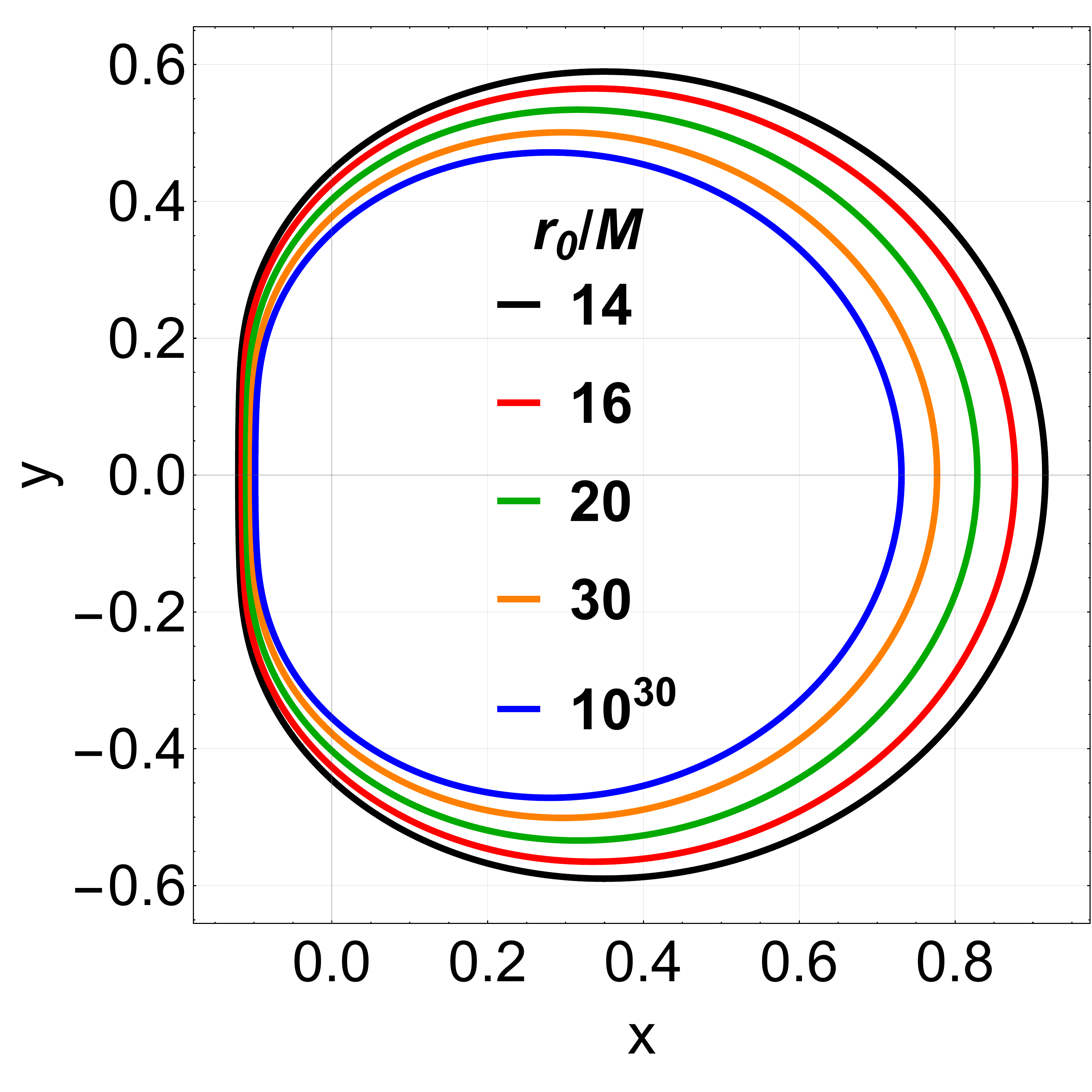}
    \\[2pt] \vspace{-0.6cm}\hspace{-3.5cm}\small{(a)}
  \end{minipage}\hfill
  \begin{minipage}[t]{0.33\linewidth}
    \centering
    \includegraphics[width=\linewidth]{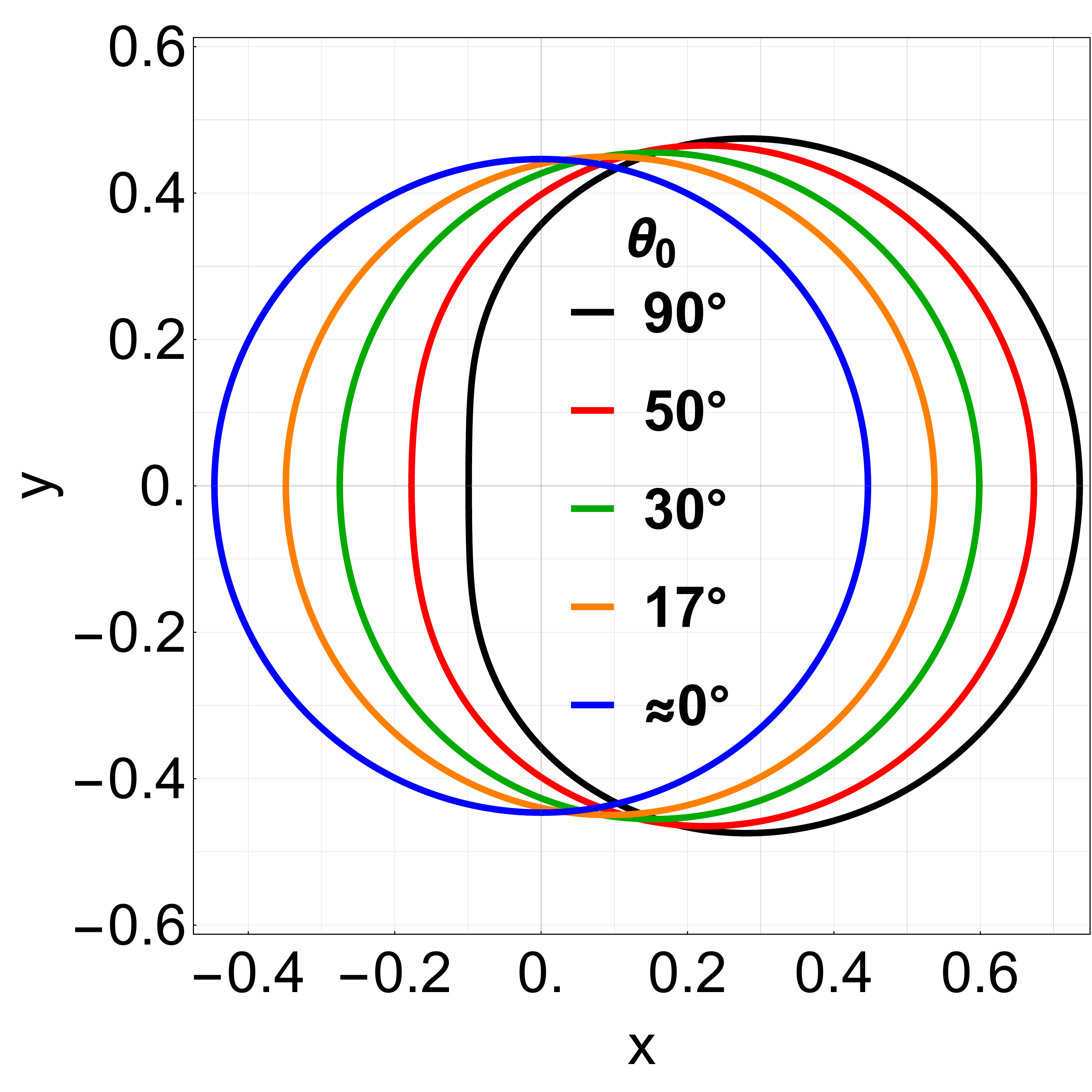}
    \\[2pt] \vspace{-0.6cm}\hspace{-3.5cm}\small{(b)}
  \end{minipage}\hfill
  \begin{minipage}[t]{0.34\linewidth}
    \centering
    \includegraphics[width=\linewidth]{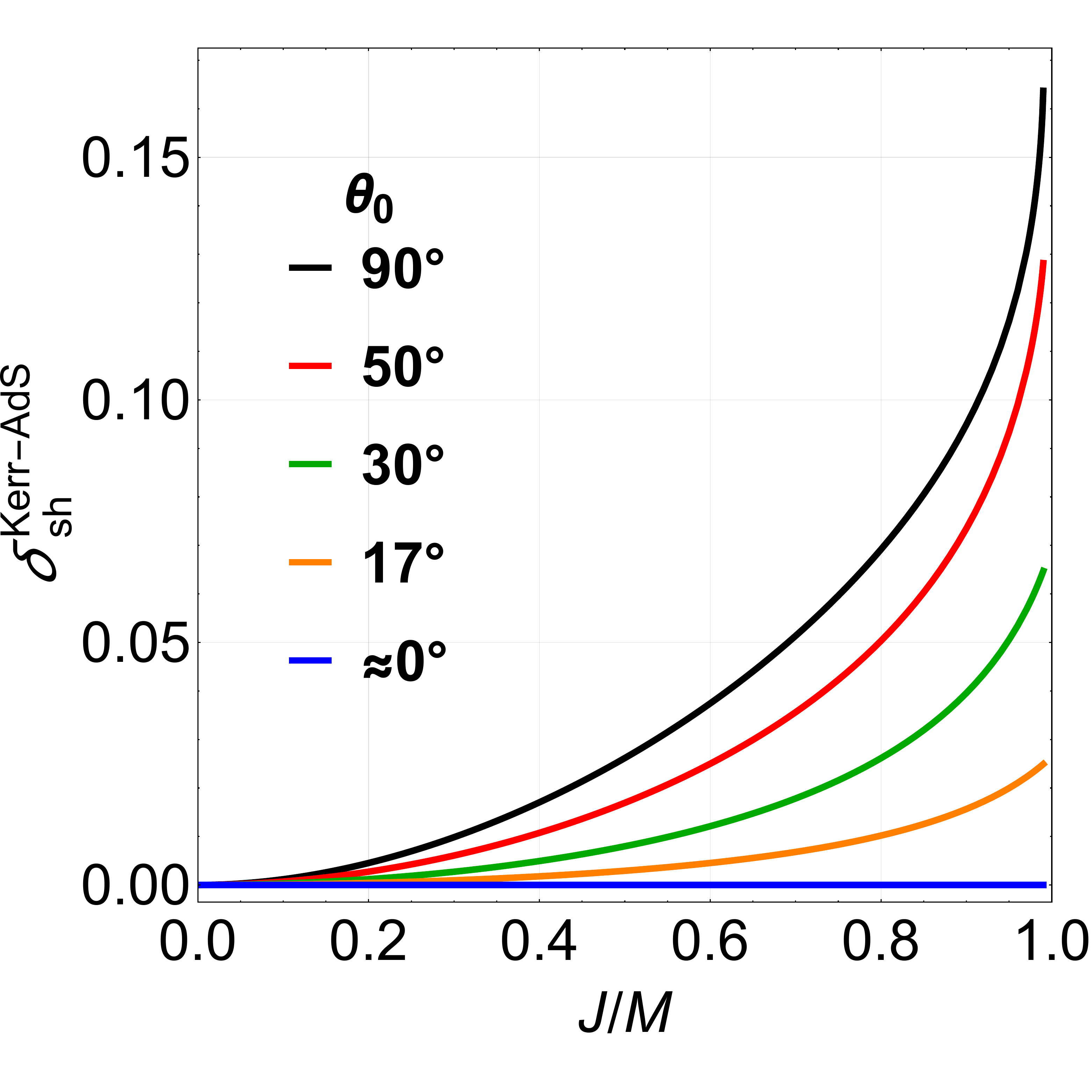}
    \\[2pt] \vspace{-0.6cm}\hspace{-3.5cm}\small{(c)}
  \end{minipage}
  \vspace{-0.8cm}
  \caption{Critical shadow boundaries for Kerr-AdS black holes in parameter space. (a) Shadow contours for distinct $r_0/M$ ($J=0.99M^2$, $\theta_0=90^\circ$, $l = 10M$). (b)  Boundaries for distinct $\theta_0$ ($J=0.99M^2$, $l = 10M$, $r_0 = 100M$). (c) Distortion parameter $\delta_{\text{sh}}^{\text{Kerr-AdS}}$ vs. $J/M$ and for distinct $\theta_0$  ($J=0.99M^2$, $l = 10M$, $r_0 = 100M$).}
  \label{fig:KerrAdSShadows2}
\end{figure*}

\begin{figure*}[h]
  \centering
  \begin{minipage}[t]{0.33\linewidth}
    \centering
    \includegraphics[width=\linewidth]{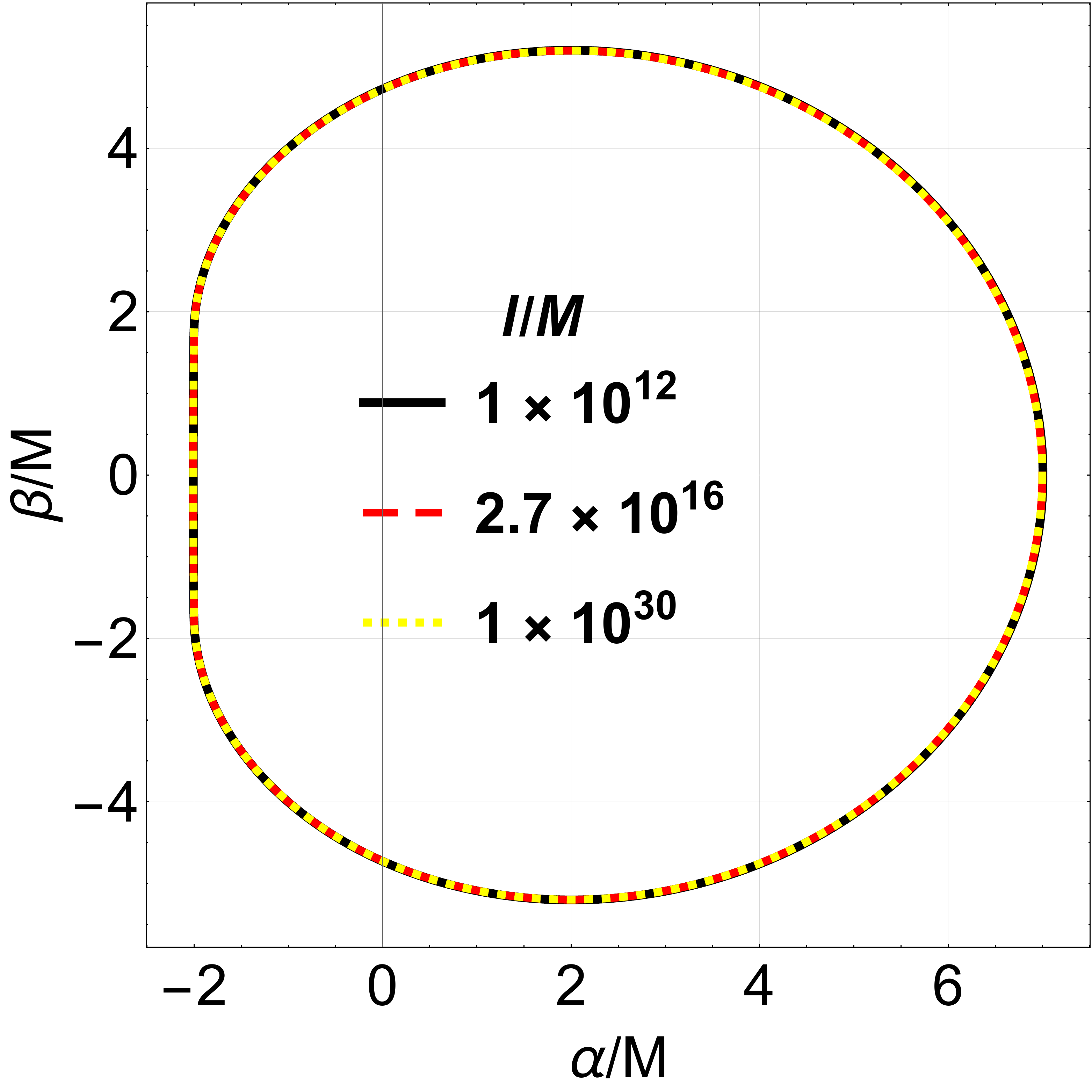}
    \\[2pt] \vspace{-0.6cm}\hspace{-3.5cm}\small{(a)}
  \end{minipage}\hfill
  \begin{minipage}[t]{0.33\linewidth}
    \centering
    \includegraphics[width=\linewidth]{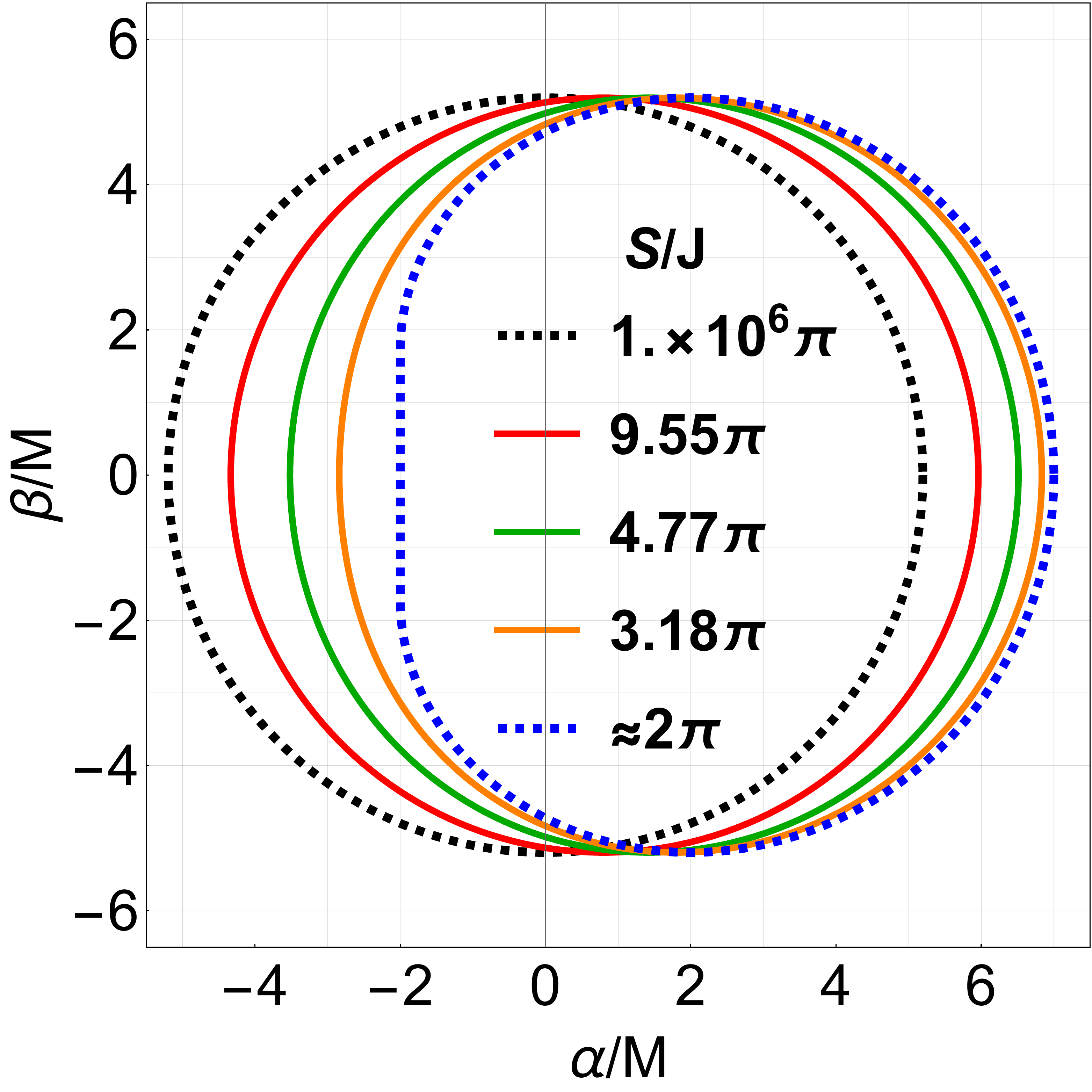}
    \\[2pt] \vspace{-0.6cm}\hspace{-3.5cm}\small{(b)}
  \end{minipage}\hfill
  \begin{minipage}[t]{0.34\linewidth}
    \centering
    \includegraphics[width=\linewidth]{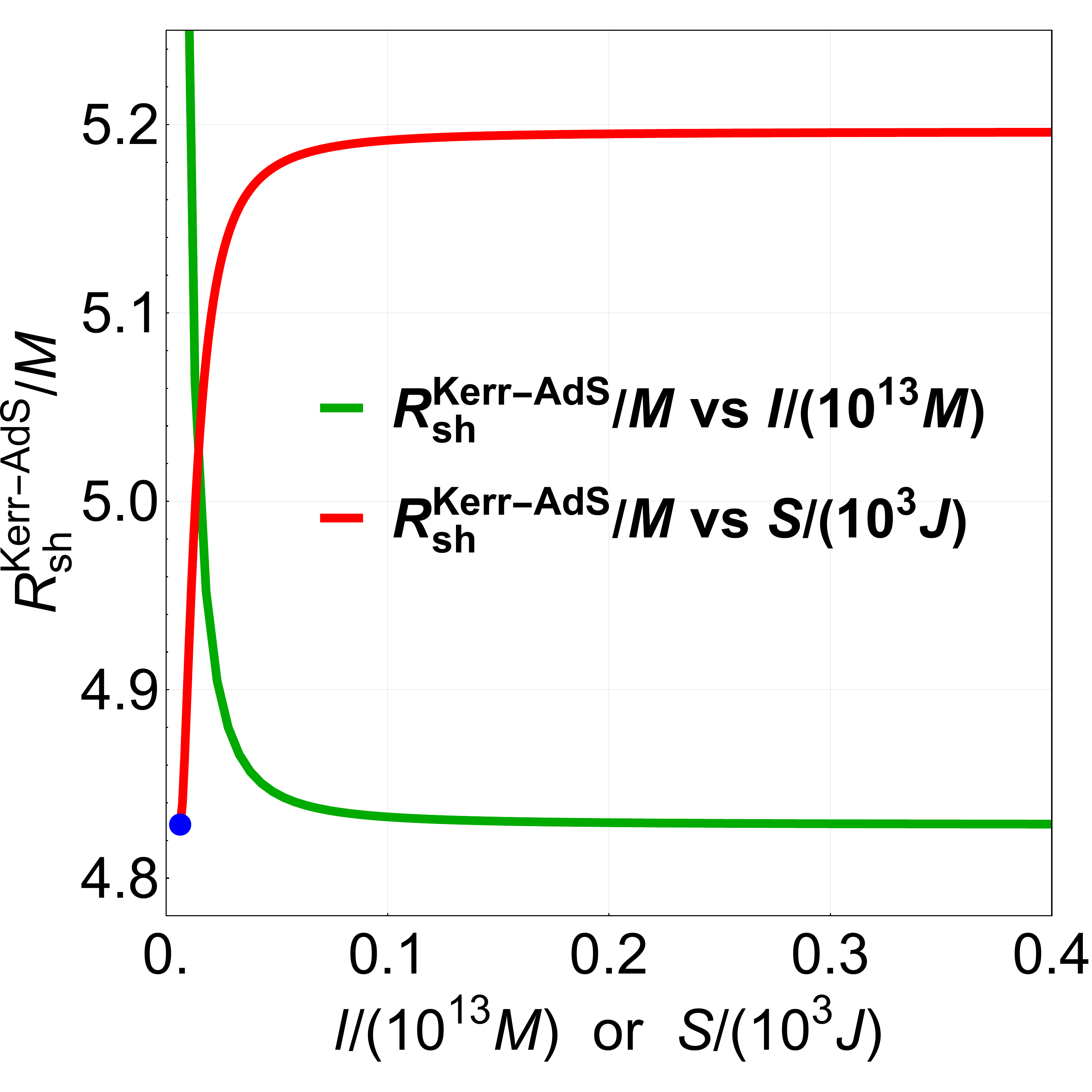}
    \\[2pt] \vspace{-0.6cm}\hspace{-3.5cm}\small{(c)}
  \end{minipage}
  \vspace{-0.8cm}
  \caption{Critical shadows for Kerr-AdS black holes in Bardeen’s coordinates of Eq.~\eqref{eq:bardeen}.  (a)~Shadow contours in parameter space for distinct $l/M$ ($J \approx M^2$, $\theta_0=90^\circ$, $r_0 = 4.1\times 10^{16} M$). (b)~Boundaries in entropy space for distinct $S/J$  ($l = 2.7 \times 10^{16} \sqrt{J}$,  $\theta_0=90^\circ$, $r_0 = 4.1\times 10^{16} \sqrt{J}$). (c)~Shadow radius $R_{\text{sh}}^{\text{Kerr-AdS}}/M$ vs. $l/M$ and $S/J$  using parameters from (a) and (b) ($\theta_0=0^\circ$). Dashed curves denote: extremal Kerr-AdS (blue, with dots), Schwarzschild-AdS limit (black), and extremal Kerr limit (yellow). }
  \label{fig:KerrAdSShadows3}
\end{figure*}

At this stage, however, the construction remains purely optical. The dependence of $\xi$, $\eta$, the celestial angles, and the shadow observables on $(m,a,l)$ provides the map from the spacetime parameters to the observer's screen. The thermodynamic and GTD interpretation of this map will be introduced only in Sec.~\ref{sec:GTDshadows}.

The preceding examples show that the notion of ``the shadow size'' must be used with care. For RN, spherical symmetry permits the critical curve to be represented by a single radius once the observer has been specified. For Kerr and Kerr--AdS, the full boundary is generally required. For the purposes of the present review, we introduce the notation
${\cal O}_{\rm sh}^{i}$, $\quad i=1,\ldots,N_{\rm sh}$, to denote a general set of independent shadow observables. Depending on the black-hole configuration, these may include the shadow radius $R_{\rm sh}$, the distortion parameter $\delta_{\rm sh}$, or other quantities characterizing the critical curve.

These observables depend both on the black-hole parameters and on quantities characterizing the observer. The observer position, four-velocity, and inclination therefore form part of the optical specification. At finite distance this dependence is explicit through the local tetrad, while in asymptotically flat applications the distant static observer provides a simplifying limiting convention \cite{PerlickTsupko2022,GrenzebachPerlickLammerzahl2014}. In any comparison with observations, additional astrophysical quantities such as mass calibration, distance, plasma properties, and image-reconstruction assumptions must also be distinguished from the intrinsic spacetime parameters \cite{EHT2019,EHT2022}.

To connect the dimensionless celestial coordinates $(x, y)$ of Eq.~\eqref{eq:KerrAdSscreenShadow} to Bardeen's standard dimensional impact parameters $(\alpha, \beta)$, we apply a linearized approximation valid for distant observers. In this regime, $x$ and $y$ correspond to the angles measured in the observer's sky and can be scaled by $r_0$. Accounting for the shift in Bardeen's convention, which aligns the center with light rays having zero angular momentum, the transformation is given by \cite{PerlickTsupko2022}
\be
\alpha = r_0 x - a \sin \theta_0, \quad \beta = r_0 y.
\label{eq:bardeen}
\ee
Crucially, in addition to requiring that the observer at $r_0$ remains within the domain of outer communications, the validity of applying this transformation to AdS spacetimes depends on the strict hierarchy $M \ll r_0 \ll l$. This means that the observer must be located sufficiently far from the black hole to be in the weak gravitational field ($M \ll r_0$), yet close enough relative to the AdS curvature radius ($r_0 \ll l$) so that cosmological curvature effects do not break the approximation. Provided this condition is met, this transformation offers a consistent way to directly compare black hole shadows in AdS spacetimes with their asymptotically flat counterparts.  

It is important to note that the shadow radius $R_{{\rm sh}}$ used throughout this work, defined in Eq.~\eqref{eq:KerrAdSshadowobservables}, is dimensionless when applied to the dimensionless celestial coordinates $(x, y)$ of Eq.~\eqref{eq:KerrAdSscreenShadow}. Conversely, if $R_{{\rm sh}}$ is computed in Eq.~\eqref{eq:KerrAdSshadowobservables} using the Bardeen coordinates for observers at infinity, as given in Eqs.~\eqref{eq:Kerrcelestialgeneral} and \eqref{eq:bardeen}, the resulting shadow radius $R_{{\rm sh}}$ has dimensions of length. In this latter case, it can be directly compared, for example, with standard values such as the Schwarzschild limit $R_{{\rm sh}}^{\rm Sch}=3\sqrt{3}\,M$.

Fig. \ref{fig:KerrAdSShadows2} displays the shadow contours of Kerr-AdS black holes using Eqs.~\eqref{eq:KerrAdSscreenShadow}, illustrating the dependence of the shadow observables on the observer's location. As the radial coordinate ($r_0$) increases, the apparent size of the shadow approaches a finite asymptotic value. Furthermore, the shadow boundary and the distortion parameter ($\delta_{\text{sh}}^{\text{Kerr-AdS}}$) for different inclination angle ($\theta_0$) reveals that the spin-induced asymmetry vanishes entirely for a polar observer. The deviation from circularity grows with the black hole spin, reaching its maximum deformation for observers located on the equatorial plane. On the other hand, Fig. \ref{fig:KerrAdSShadows3} illustrates the critical shadows of Kerr-AdS black holes in Bardeen's coordinates using Eqs.~\eqref{eq:bardeen}, mapping shadow contours in parameter and entropy spaces, and the corresponding shadow radius behaviors. As shown in Fig. \ref{fig:KerrAdSShadows3}(a), the shadow contours increasingly overlap as the AdS radius increases; for $l \ge 10^{12}M$, they become practically indistinguishable from the asymptotically flat limit ($l = 10^{30}M$), rendering the shadow almost identical to the standard Kerr case.

Furthermore, we observe that smaller AdS radii relative to the observer's position ($r_0$) yield larger apparent shadow radii. However, this specific regime indicates that the transformation to Bardeen coordinates breaks down in the immediate vicinity of the observer, limiting the coordinate system's validity at those scales.

This separation completes the purely optical part of the review. RN provides a one-radius spherical example, Kerr introduces a photon region and an asymmetric multi-observable shadow, and Kerr--AdS adds the effects of a non-Minkowskian asymptotic geometry and a finite observer prescription. In the next section, these optical observables will be related to the equilibrium states discussed previously. The crucial point will be that a shadow observable can serve as an alternative parameter for a thermodynamic family only when the corresponding optical map is sufficiently regular; it is not promoted to an independent thermodynamic degree of freedom, nor is it tied a priori to any one of the three GTD metrics.

\section{Geometrothermodynamics of black-hole shadows}
\label{sec:GTDshadows}
The preceding sections deliberately separated the thermodynamic, GTD, and optical descriptions. We now combine them. The central point is that the shadow is not promoted to an additional thermodynamic degree of freedom. Instead, the same macroscopic black-hole state that determines the fundamental equation also determines the spacetime and, once the observer has been fixed, its critical shadow curve. The GTD--shadow correspondence therefore associates thermodynamic and GTD information already defined on the equilibrium space with the optical image of the same state.

The thermodynamic and GTD analyses can be performed in either order. What is essential is that the GTD equilibrium conditions and curvature structures be compared with the independently defined thermodynamic properties of the system before an optical interpretation is assigned. Once this identification has been made, the spacetime parameters of the same equilibrium state are inserted into the shadow construction of Sec.~\ref{sec:BHshadows}. A curvature singularity, a response-function divergence, an extremal state, and a curvature zero must therefore retain their distinct thermodynamic meanings even when they are represented by shadow observables.

The explicit GTD--shadow program developed in Refs.~\cite{LadinoRomeroQuevedo2024,LadinoRomeroQuevedo2025,LadinoRomeroQuevedo2026} provides the basis for the discussion below. The RN--AdS analysis of Ref.~\cite{LadinoRomeroQuevedo2024} established the methodology for combining quasihomogeneous GTD, phase structure, and finite-distance shadows. The Kerr--AdS extension of Ref.~\cite{LadinoRomeroQuevedo2025} incorporated rotation, an asymmetric critical curve, and ensemble dependence. The asymptotically flat RN and Kerr analysis of Ref.~\cite{LadinoRomeroQuevedo2026} then connected Davies spinodal instabilities and GTD curvature structures with definite shadow configurations. In the present review these published results are reorganized using the three GTD metrics constructed from the same fundamental equation in Sec.~\ref{sec:GTD}. This distinction is important: the optical markers obtained in the published calculations remain properties of the same black-hole states, whereas the new three-metric analysis gives a more systematic GTD classification of those states.

\subsection{Optical representation of equilibrium states}

Let ${\cal E}$ denote the equilibrium space with thermodynamic coordinates $E^a$, and let the set of theoretical shadow observables introduced in Sec.~\ref{subsec:shadowobservables} be denoted by ${\cal O}_{\rm sh}^i$. Once the observer configuration is fixed, the spacetime and null-geodesic calculations define an optical map
\be
{\cal O}_{\rm sh}^{i}
=
{\cal O}_{\rm sh}^{i}(E^a),
\qquad
i=1,\ldots,N_{\rm sh}.
\label{eq:opticalmap}
\ee
The dependence is indirect. A thermodynamic state determines the conserved black-hole parameters, these parameters determine the spacetime metric, and the metric determines the critical null geodesics and their projection onto the observer's screen.

For a one-dimensional thermodynamic branch described directly by one of its physical thermodynamic variables, a shadow observable can be used locally as an alternative label whenever its derivative along that branch does not vanish. In that case any thermodynamic or GTD scalar can be represented as a function of the chosen shadow observable. This is a reparametrization of a physical branch, not the introduction of a new thermodynamic coordinate.

For a multidimensional equilibrium family, a single shadow radius is generally insufficient. The optical Jacobian is
\be
{\cal J}^{i}{}_{a}
=
{\cal O}^{i}_{{\rm sh},a}.
\label{eq:opticalJacobian}
\ee
If the equilibrium family has dimension $n$, a necessary local condition for a set of $N_{\rm sh}$ optical observables to reconstruct all thermodynamic coordinates is $ N_{\rm sh}\geq n$. When, in addition,
\be
{\rm rank}\,{\cal J}=n,
\label{eq:observabilitycondition}
\ee
the chosen optical observables contain locally enough independent information to parametrize the thermodynamic state after the observer and other nuisance parameters have been fixed. If the rank is smaller than $n$, an optical degeneracy remains. Eq.~\eqref{eq:observabilitycondition} is a local diagnostic motivated by the existing shadow-thermodynamic constructions; it should not be interpreted as a theorem of global observability. Maximal rank does not guarantee global injectivity, and different thermodynamic branches may still have the same optical image.

For any of the three GTD metrics, the scalar curvature remains a scalar on ${\cal E}$. If the optical map is regular on the branch under consideration, one may represent
\be
{\cal R}^{A}_{\rm GTD}(E^a)
\longrightarrow
{\cal R}^{A}_{\rm GTD}
\left({\cal O}^{i}_{\rm sh}\right),
\qquad
A=I,II,III.
\label{eq:GTDopticalrepresentation}
\ee
This notation denotes composition with the optical map. It does not identify the equilibrium manifold with the observer's screen, and the GTD curvature is not a spacetime curvature.

\subsection{Reissner--Nordstr\"om black hole}

\begin{figure*}[h]
  \centering
  \begin{minipage}[t]{0.333\linewidth}
    \centering
    \includegraphics[width=\linewidth]{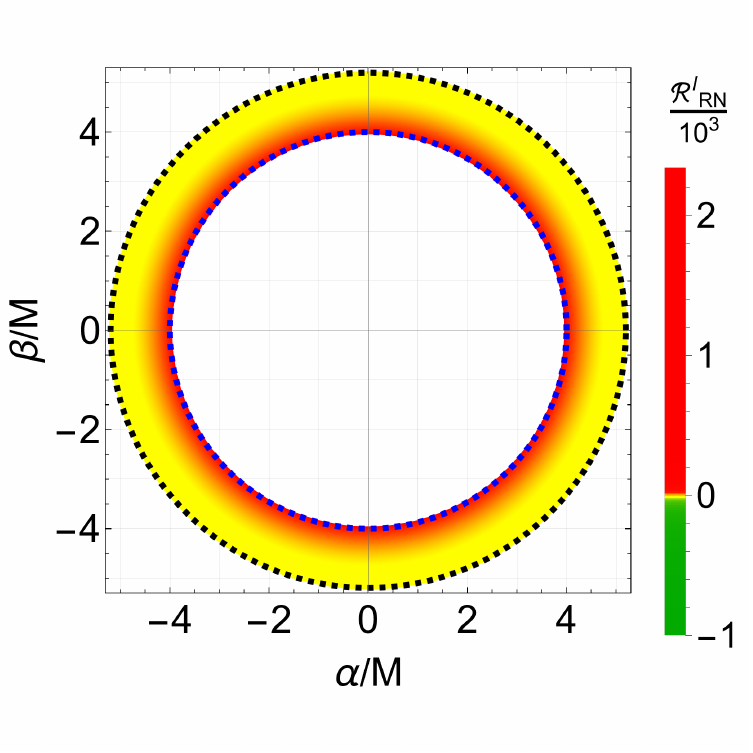}
    \\[2pt] \vspace{-0.6cm}\hspace{-3.5cm}\small{(a)}
  \end{minipage}\hfill
  \begin{minipage}[t]{0.333\linewidth}
    \centering
    \includegraphics[width=\linewidth]{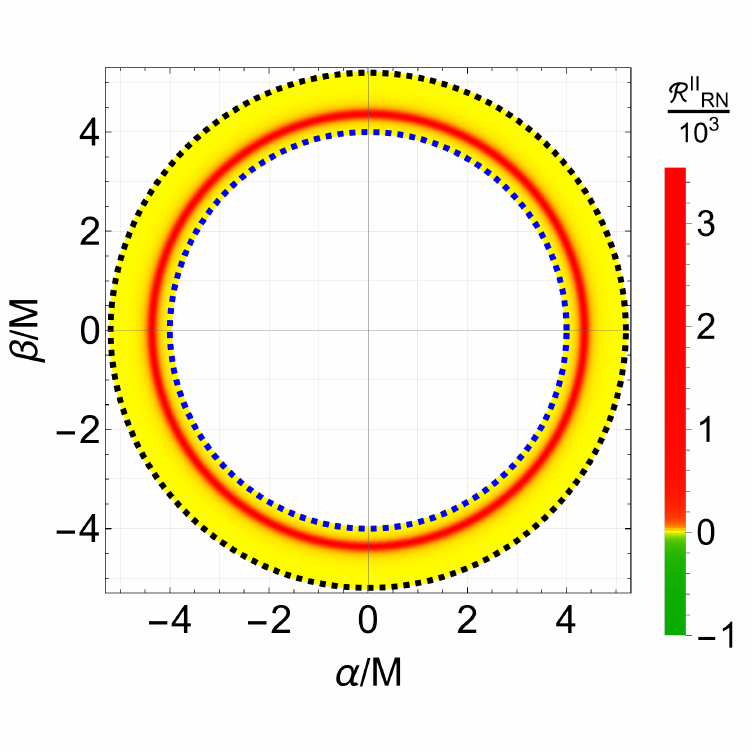}
    \\[2pt] \vspace{-0.6cm}\hspace{-3.5cm}\small{(b)}
  \end{minipage}\hfill
  \begin{minipage}[t]{0.333\linewidth}
    \centering
    \includegraphics[width=\linewidth]{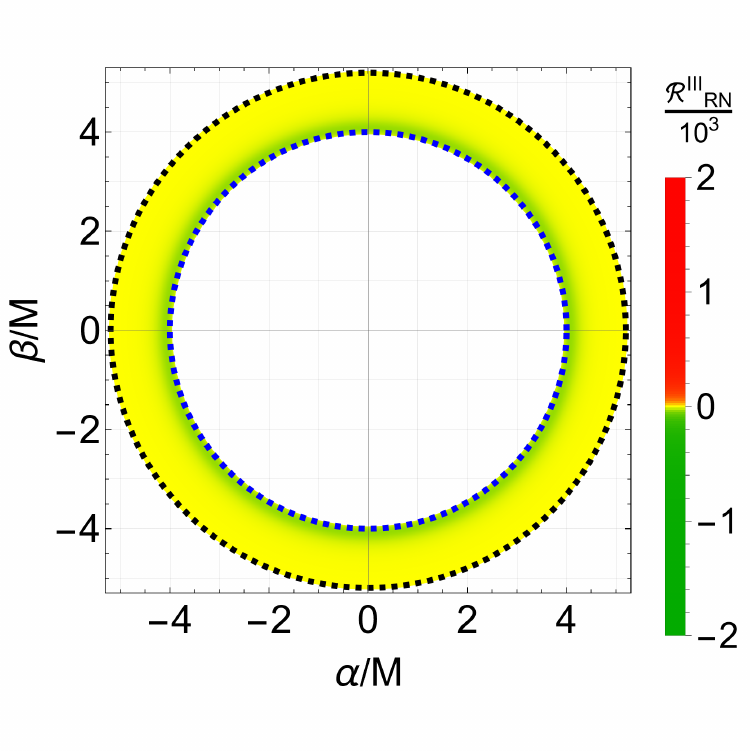}
    \\[2pt] \vspace{-0.6cm}\hspace{-3.5cm}\small{(c)}
  \end{minipage}
  \vspace{-0.8cm}
  \caption{(a)~${\cal R}_{\rm RN}^{I}$, (b)~${\cal R}_{\rm RN}^{II}$, and (c)~${\cal R}_{\rm RN}^{III}$ across the shadow’s profile of RN black holes ($r_0 \to \infty$). Dashed curves denote: extremal limit (blue) and Schwarzschild limit (black). }
  \label{fig:GTD1}
\end{figure*}

The RN case provides the simplest analytic realization of the correspondence. Sec.~\ref{sec:GTD} showed that the three GTD metrics constructed from the single fundamental equation $M(S,Q)$ distinguish three different thermodynamic structures. Condition~$I$, Eq.~\eqref{eq:RNI}, is $S_{ext}=\pi Q^2$ and coincides with extremality. Condition~$II$, Eq.~\eqref{eq:RNII}, is $S_D=3\pi Q^2$ and reproduces the fixed-charge Davies instability. Condition~$III$, Eq.~\eqref{eq:RNIII}, is $Q=0$ and identifies the Schwarzschild limit. The optical construction of Sec.~\ref{subsec:RNshadow} maps these three states into three different points of the RN shadow family. Consider first the Davies state (D-state). Substitution of $S_D=3\pi Q^2$ into the fundamental equation, Eq.~\eqref{eq:MRN}, gives directly
\be
\left.\frac{|Q|}{M}\right|_{D-state}
=
\frac{\sqrt{3}}{2}.
\label{eq:RNDaviescharge}
\ee
In the three-metric formulation of Sec.~\ref{sec:GTD}, the complete chain is
\be
C_Q^{-1}=0
\quad\Longleftrightarrow\quad
S_D=3\pi Q^2
\quad\Longleftrightarrow\quad
{\cal R}_{\rm RN}^{II}\rightarrow\infty
\quad\Longrightarrow\quad
R_{\rm sh}^{\rm RN}
=
R_{{\rm sh},D}^{\rm RN}.
\label{eq:RNDavieschain}
\ee
The final implication is an optical assignment of the same equilibrium state; it does not mean that the shadow produces the instability. The first GTD metric selects a different state. Condition~$I$, Eq.~\eqref{eq:RNI}, gives the extremal state, and as shown already in Sec.~\ref{subsec:RNshadow}, this  state has $r_{\rm ph}=2M$ and
\be
R_{{\rm sh},\rm ext}^{\rm RN}=4M.
\label{eq:RNextshadowV}
\ee
The curvature ${\cal R}_{\rm RN}^{I}$ diverges there, while the isothermal electric susceptibility and the inverse mixed response identified in Sec.~\ref{sec:GTD} are also singular. Thus, the first and second GTD metrics assign different geometric singularities to two thermodynamically distinct states: extremality and the Davies spinodal instability.

Condition~$III$, Eq.~\eqref{eq:RNIII}, is $Q=0$. It is not a curvature singularity of ${\cal R}_{\rm RN}^{III}$; rather, $g^{III}$ becomes degenerate there and the direct thermo-electric coupling vanishes. Optically this is the Schwarzschild endpoint,
\be
R_{{\rm sh},Q=0}^{\rm RN}=3\sqrt{3}\,M.
\label{eq:RNSchwarzschildshadowV}
\ee
The RN optical family therefore contains three distinguished markers with different physical meanings: the Schwarzschild limit at $3\sqrt{3}M$, the Davies instability at $R_{{\rm sh},D}^{\rm RN}$, and extremality at $4M$. The shadow radius varies smoothly between these states, so none of these optical values is itself an optical singularity.

This formulation differs from the potential--metric organization used in Ref.~\cite{LadinoRomeroQuevedo2026}, where the Legendre-transformed first metric and the mass-representation second metric were chosen so that both reproduced the Davies point.  The present review instead keeps the common fundamental equation $M(S,Q)$ for all three GTD metrics and thereby separates the extremal, Davies, and Schwarzschild structures.

Fig. \ref{fig:GTD1} presents the behavior of three GTD curvature scalars across the shadow profile of RN black holes for an observer at infinity. The first scalar, ${\cal R}_{\rm RN}^{I}$, accurately diverges at the extremal black hole limit, $R_{{\rm sh},\rm ext}^{\rm RN}=4M$. In contrast, the second scalar, ${\cal R}_{\rm RN}^{II}$, diverges precisely at the Davies point, effectively capturing the system's thermodynamic critical spinodal point at $R_{{\rm sh},D}^{\rm RN}$. The third scalar, ${\cal R}_{\rm RN}^{III}$, remains completely regular across the parameter space. Finally, none of the three GTD scalars in the RN spacetime currently exhibit physically valid regions of zero thermodynamic curvature, which correspond to non-interacting states (N-state), that satisfy the required physical conditions $S > 0$ and $T > 0$.

\subsection{Kerr black hole}

\begin{figure*}[h]
  \centering
  \begin{minipage}[t]{0.333\linewidth}
    \centering
    \includegraphics[width=\linewidth]{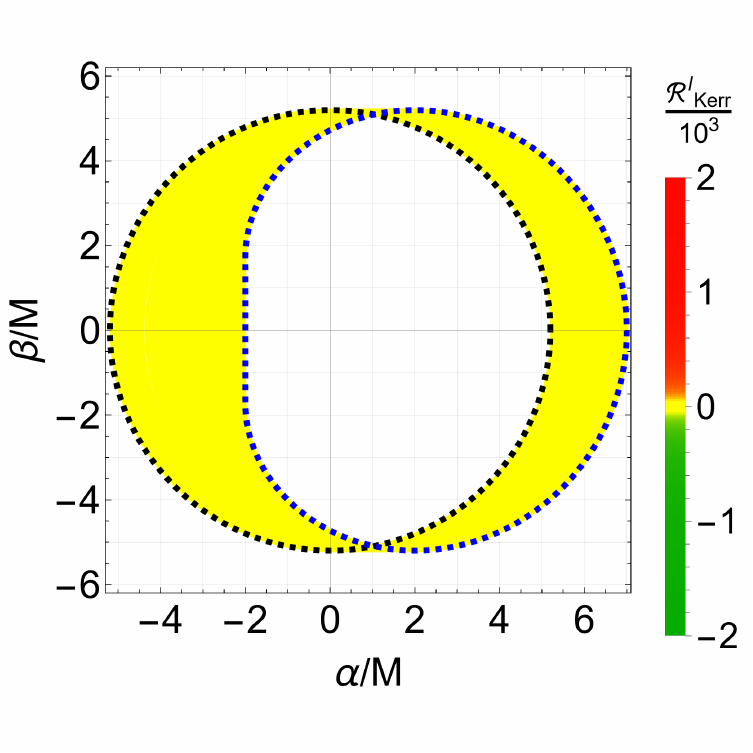}
    \\[2pt] \vspace{-0.6cm}\hspace{-3.5cm}\small{(a)}
  \end{minipage}\hfill
  \begin{minipage}[t]{0.333\linewidth}
    \centering
    \includegraphics[width=\linewidth]{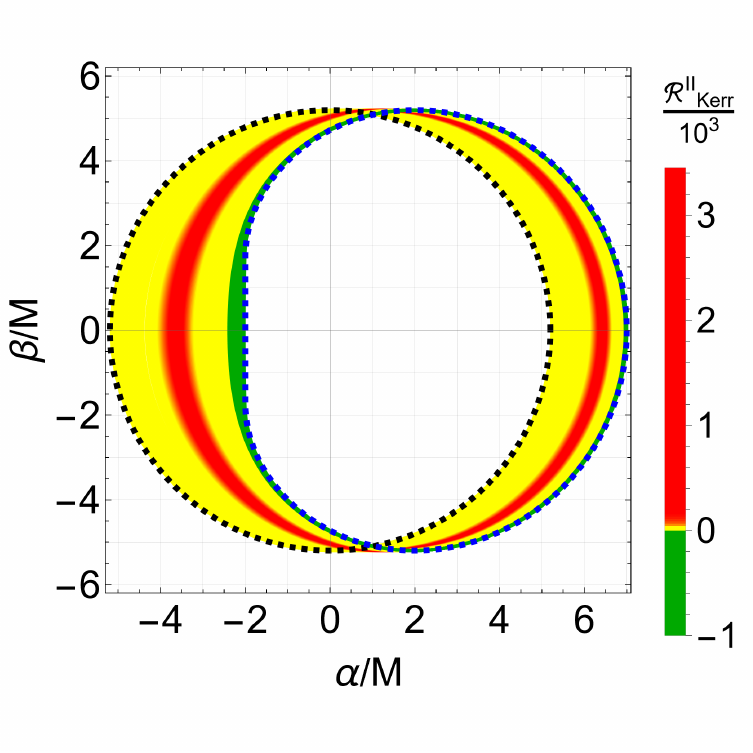}
    \\[2pt] \vspace{-0.6cm}\hspace{-3.5cm}\small{(b)}
  \end{minipage}\hfill
  \begin{minipage}[t]{0.333\linewidth}
    \centering
    \includegraphics[width=\linewidth]{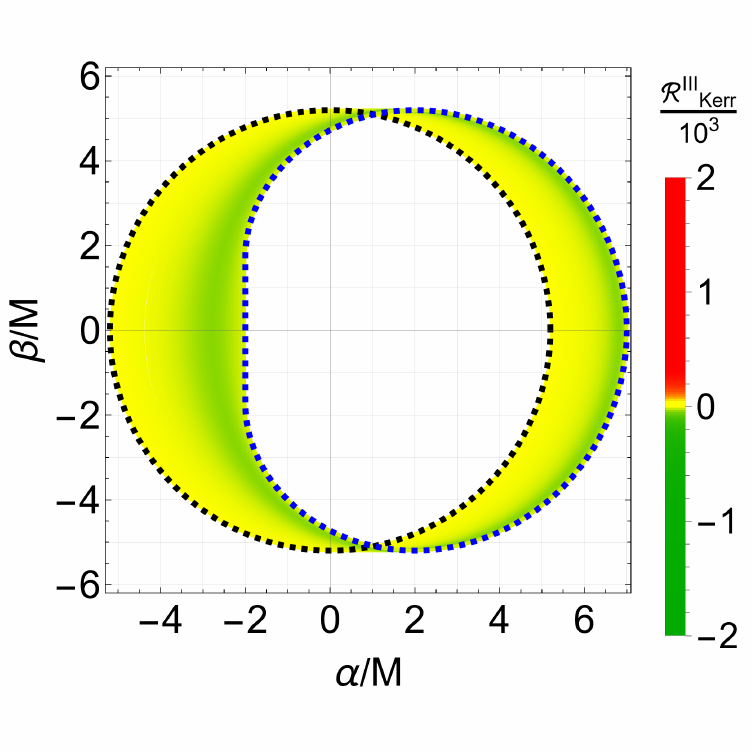}
    \\[2pt] \vspace{-0.6cm}\hspace{-3.5cm}\small{(c)}
  \end{minipage}
  \vspace{-0.8cm}
  \caption{(a)~${\cal R}_{\rm Kerr}^{I}$, (b)~${\cal R}_{\rm Kerr}^{II}$, and (b)~${\cal R}_{\rm Kerr}^{III}$ across the shadow’s profile of Kerr black holes  (  $\theta_0=90^\circ$, $r_0 \to \infty$). Dashed curves denote: extremal limit (blue) and Schwarzschild limit (black). }
  \label{fig:GTD2}
\end{figure*}

For Kerr, the same procedure must be applied to a nonspherical shadow. Sec.~\ref{sec:GTD} showed that condition~$I$, Eq.~\eqref{eq:KerrIeval}, has no physical solution and that the first GTD metric constructed from $M(S,J)$ is flat, i.e.,
${\cal R}_{\rm Kerr}^{I}=0$. Condition~$II$, Eq.~\eqref{eq:KerrII}, determines the fixed-$J$ Davies instability, where $C_J$ diverges and ${\cal R}_{\rm Kerr}^{II}$ has a singularity.  Condition~$III$, Eq.~\eqref{eq:KerrIII}, gives $J=0$, the Schwarzschild limit, where the direct thermo-rotational coupling vanishes and $g^{III}$ becomes degenerate. Moreover, evaluating the D-state, Eq.~(\ref{eq:KerrDavies}), in the Kerr fundamental equation, Eq.~\eqref{eq:MKerr}, gives
\be
\left.\frac{|J|}{M^2}\right|_{D-state}
=
\sqrt{2\sqrt{3}-3}
\simeq0.681.
\label{eq:KerrDaviesSpinV}
\ee
The optical image of this state is obtained by setting $a=J/M$ in the critical impact parameters Eqs.~\eqref{eq:Kerrxi}-\eqref{eq:KerrEta}, and projecting the resulting critical curve with Eq.~\eqref{eq:Kerrcelestialgeneral}. No additional radial variable is required in the thermodynamic discussion: $r_{\rm sph}$ retains its role from Sec.~\ref{subsec:Kerrshadow} as the parameter along each individual critical curve. For the fixed observer and the shadow-size convention adopted in Ref.~\cite{LadinoRomeroQuevedo2026}, the D-state is represented optically by the characteristic shadow radius $R_{{\rm sh},D}^{\rm Kerr}$.
Furthermore, as discussed in Sec.~\ref{sec:zero_temperature}, the second GTD scalar curvature also contains a physical zero, Eq.~\eqref{xzerokerr}. This condition represents an N-state characterized by the relation
\be
\left.\frac{|J|}{M^2}\right|_{N-state}
\simeq 0.974.
\label{eq:KerrCurvatureZeroShadowV}
\ee
The N-state must therefore remain distinct from the Davies marker. Finally, the third GTD metric provides another type of information. Its curvature has no Davies point in the physical Kerr domain, but the metric becomes degenerate at $J=0$, where the thermo-rotational coupling disappears. The optical image is the Schwarzschild critical curve with radius $3\sqrt{3}M$. In contrast, extremality, $S_{ext}^2=4\pi^2J^2$, is a physical boundary of the Kerr family but is not one of the three GTD equilibrium conditions obtained from $M(S,J)$ in Sec.~\ref{sec:GTD}. To visualize these features, Fig.~\ref{fig:GTD2} illustrates the behavior of the three GTD curvature scalars mapped across the Kerr black hole shadow profile. As expected, the first scalar, ${\cal R}_{\rm Kerr}^{I}$, is identically zero, corresponding to a flat GTD geometry. The second scalar, ${\cal R}_{\rm Kerr}^{II}$, correctly diverges at the Davies point at $R_{{\rm sh},D}^{\rm Kerr}$ and additionally features a curvature zero at $R_{{\rm sh},N}^{\rm Kerr}$, which identifies a N-state. Finally, the third scalar, ${\cal R}_{\rm Kerr}^{III}$, remains entirely regular, exhibiting neither divergences nor zeros across the thermodynamic equilibrium space.

As in RN, the organization adopted here should be distinguished from the published potential--metric construction of Ref.~\cite{LadinoRomeroQuevedo2026}, in which  the first metric was constructed from a Legendre-transformed potential and, together with $g^{II}[M]$, reproduced the fixed-$J$ Davies point. In the common-potential construction of this review, $g^I[M]$ is flat and only $g^{II}[M]$ carries the D-state. 

The Kerr example also makes clear why one shadow scalar is not sufficient in general. At fixed inclination and along a one-parameter spin sequence, a characteristic size can label the states used above. If the inclination or additional black-hole parameters are unknown, size and shape changes can be degenerate. The full critical curve, or a sufficiently large set of independent shadow observables, must then be used in accordance with Eq.~(\ref{eq:observabilitycondition}).

\subsection{Kerr--AdS black hole}

Kerr--AdS is the higher-dimensional example. Its equilibrium space has coordinates $(S,J,l)$, and Sec.~\ref{sec:GTD} showed that quasihomogeneity requires all three variables. The three GTD metrics are therefore constructed from the same fundamental equation $M(S,J,l)$. In the regular physical domain analyzed in Ref.~\cite{LadinoRomeroQuevedo2025}, condition~$I$, Eq.~\eqref{eq:KerrAdSI}, does not generate a generic physical singularity and the simultaneous conditions associated with $g^{III}$, Eqs.~\eqref{eq:KerrAdSIIIa}-\eqref{eq:KerrAdSIIId}, do not produce an additional admissible equilibrium locus. The physically relevant local structure is contained in condition~$II$, Eq.~\eqref{eq:KerrAdSII}.
%
As emphasized in Sec.~\ref{Kerr-AdS sec}, formal roots must be restricted to the physical thermodynamic domain. In the canonical domain considered in Ref.~\cite{LadinoRomeroQuevedo2025}, the thermal branch $M_{,SS}=0$ supplies the relevant local critical structure, whereas the solutions of the coupled branch occur outside the rotational domain retained in that analysis. The optical map must use exactly the Kerr--AdS conventions established in Sec.~\ref{subsec:KerrAdSshadow}. A thermodynamic state $(S,J,l)$ first determines $M=M(S,J,l)$. The spacetime rotation parameter and metric parameter are then
\be
a=\frac{J}{M},
\qquad
\Xi=1-\frac{a^2}{l^2},
\qquad
m=M\Xi^2.
\label{eq:KerrAdSThermoToMetricV}
\ee
These are the parameters entering $\Delta_r$ and the critical photon equations of Sec.~\ref{subsec:KerrAdSshadow}. In particular, no rescaled impact parameter is introduced. The same quantities
$ 
\xi=\frac{L}{E},
\ 
\eta=\frac{{\cal Q}}{E^2}
$ 
are used throughout the review. The critical values $\xi(r_{\rm sph})$ and $\eta(r_{\rm sph})$ are those of Eqs.~\eqref{eq:KerrAdSxicriticalShadow} and \eqref{eq:KerrAdSetacriticalShadow}. Their finite-distance projection is described by the local celestial angles $(\rho,\sigma)$ of Eqs.~\eqref{eq:KerrAdSrhoShadow} and \eqref{eq:KerrAdSsigmaShadow}, followed by the stereographic screen coordinates $(x,y)$ of Eq.~\eqref{eq:KerrAdSscreenShadow}. The symbols $(x,y)$ are used only for the screen of the observer and never as thermodynamic reparametrizations.

\begin{figure*}[h]
  \centering
  \begin{minipage}[t]{0.333\linewidth}
    \centering
    \includegraphics[width=\linewidth]{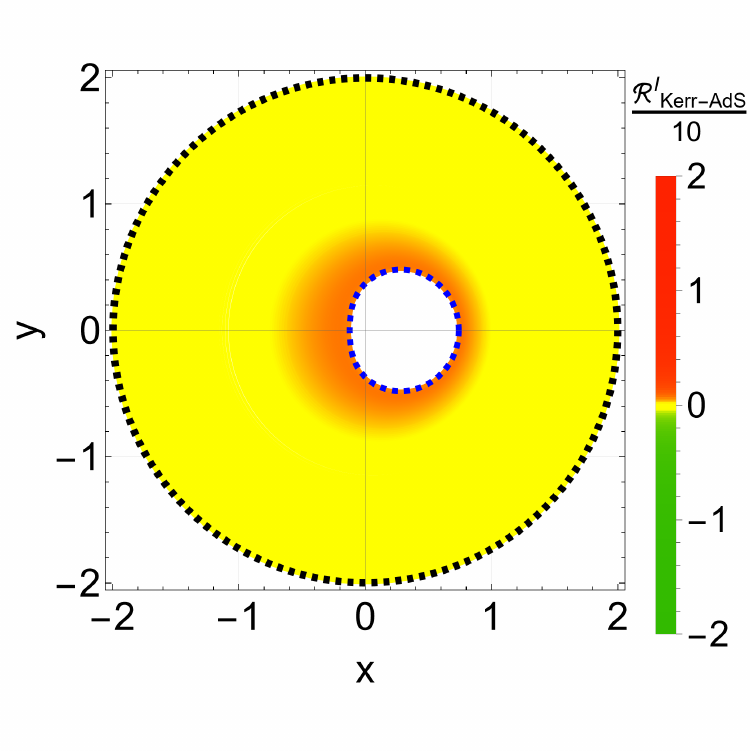}
    \\[2pt] \vspace{-0.6cm}\hspace{-3.5cm}\small{(a)}
  \end{minipage}\hfill
  \begin{minipage}[t]{0.333\linewidth}
    \centering
    \includegraphics[width=\linewidth]{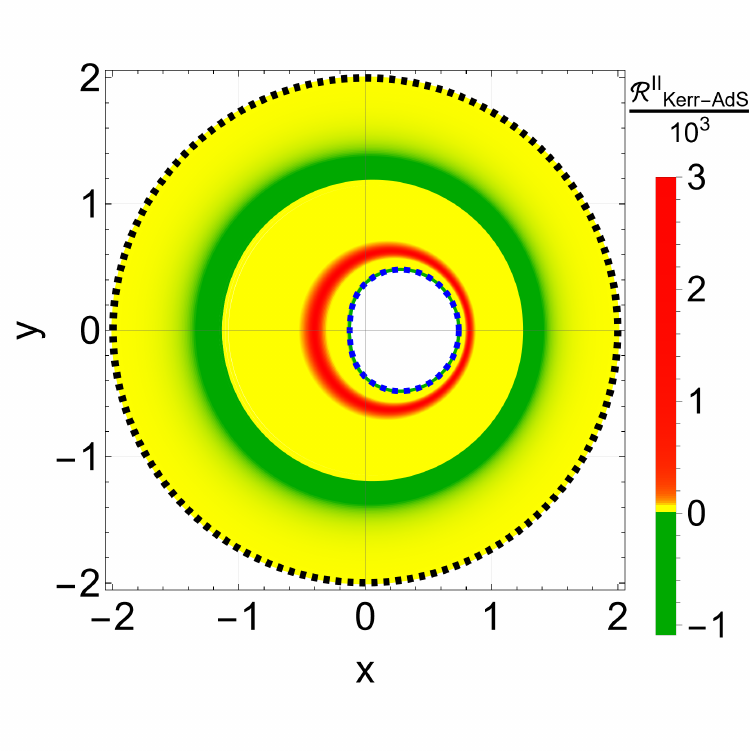}
    \\[2pt] \vspace{-0.6cm}\hspace{-3.5cm}\small{(b)}
  \end{minipage}\hfill
  \begin{minipage}[t]{0.333\linewidth}
    \centering
    \includegraphics[width=\linewidth]{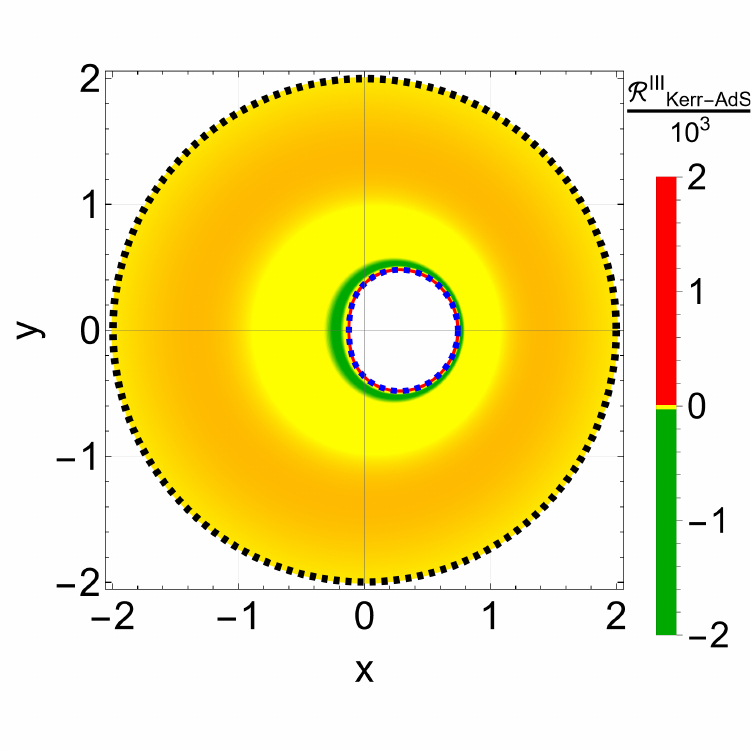}
    \\[2pt] \vspace{-0.6cm}\hspace{-3.5cm}\small{(c)}
  \end{minipage}\hfill
   \begin{minipage}[t]{0.333\linewidth}
    \centering
    \includegraphics[width=\linewidth]{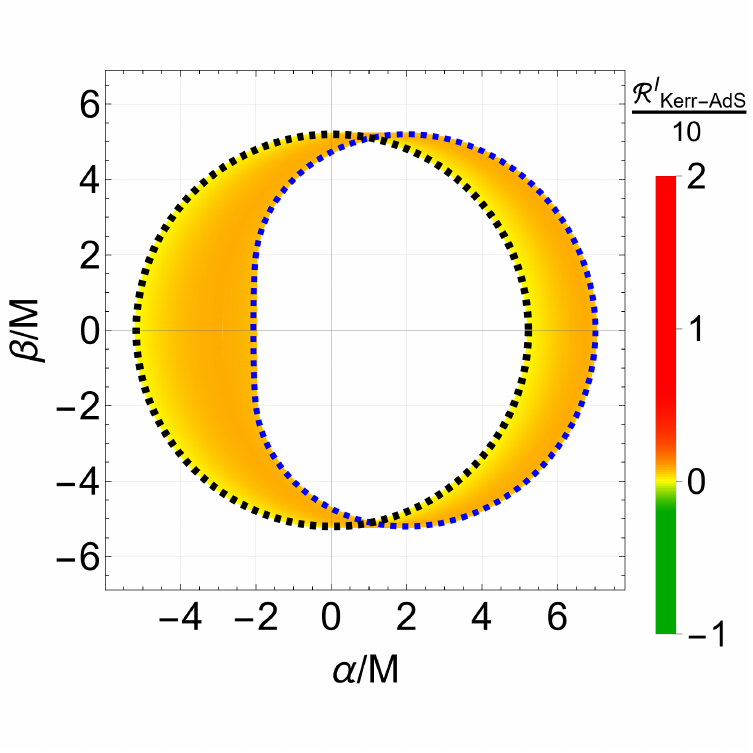}
    \\[2pt] \vspace{-0.6cm}\hspace{-3.5cm}\small{(a)}
  \end{minipage}\hfill
  \begin{minipage}[t]{0.333\linewidth}
    \centering
    \includegraphics[width=\linewidth]{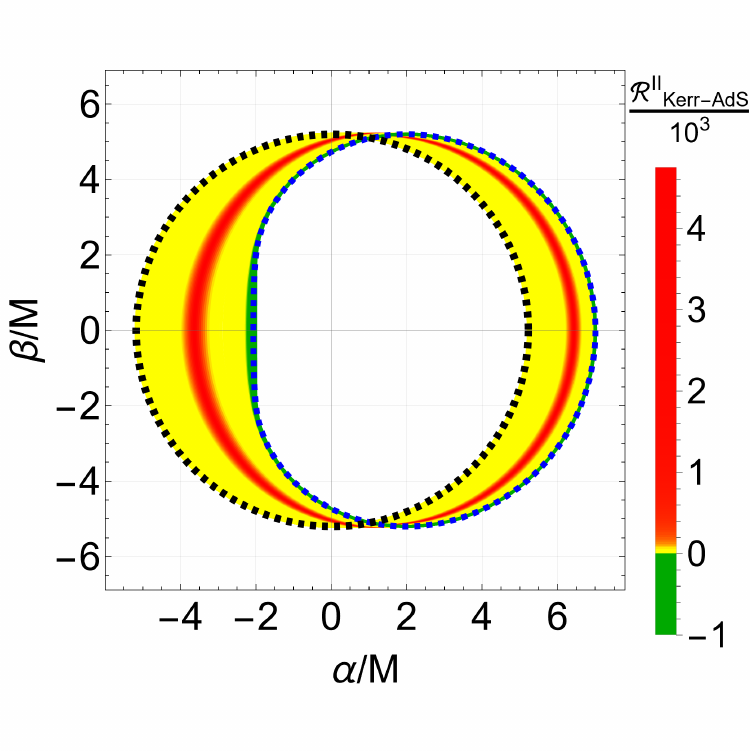}
    \\[2pt] \vspace{-0.6cm}\hspace{-3.5cm}\small{(b)}
  \end{minipage}\hfill
  \begin{minipage}[t]{0.333\linewidth}
    \centering
    \includegraphics[width=\linewidth]{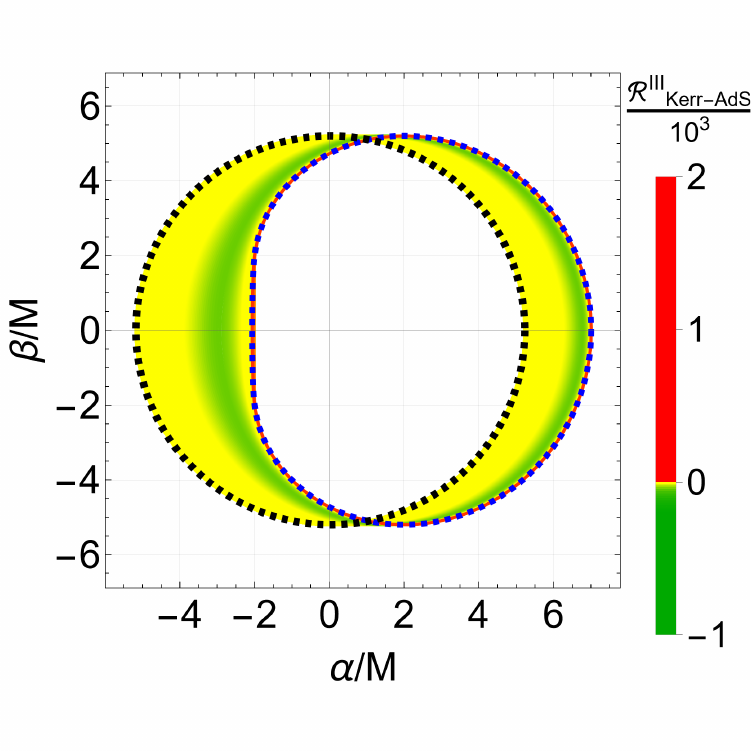}
    \\[2pt] \vspace{-0.6cm}\hspace{-3.5cm}\small{(c)}
  \end{minipage}
  \vspace{-0.8cm}
  \caption{(a)~${\cal R}_{\rm Kerr-AdS}^{I}$, (b)~${\cal R}_{\rm Kerr-AdS}^{II}$, and (c)~${\cal R}_{\rm Kerr-AdS}^{III}$ across the shadow’s profile of Kerr-AdS black holes ($\theta_0=90^\circ$, $r_0 = 100 \sqrt{J}$, $l = 10 \sqrt{J}$). (d)~${\cal R}_{\rm Kerr-AdS}^{I}$, (e)~${\cal R}_{\rm Kerr-AdS}^{II}$, and (f)~${\cal R}_{\rm Kerr-AdS}^{III}$ across the shadow’s profile of Kerr-AdS black holes in Bardeen’s coordinates of Eq.~\eqref{eq:bardeen},and in entropy space ($l = 1 \times 10^{12} \sqrt{J}$,  $\theta_0=90^\circ$, $r_0 = 4.1\times 10^{16} \sqrt{J}$). Dashed curves denote: extremal limit (blue) and photon region limit (black). }
  \label{fig:GTD3}
\end{figure*}

For a fixed observer, the complete correspondence begins with the thermodynamic state specified by $(S,J,l)$, which determines the physical and metric parameters $(M,a,m,l)$. These parameters determine the critical impact parameters $\xi(r_{\rm sph})$ and $\eta(r_{\rm sph})$ associated with the spherical photon orbits. The impact parameters then determine the celestial angles $(\rho,\sigma)$ and, finally, the position $(x,y)$ of the shadow boundary on the observer's screen.
Every thermodynamic state determines a complete critical curve on the screen. Consequently, a physical zero of Eq.~(\ref{eq:KerrAdSIIa}) is represented by a definite shadow profile rather than, in general, by a unique radius.

For practical comparisons one may compress the critical curve into the size and distortion observables $R_{\rm sh}$ and $\delta_{\rm sh}$ defined in Eq.~\eqref{eq:KerrAdSshadowobservables}, as was done in Ref.~\cite{LadinoRomeroQuevedo2025}. Such a reduction is useful for plotting thermodynamic branches, but it does not change the dimensionality of the equilibrium space. In an inverse problem, a single $R_{\rm sh}$ cannot generically reconstruct $(S,J,l)$; several independent optical observables are required, and their Jacobian must satisfy Eq.~(\ref{eq:observabilitycondition}) after the observer variables have been fixed.

The canonical and grand-canonical ensembles must also remain distinct. At fixed $J$, the Kerr--AdS thermodynamics contains the small/large black-hole structure, including the local spinodal branches and the critical state, Eq.~\eqref{eq:KerrAdScritical}. As discussed in Sec.~\ref{sec:GTD}, Legendre invariance of the GTD construction does not imply equivalence between different ensembles. Although the optical profile may represent states in either ensemble, their thermodynamic interpretation must be established independently.

Fig.~\ref{fig:GTD3} illustrates the behavior of the three GTD curvature scalars mapped across the shadow profiles of Kerr--AdS black holes in two distinct regimes. Figs.~\ref{fig:GTD3}(a)--(c) depict the projections onto the dimensionless observer screen coordinates $(x,y)$ for a small AdS scale ($l = 10\sqrt{J}$). In this regime, the first scalar ${\cal R}_{\rm Kerr-AdS}^{I}$ is no longer flat or identically zero as in the Kerr case, remaining smooth with neither zeros nor divergences across the shadow profile. The second scalar ${\cal R}_{\rm Kerr-AdS}^{II}$ diverges precisely at the D-states, $(S_{min},T_{min})$ and $(S_{m},T_{m})$, while exhibiting a zero-curvature contour identifies an N-state at the shadow radius $R_{{\rm sh},N}^{\rm Kerr-AdS}$. The third scalar ${\cal R}_{\rm Kerr-AdS}^{III}$ exhibits no divergences but has two distinct zeros associated with different thermodynamic N-states.

On the other hand, Figs.~\ref{fig:GTD3}(d)--(f) showcase the scalar behaviors in Bardeen coordinates $(\alpha/M, \beta/M)$ in the asymptotically flat regime ($l = 1 \times 10^{12} \sqrt{J}$). Here, ${\cal R}_{\rm Kerr-AdS}^{I}$ preserves its flat geometry. The second scalar ${\cal R}_{\rm Kerr-AdS}^{II}$ closely resembles its flat Kerr counterpart, as shown in Fig.~\ref{fig:GTD2}.  Finally, ${\cal R}_{\rm Kerr-AdS}^{III}$ likewise mirrors the divergence free behavior of pure Kerr while retaining an N-state. 

Therefore, the role of GTD in Kerr--AdS is therefore organizational as well as geometric. The three-dimensional quasihomogeneous fundamental equation fixes the GTD construction, condition~$II$, Eq.~\eqref{eq:KerrAdSII}, separates the thermal and coupled-response sectors, and the physical-domain restrictions determine which formal geometric structures are admissible. The shadow then supplies an observer-dependent representation of those already identified states.


\subsection{Comparison with thermodynamic geometry}

Previous studies established important connections between black-hole optics and thermodynamics by showing that photon-orbit quantities and shadow observables can reflect thermodynamic branches and critical behavior in AdS black holes, while thermodynamic geometry provided a geometric characterization of equilibrium states \cite{WeiLiu2018,WeiLiuWang2019,ZhangGuo2020}. GTD extends this connection by providing a Legendre-invariant geometric framework in which the thermodynamic structure of the equilibrium space can be systematically related to optical observables. From this perspective, the central issue is to determine how the structures identified by the different GTD metrics are represented in the optical space and what additional thermodynamic information can be extracted from this correspondence.

For the asymptotically flat RN black hole, the standard entropy-based thermodynamic geometry is flat in the usual two-dimensional state space and does not reproduce the fixed-charge Davies instability \cite{Ruppeiner2008}. In the common-potential GTD construction used here, $g^I[M]$ is singular at extremality, $g^{II}[M]$ is singular at the Davies state, and $g^{III}[M]$ becomes degenerate in the Schwarzschild limit. The same RN shadow family therefore represents three distinct GTD structures rather than a single curvature singularity.

For the Kerr black hole, thermodynamic geometry is non-flat, but its singularity structure depends on the thermodynamic state space and ensemble and does not generically reproduce the fixed-$J$ Davies curve in the standard Kerr--Newman construction \cite{Ruppeiner2008}. The present GTD description again separates the three metric families. The first metric constructed from $M(S,J)$ is flat, the second reproduces the Davies divergence of $C_J$ and also contains an N-state. The third metric is curved but has no Davies point and becomes degenerate at the Schwarzschild limit. The shadow markers associated with the D-states and N-states are therefore attached specifically to the structures of $g^{II}[M]$ in the organization adopted in this review.

For Kerr--AdS, the comparison is more balanced. Thermodynamic geometry can reproduce rotating AdS instabilities, critical behavior, and effective interaction sectors once the appropriate fluctuation space and ensemble are specified \cite{Sahay2017,WeiLiu2021,DehyadegariSheykhi2021}. The distinctive GTD feature is structural: the complete quasihomogeneous fundamental equation is embedded in a Legendre-invariant phase-space construction, the AdS radius $l$ participates explicitly in the equilibrium geometry, and condition~$II$, Eq.~\eqref{eq:KerrAdSII}, separates different response sectors within the same three-dimensional state space \cite{LadinoRomeroQuevedo2025}.

Consequently, the black hole shadows and their observables associated with GTD behaviors (such as singularities that coincide with Davies points or thermodynamic critical points) should be identical when analyzed from the perspective of thermodynamic geometry. 
For a fixed spacetime state, the shadow itself is independent of the thermodynamic-geometric framework adopted. The distinction between GTD and thermodynamic geometry therefore lies in the thermodynamic-geometric classification assigned to that state, including the curvature structure and its interpretation in terms of effective microscopic interactions.
Indeed, as highlighted in Ref. \cite{LadinoRomeroQuevedo2025}, the microscopic interactions of Kerr-AdS black holes differ significantly: thermodynamic geometry reveals different zero-curvature points and starkly contrasting attractive and repulsive regimes for both small and large black hole phases compared to GTD. This discrepancy stems primarily from the distinct choice of thermodynamic variables used to geometrically describe the system.

Thus, GTD does not replace the shadow map developed in ordinary shadow thermodynamics. It supplies a Legendre-invariant thermodynamic-geometric classification of the state that is being mapped. The strongest statements are those for which a GTD condition or curvature singularity can be identified with a concrete thermodynamic property and then assigned an optical image. Curvature signs and zeros provide additional geometric information, but their interpretation in terms of effective microscopic attraction, repulsion, or non-interaction remains metric dependent. The shadow does not remove this dependence, since all GTD descriptions refer to the same underlying spacetime state. It provides an optical representation of these states without altering their thermodynamic meaning. This distinction is essential for the SM diagrams discussed in Sec.~\ref{sec:SMdiagrams}.


\section{Shadow--microstructure diagrams}
\label{sec:SMdiagrams}

The GTD--shadow correspondence of Sec.~\ref{sec:GTDshadows} assigns optical configurations to thermodynamic states whose GTD properties have already been identified. SM diagrams organize this information in a common representation. The terminology was introduced explicitly for asymptotically flat RN and Kerr black holes in Ref.~\cite{LadinoRomeroQuevedo2026}, while closely related curvature--shadow constructions had previously been developed for RN--AdS and Kerr--AdS black holes in Refs.~\cite{LadinoRomeroQuevedo2024,LadinoRomeroQuevedo2025}.

The term ``microstructure'' requires clarification. In the present
context, it does not refer to an explicitly identified set of microscopic
degrees of freedom of the black hole. Rather, it denotes the effective
microscopic interpretation conventionally associated with thermodynamic
curvature \cite{WeiLiuMann2019Ruppeiner}. In this phenomenological interpretation, the sign and
singular structure of the curvature are used to characterize effective
interaction sectors of the underlying thermodynamic system. Accordingly,
a SM diagram does not provide a direct image of
black-hole microscopic constituents. It represents, in optical variables,
the GTD curvature structure assigned to an equilibrium state. Establishing
a statistical-mechanical foundation for this interpretation remains an
open problem.

The SM diagrams introduced in Ref.~\cite{LadinoRomeroQuevedo2026} were constructed by considering GTD metrics associated with different thermodynamic potentials. In the present review, we adopt an alternative and complementary approach in which the three GTD metrics are constructed from the same fundamental equation. This common-potential formulation provides a uniform setting for comparing the thermodynamic structures described by the different metrics. The resulting RN and Kerr diagrams therefore offer a complementary characterization of the same black-hole states, while their optical locations remain determined by the corresponding spacetime and observer configuration.

Thus, an SM diagram represents, in optical variables, three types of information established independently: the thermodynamic branch and its stability properties, the curvature or degeneracy structure of a selected GTD metric, and the shadow observable or set of observables associated with the same spacetime state.

\subsection{Definition and interpretation}
\label{subsec:SMdefinition}

For a physical branch of black-hole states, let a macroscopic parameter be denoted generically by $\lambda$, and let ${\cal O}_{\rm sh}$ be one of the theoretical shadow observables introduced in Sec.~\ref{subsec:shadowobservables}. For a selected GTD metric $g^A$, with $A=I,II,III$, one may write
\be
{\cal R}_{\rm GTD}^{A}
=
{\cal R}_{\rm GTD}^{A}(\lambda),
\qquad
{\cal O}_{\rm sh}
=
{\cal O}_{\rm sh}(\lambda).
\label{eq:SMparametric}
\ee
Whenever the optical map is locally regular, the parameter can be eliminated and the same GTD scalar can be represented as ${\cal R}_{\rm GTD}^{A}({\cal O}_{\rm sh})$. For a multidimensional thermodynamic family this representation must be generalized to a sufficient set of optical observables, and the local rank condition Eq.~\eqref{eq:observabilitycondition} determines whether those observables contain enough information to reconstruct the thermodynamic state.

As in thermodynamic geometry, the sign convention used in the GTD--shadow literature associates positive thermodynamic curvature with an effective repulsive interaction, negative curvature with an effective attractive interaction, and vanishing curvature with an effectively non-interacting configuration \cite{LadinoRomeroQuevedo2024,LadinoRomeroQuevedo2025,LadinoRomeroQuevedo2026}. These interpretations are phenomenological. They characterize correlations encoded by the equilibrium geometry and should not be identified with literal microscopic force laws. 
 \begin{table}[ht!]
\centering
\caption{Classification of microscopic thermodynamic phases.}
\label{tab:phaseclassification}
\begin{tabular}{c c c}
\hline
\textbf{Label} & \textbf{Scale} & \textbf{Interaction} \\
\hline
AS & Small & Attractive \\
NS & Small & Non-interactive \\
RS & Small & Repulsive \\
\hline
AI & Intermediate & Attractive \\
NI & Intermediate & Non-interactive \\
RI & Intermediate & Repulsive \\
\hline
AL & Large & Attractive \\
NL & Large & Non-interactive \\
RL & Large & Repulsive \\
\hline
\multicolumn{3}{c}{
$\mathrm{A}:~\mathcal{R}<0,\qquad
 \mathrm{N}:~\mathcal{R}=0,\qquad
 \mathrm{R}:~\mathcal{R}>0$
} \\
\hline
\end{tabular}
\end{table}

A curvature singularity has a different status from a curvature sign or a curvature zero. Its physical meaning must be established by comparison with a thermodynamic structure such as a response-function divergence, a stability boundary, extremality, or another distinguished equilibrium locus.  An SM diagram must therefore retain the thermodynamic meaning attached to every special optical marker. In Ref.~\cite{LadinoRomeroQuevedo2026}, compact labels were introduced to characterize thermodynamic branches according to their curvature signs. This terminology is summarized in Table~\ref{tab:phaseclassification}.

Next, we describe the thermodynamic branch and curvature sign associated with each GTD metric separately. This notation facilitates comparison among the three metric structures and allows the D-states, N-states, and their corresponding curvature properties to be identified explicitly. In Figs.~\ref{fig:GTD-SMD1}, \ref{fig:GTD-SMD2}, and \ref{fig:GTD-SMD3}, we present the corresponding SM diagrams, detailing the microscopic thermodynamic phases and their respective constraints. Notably, for asymptotically flat RN and Kerr black holes, there is no natural external length scale against which the horizon radius can be compared to define “small” and “large” branches. These labels are therefore used conventionally, in analogy with the corresponding phases in AdS, where the AdS radius $l$ provides a natural reference scale \cite{KubiznakMannTeo2017}.


\subsection{Reissner--Nordstr\"om black hole}
\label{subsec:SMRN}

\begin{figure*}[h]
  \centering
  \begin{minipage}[t]{0.685\linewidth}
    \centering
    \includegraphics[width=\linewidth]{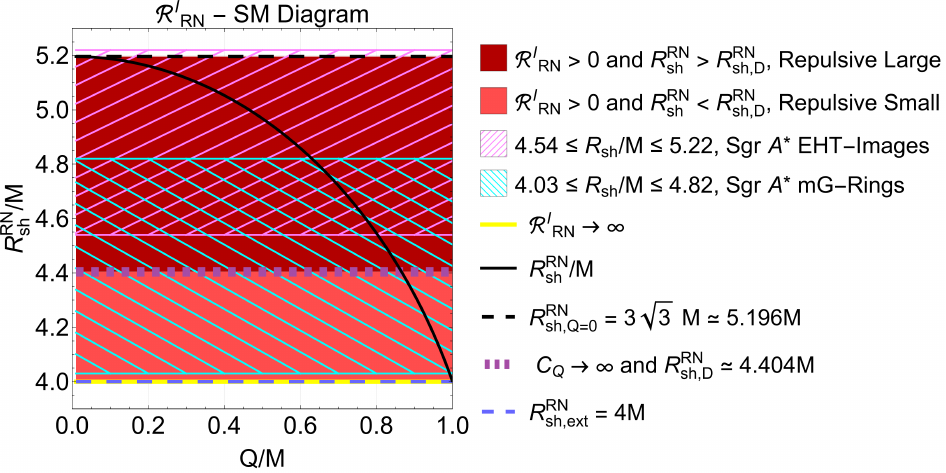}
    \\[2pt] \vspace{-0.6cm}\hspace{-9cm}\small{(a)}
  \end{minipage}\hfill
  \begin{minipage}[t]{0.315\linewidth}
    \centering
    \includegraphics[width=\linewidth]{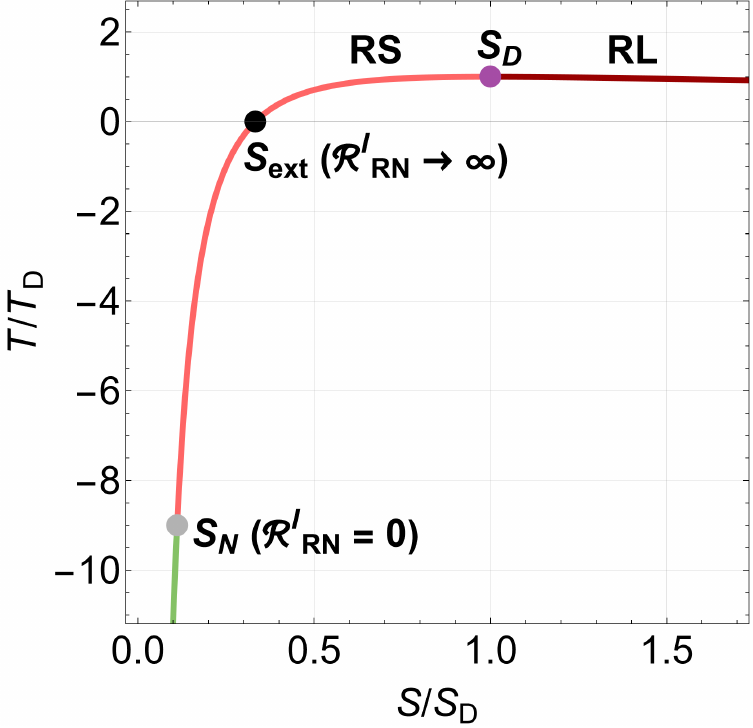}
    \\[2pt] \vspace{-0.6cm}\hspace{-3.5cm}\small{(b)}
  \end{minipage}\hfill
    \begin{minipage}[t]{0.685\linewidth}
    \centering
    \includegraphics[width=\linewidth]{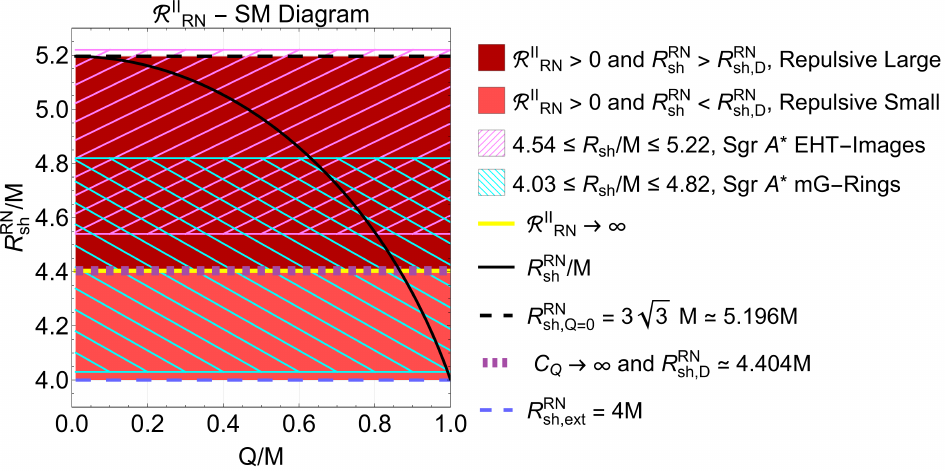}
    \\[2pt] \vspace{-0.6cm}\hspace{-9cm}\small{(c)}
  \end{minipage}\hfill
  \begin{minipage}[t]{0.315\linewidth}
    \centering
    \includegraphics[width=\linewidth]{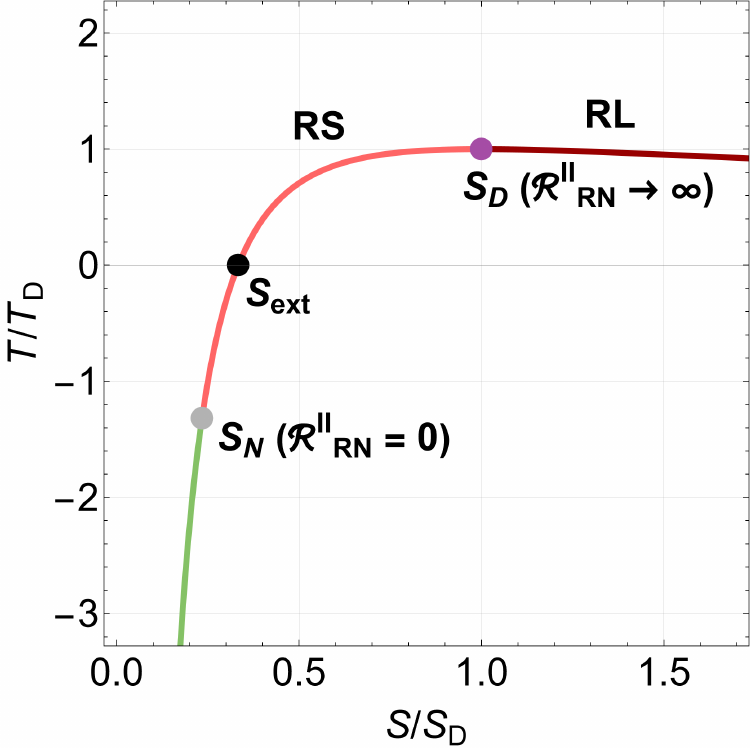}
    \\[2pt] \vspace{-0.6cm}\hspace{-3.5cm}\small{(d)}
  \end{minipage}\hfill
    \begin{minipage}[t]{0.685\linewidth}
    \centering
    \includegraphics[width=\linewidth]{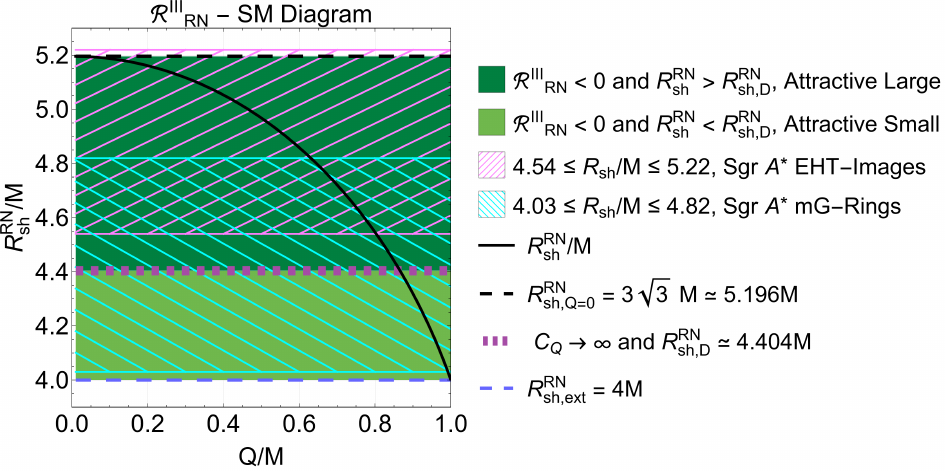}
    \\[2pt] \vspace{-0.6cm}\hspace{-9cm}\small{(e)}
  \end{minipage}\hfill
  \begin{minipage}[t]{0.315\linewidth}
    \centering
    \includegraphics[width=\linewidth]{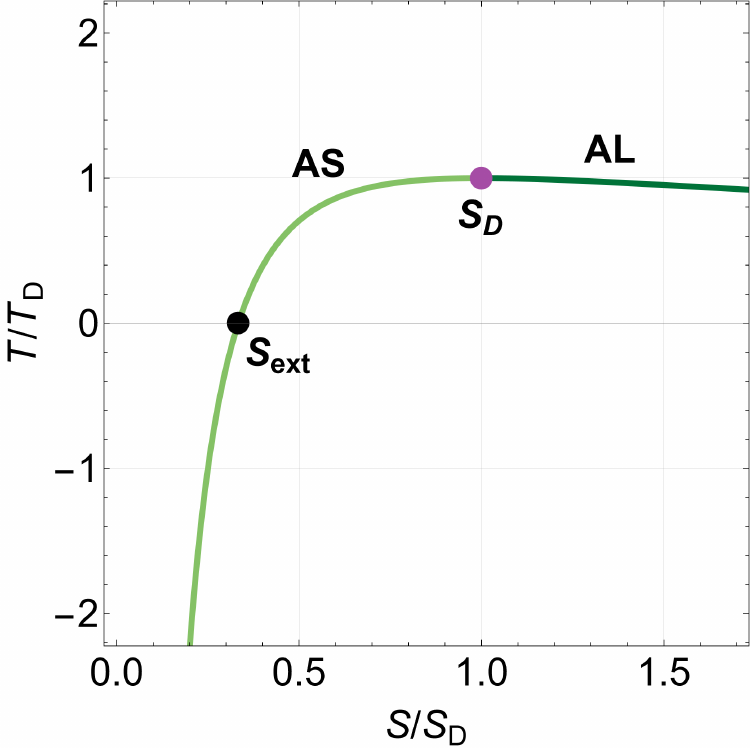}
    \\[2pt] \vspace{-0.6cm}\hspace{-3.5cm}\small{(f)}
  \end{minipage}
  \vspace{-0.7cm}
  \caption{SM diagrams for RN black holes using $R_{{\rm sh}}^{\rm RN}/M$ ($r_0 \to \infty$) and $Q/M$, and their corresponding $T/T_D$ vs. $S/S_D$ behaviors. (a)~and~(b)~for ${\cal R}_{\rm RN}^{I}$. (c)~and~(d)~for ${\cal R}_{\rm RN}^{II}$. (e)~and~(f)~for ${\cal R}_{\rm RN}^{III}$.}
  \label{fig:GTD-SMD1}
\end{figure*}

For the RN black hole, the optical family is one dimensional once the mass scale and observer convention are fixed. The asymptotic shadow radius is a monotonic function of $|Q|/M$, decreasing from $3\sqrt{3}M$ at $Q=0$ to $R_{{\rm sh},ext}^{\rm RN}=4M$ at extremality. Sec.~\ref{sec:GTDshadows} identified three distinguished states,
\be
Q=0,
\qquad
S_D=3\pi Q^2,
\qquad
S_{ext}=\pi Q^2,
\label{eq:SMRNthree}
\ee
corresponding respectively to the Schwarzschild limit, the Davies spinodal instability, and extremality.
Moreover, the D-state satisfies
\be
\left.\frac{|Q|}{M}\right|_{D-state}
=
\frac{\sqrt{3}}{2},
\qquad r_{{\rm ph},D}
=
\frac{3+\sqrt{3}}{2}\,M,
\qquad
R_{{\rm sh},D}^{\rm RN}
=
\sqrt{9+6\sqrt{3}}\,M
\simeq4.404\,M,
\label{eq:SMRNDavies}
\ee
and is a singularity  of ${\cal R}_{\rm RN}^{II}$. The extremal state has $|Q|/M=1$ and $R_{{\rm sh},ext}^{\rm RN}=4M$ and corresponds to  a singularity  of ${\cal R}_{\rm RN}^{I}$. The Schwarzschild limit has $R_{\rm sh}^{\rm RN}=3\sqrt{3}M$ and is the degeneracy locus of $g^{III}$. The curvature signs provide additional information. For positive $\beta_M$ and throughout the charged physical domain, the explicit expressions of Sec.~\ref{subsec RNGTD} imply
\be
{\cal R}_{\rm RN}^{I}>0,
\qquad
{\cal R}_{\rm RN}^{II}>0,
\qquad
{\cal R}_{\rm RN}^{III}<0,
\label{eq:SMRNsigns}
\ee
away from the corresponding singular or degenerate loci. Fig. \ref{fig:GTD-SMD1} presents the SM diagrams for RN black holes in the parameter space of $R_{\rm sh}^{\rm RN}/M$ versus $Q/M$, alongside their corresponding thermodynamic behaviors of $T/T_D$ versus $S/S_D$. The analysis is divided into three distinct thermodynamic curvature scalars ${\cal R}_{\rm RN}^{I}$, ${\cal R}_{\rm RN}^{II}$, and ${\cal R}_{\rm RN}^{III}$, revealing different thermodynamic microscopic phases depending on the shadow observation regions. For the scalar ${\cal R}_{\rm RN}^{I}$, the microstructure is dominated by repulsive interactions, giving rise to two phases: the RL phase when $R_{\rm sh}^{\rm RN} > R_{\rm sh,D}^{\rm RN}$, and the RS phase when $R_{\rm sh}^{\rm RN} < R_{\rm sh,D}^{\rm RN}$. A similar behavior is observed for the scalar ${\cal R}_{\rm RN}^{II}$, where the RL and RS microscopic phases also emerge exclusively. Thus, the first and second metrics assign an effective repulsive character to the charged RN states, whereas the third metric (${\cal R}_{\rm RN}^{III}$) assigns an effective attractive character. For this third scalar, interactions are purely attractive, resulting in the AL phase for $R_{\rm sh}^{\rm RN} > R_{\rm sh,D}^{\rm RN}$ and the AS phase for $R_{\rm sh}^{\rm RN} < R_{\rm sh,D}^{\rm RN}$.

 The same shadow therefore has different effective microscopic labels depending on the GTD metric, even though the thermodynamic locations of extremality, Davies instability, and the Schwarzschild reduction are unambiguous. These diagrams also integrate current astronomical observation bounds (such as the EHT-Images and mG-Rings bands of Sgr A*), linking the observables of the RN black hole shadows to the repulsive or attractive regime of their underlying thermodynamic microscopic phases.

\subsection{Kerr black hole}
\label{subsec:SMKerr}

\begin{figure*}[h]
  \centering
  \begin{minipage}[t]{0.685\linewidth}
    \centering
    \includegraphics[width=\linewidth]{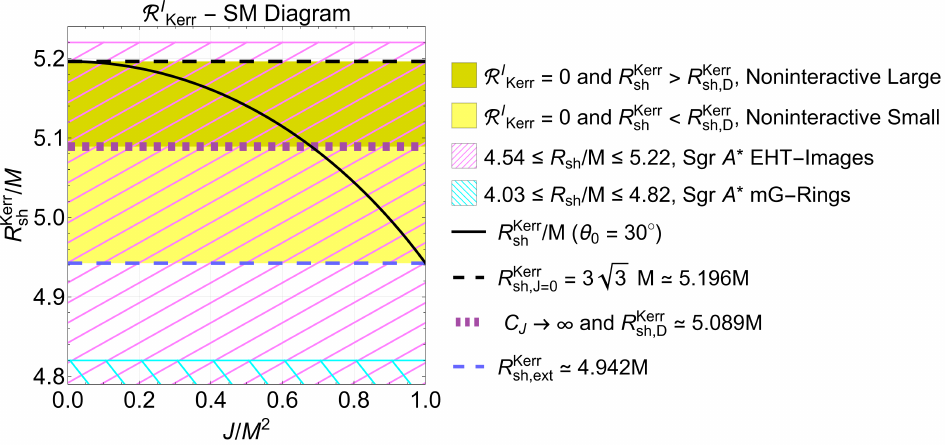}
    \\[2pt] \vspace{-0.6cm}\hspace{-9cm}\small{(a)}
  \end{minipage}\hfill
  \begin{minipage}[t]{0.315\linewidth}
    \centering
    \includegraphics[width=\linewidth]{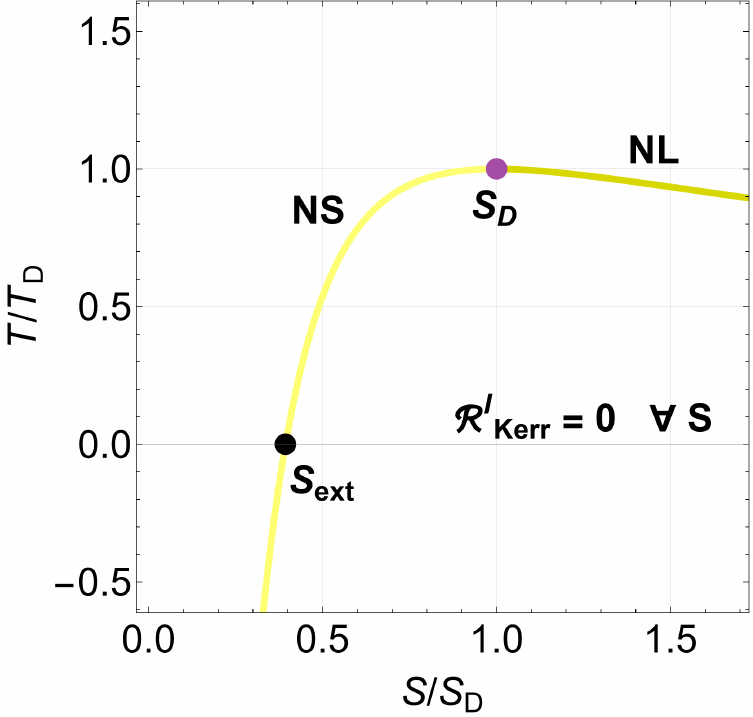}
    \\[2pt] \vspace{-0.6cm}\hspace{-3.5cm}\small{(b)}
  \end{minipage}\hfill
    \begin{minipage}[t]{0.685\linewidth}
    \centering
    \includegraphics[width=\linewidth]{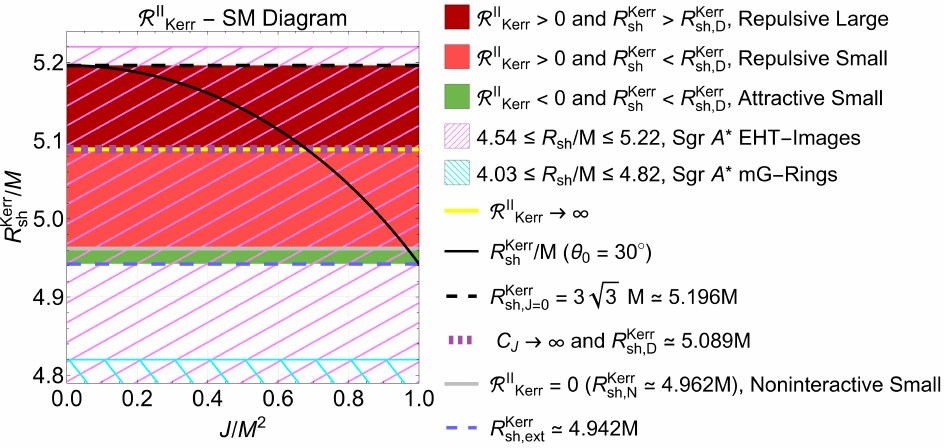}
    \\[2pt] \vspace{-0.6cm}\hspace{-9cm}\small{(c)}
  \end{minipage}\hfill
  \begin{minipage}[t]{0.315\linewidth}
    \centering
    \includegraphics[width=\linewidth]{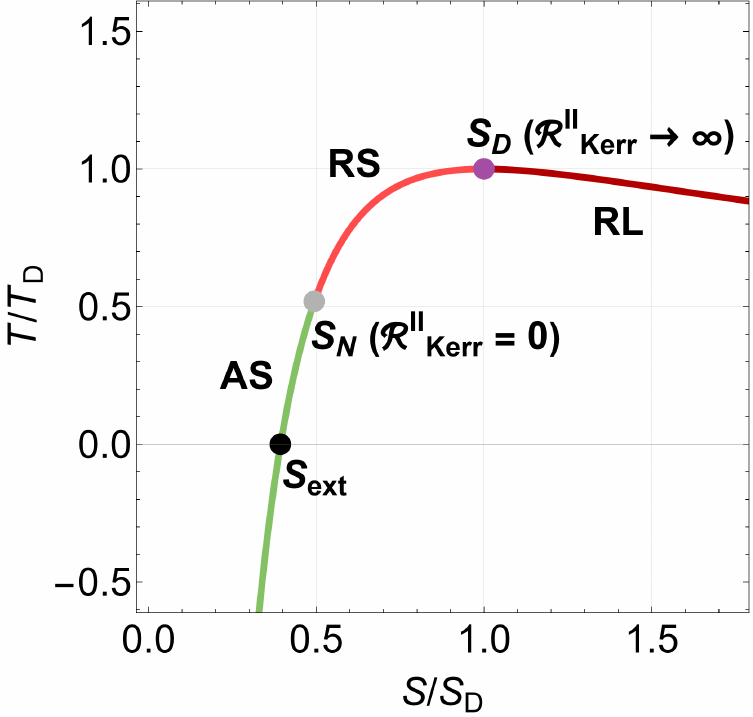}
    \\[2pt] \vspace{-0.6cm}\hspace{-3.5cm}\small{(d)}
  \end{minipage}\hfill
    \begin{minipage}[t]{0.685\linewidth}
    \centering
    \includegraphics[width=\linewidth]{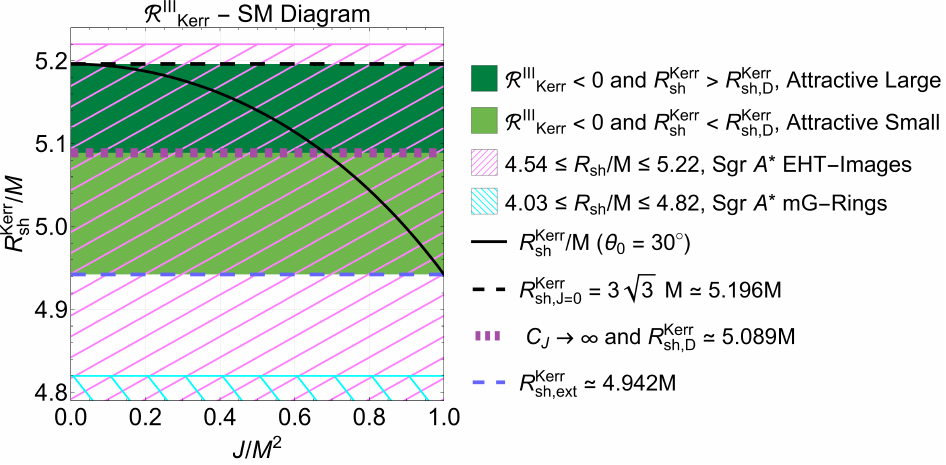}
    \\[2pt] \vspace{-0.6cm}\hspace{-9cm}\small{(e)}
  \end{minipage}\hfill
  \begin{minipage}[t]{0.315\linewidth}
    \centering
    \includegraphics[width=\linewidth]{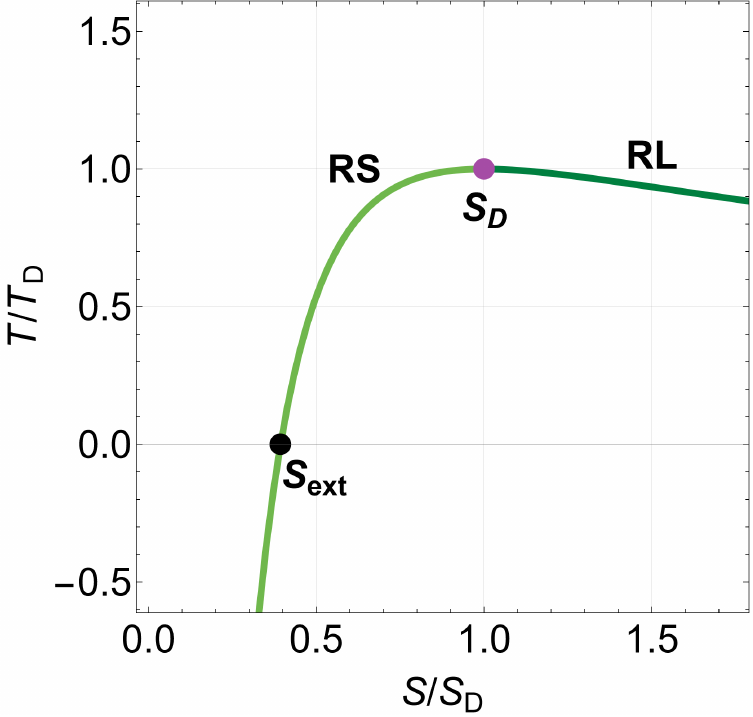}
    \\[2pt] \vspace{-0.6cm}\hspace{-3.5cm}\small{(f)}
  \end{minipage}
  \vspace{-0.7cm}
  \caption{Shadow Microscopic diagrams for Kerr black holes using $R_{{\rm sh}}^{\rm Kerr}/M$ ($r_0 \to \infty$) and $J/M^2$, and their corresponding $T/T_D$ vs. $S/S_D$ behaviors. (a)~and~(b)~for ${\cal R}_{\rm Kerr}^{I}$. (c)~and~(d)~for ${\cal R}_{\rm Kerr}^{II}$. (e)~and~(f)~for ${\cal R}_{\rm Kerr}^{III}$.}
  \label{fig:GTD-SMD2}
\end{figure*}

For Kerr the complete shadow is asymmetric, so an SM diagram based on one characteristic radius is necessarily a reduced representation. For the one-parameter shadow sequence, in contrast to the analogous work in Ref.~\cite{LadinoRomeroQuevedo2026}, we set the observer inclination for Sagittarius A* to $\theta_O=30^\circ$ (inclinations $\theta_O \lesssim 50^\circ$ are widely accepted in the literature for this source) instead of $46^\circ$. Furthermore, we adopt a different shadow-size convention: for all rotating shadow analyses in this work, the characteristic radius is measured using the four-point reference circle defined in Eq.~\eqref{eq:KerrAdSshadowobservables}. The shadow radius in this case is a monotonic function of the dimensionless angular momentum $\vert{}J\vert{}/M^2$, decreasing from $3\sqrt{3}M$ at $J=0$ to its minimum characteristic size at extremality, $R_{{\rm sh},ext}^{\rm Kerr} \simeq 4.942M$, for an inclination angle of $30^\circ$. The D-state, Eq.~\eqref{eq:KerrDaviesSpinV},  then corresponds to
\be
\left.\frac{|J|}{M^2}\right|_{D-state}
=
\sqrt{2\sqrt{3}-3}
\simeq0.681,
\qquad
R_{{\rm sh},D}^{\rm Kerr}
\simeq5.089\,M.
\label{eq:SMKerrDavies}
\ee
Additionally, the optical representation of the Kerr N-state can be obtained from Eqs.~\eqref{xzerokerr} and \eqref{eq:KerrCurvatureZeroShadowV} as
\be
\left.\frac{|J|}{M^2}\right|_{N-state}\simeq0.974,
\qquad
R_{{\rm sh},N}^{\rm Kerr}
\simeq4.962,M.
\label{eq:SMKerrzeroShadow}
\ee
The SM diagrams for Kerr black holes in the $R_{\rm sh}^{\rm Kerr}/M$ versus $J/M^2$ plane, together with the corresponding $T/T_D$ versus $S/S_D$ profiles, are shown in Fig.~\ref{fig:GTD-SMD2}. For the first metric, ${\cal R}_{\rm Kerr}^{I}=0$ throughout the Kerr family, yielding NL and NS sectors separated by the D-state shadow radius, as shown in Fig.~\ref{fig:GTD-SMD2}(a). For $g^{II}$, the curvature is positive from the Schwarzschild limit to the N-state, except at the D-state, and becomes negative toward extremality, corresponding to effective repulsive and attractive sectors, respectively, as illustrated in Fig.~\ref{fig:GTD-SMD2}(c). Finally, ${\cal R}_{\rm Kerr}^{III}<0$ throughout the rotating domain and becomes degenerate at $J=0$, assigning an effective attractive character to rotating states while distinguishing the Schwarzschild limit through metric degeneracy, as shown in Fig.~\ref{fig:GTD-SMD2}(e).


\subsection{Kerr--AdS black hole}
\label{subsec:SMKerrAdS}

The Kerr--AdS case requires a multidimensional generalization of the SM
construction. As established in Sec.~\ref{Kerr-AdS sec}, the equilibrium space
is described by the thermodynamic variables $(S,J,l)$. The optical description is also intrinsically multidimensional because the shadow is an asymmetric
critical curve characterized by the finite-observer construction of
Sec.~\ref{subsec:KerrAdSshadow}. Consequently, a single shadow radius
cannot generically parametrize the complete equilibrium state.

The analysis of the three GTD metrics in Sec.~\ref{subsec:KerrAdSshadow} shows that
their corresponding equilibrium conditions probe different sectors of
the Kerr--AdS thermodynamic structure. Condition~$I$, Eq.~\eqref{eq:KerrAdSI}, involves the
determinant of the complete Hessian of the fundamental equation and does
not produce an additional physical singular locus in the domain under
consideration. Condition~$II$, Eq.~\eqref{eq:KerrAdSII}, as discussed previously, contains two factors associated with distinct response sectors. The thermal branch $M_{,SS}=0$ determines the canonical spinodal structure, while the second factor probes instabilities in the $(J,l)$ sector. As shown in Sec.~\ref{Kerr-AdS sec}, however, this second branch does not yield an additional admissible singular locus in the physical domain considered.  Condition~$III$, Eqs.~\eqref{eq:KerrAdSIIIa}--\eqref{eq:KerrAdSIIId}, probes mixed-coupling degeneracies and likewise does not yield an additional physical singular locus in the domain considered. Thus, among the three GTD metrics, the physically relevant local singular structure of the Kerr--AdS black hole is encoded in the thermal branch given by $M_{,SS}=0$.

This result provides the thermodynamic basis for the Kerr--AdS SM
construction. Each physical state on the thermal branch can be mapped
into optical space using the procedure developed in Sec.~\ref{sec:BHshadows}. A state
$(S,J,l)$ determines $M$, $a=J/M$, $\Xi$, and $m=M\Xi^2$. These
quantities determine the critical impact parameters
$\xi(r_{\rm sph})$ and $\eta(r_{\rm sph})$, followed by the local
celestial angles $(\rho,\sigma)$ and finally by the screen coordinates
$(x,y)$. The symbols $(x,y)$ are used exclusively as screen
coordinates. For compact representations, the complete critical curve
may be characterized by $R_{\rm sh}$, $\delta_{\rm sh}$, and further
independent shadow observables as required.

The resulting SM representation must also preserve the thermodynamic
distinction among the structures occurring within a selected ensemble.
In the canonical ensemble, the zeros of $M_{,SS}$ determine the local
spinodal boundaries through the divergence of $C_{lJ}$. The critical
point corresponds to the limiting state at which the relevant spinodal
branches merge and therefore has a different thermodynamic status from
a generic point on either spinodal branch. The first-order small/large
black-hole transition is instead a global thermodynamic phenomenon and
is not determined solely by the local condition
$M_{,SS}=0$. In the grand-canonical ensemble the global phase structure
is different and includes the Hawking--Page transition
\cite{CaldarelliCognolaKlemm2000,LadinoRomeroQuevedo2025}. Legendre
invariance of the GTD formalism does not imply ensemble independence.

A Kerr--AdS SM diagram must therefore combine several levels of
information. The thermodynamic state is specified in the equilibrium
space $(S,J,l)$, the selected GTD metric determines the corresponding
geometric structure, and the optical map assigns that state to a
particular shadow configuration. In particular, the spinodal
singularities associated with ${\cal R}^{II}$ can be represented by
their corresponding shadow configurations, while the critical point and the global first-order and Hawking--Page transitions must retain their distinct thermodynamic identification.

\begin{figure*}[h]
  \centering
  \begin{minipage}[t]{0.685\linewidth}
    \centering
    \includegraphics[width=\linewidth]{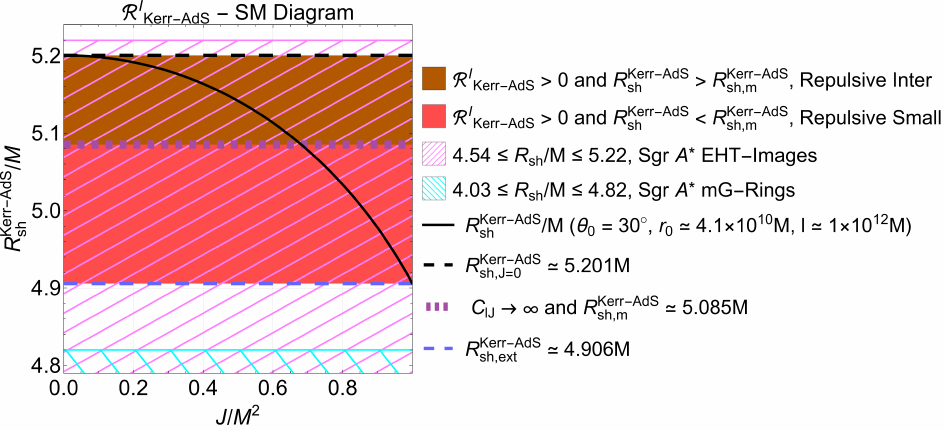}
    \\[2pt] \vspace{-0.4cm}\hspace{-9cm}\small{(a)}
  \end{minipage}\hfill
  \begin{minipage}[t]{0.315\linewidth}
    \centering
    \includegraphics[width=\linewidth]{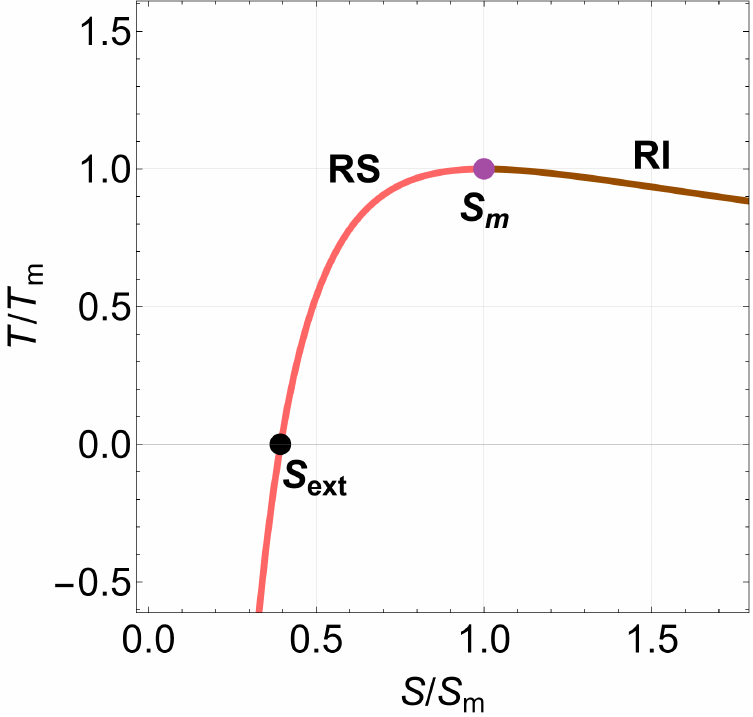}
    \\[2pt] \vspace{-0.4cm}\hspace{-3.5cm}\small{(b)}
  \end{minipage}\hfill
    \begin{minipage}[t]{0.685\linewidth}
    \centering
    \includegraphics[width=\linewidth]{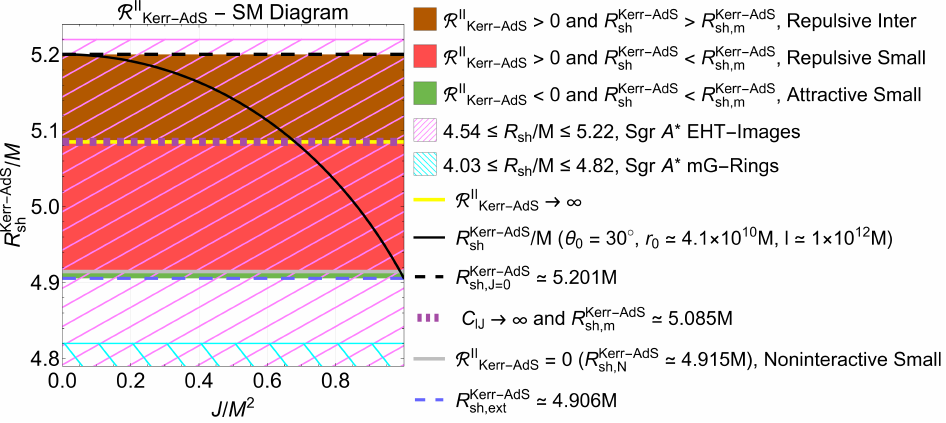}
    \\[2pt] \vspace{-0.4cm}\hspace{-9cm}\small{(c)}
  \end{minipage}\hfill
  \begin{minipage}[t]{0.315\linewidth}
    \centering
    \includegraphics[width=\linewidth]{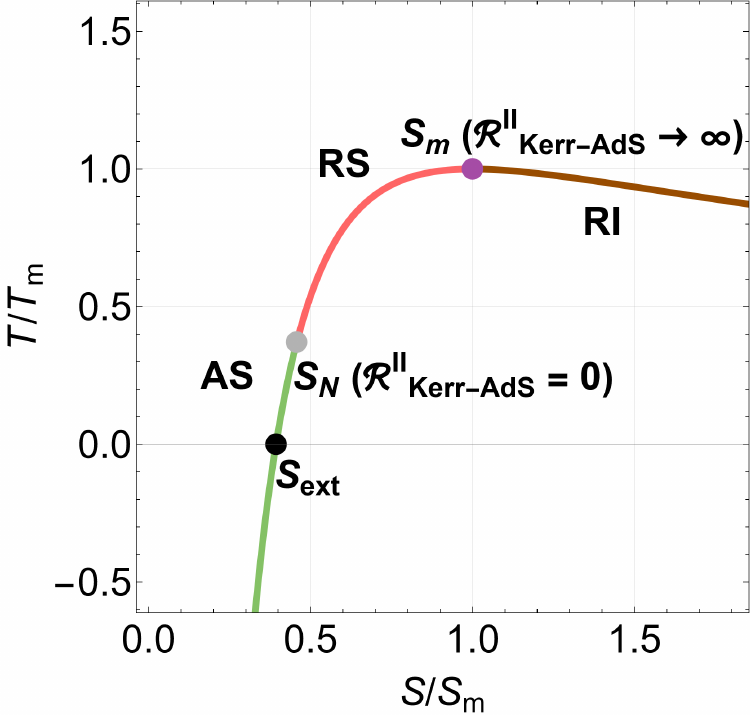}
    \\[2pt] \vspace{-0.4cm}\hspace{-3.5cm}\small{(d)}
  \end{minipage}\hfill
    \begin{minipage}[t]{0.685\linewidth}
    \centering
    \includegraphics[width=\linewidth]{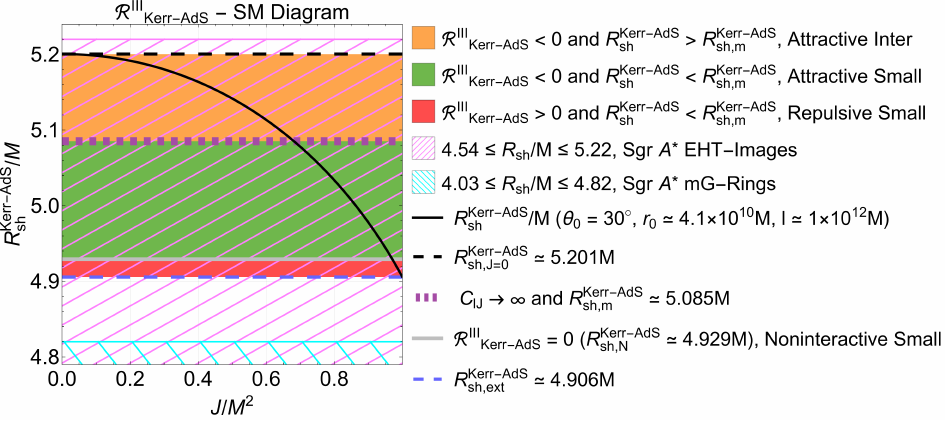}
    \\[2pt] \vspace{-0.4cm}\hspace{-9cm}\small{(e)}
  \end{minipage}\hfill
  \begin{minipage}[t]{0.315\linewidth}
    \centering
    \includegraphics[width=\linewidth]{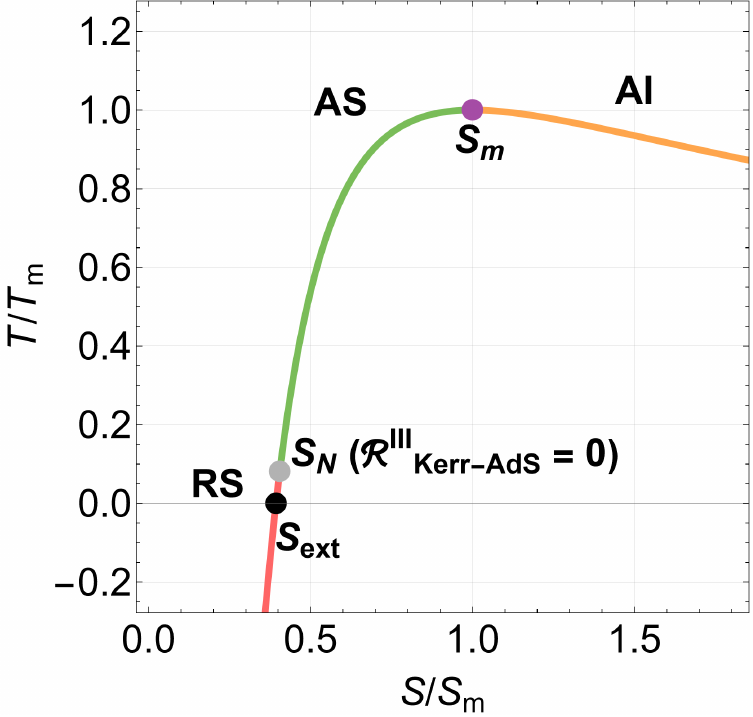}
    \\[2pt] \vspace{-0.4cm}\hspace{-3.5cm}\small{(f)}
  \end{minipage}
  \vspace{-0.7cm}
  \caption{Shadow Microscopic diagrams for Kerr-AdS black holes using $R_{{\rm sh}}^{\rm Kerr-AdS}/M$ and $J/M^2$, and their corresponding $T/T_m$ vs. $S/S_m$ behaviors. (a)~and~(b)~for ${\cal R}_{\rm Kerr-AdS}^{I}$. (c)~and~(d)~for ${\cal R}_{\rm Kerr-AdS}^{II}$. (e)~and~(f)~for ${\cal R}_{\rm Kerr-AdS}^{III}$.}
  \label{fig:GTD-SMD3}
\end{figure*}
Furthermore, as shown in Sec.~\ref{sec:zero_temperature}, Kerr--AdS black holes exhibit two branches of N-states and two spinodal points. We now focus on the asymptotically flat regime, characterized by a large AdS radius, $l = 1 \times 10^{12}M$. Fig.~\ref{fig:xext_xzero_KerrAdS}(b) shows that increasing $l$ increases the separation $\Delta x_{\rm zero}$ between the N-states. As a result, $x_{zero}^{+}$, corresponding to the N-state in the large black hole regime, becomes optically undetectable within the photon sphere. A similar effect occurs for the D-state associated with the Davies point $(S_{min},T_{min})$. Therefore, in the SM diagram shown in Fig.~\ref{fig:GTD-SMD3}, we instead use the Davies point $(S_m,T_m)$ and the N-state corresponding to $x_{zero}^{-}$, which lies in the stable small black hole regime. Assuming fixed observer parameters of $\theta_O = 30^\circ$ and $r_O = 4.1 \times 10^{10}M$, the non-rotating limit yields a shadow radius of $R_{\rm sh, J=0}^{\rm Kerr-AdS} \simeq 5.201M$. Therefore, the D-state corresponding to the point $(S_m,T_m)$ is mapped to the optical observable
\be
    R_{{\rm sh},m}^{\rm Kerr-AdS} \simeq 5.085M.
\ee
Similarly, the extremal Kerr--AdS state corresponds to a shadow radius of $R_{{\rm sh},ext}^{\rm Kerr-AdS} \simeq 4.906M$, while the N-state associated with the small black hole regime is represented optically by
\be
R_{{\rm sh},N}^{II} \simeq 4.915M.
\ee
Notice that the third metric also identifies a different N-state, defined by ${\cal R}_{\rm Kerr-AdS}^{III} = 0$, with an observable
\be
R_{{\rm sh},N}^{III} \simeq 4.929M.
\ee
These specific shadow radii function as optical thresholds that mark the transition between attractive and repulsive microscopic interactions in small black hole states, with the precise boundary dictated by the selected thermodynamic metric. 

\begin{table*}[t]
\centering
\caption{Summary of optical representation of distinguished thermodynamic states for RN, Kerr, and Kerr--AdS black holes, listing the GTD metric and shadow radius $R_{\rm sh}/M$ for each state (parameter configurations as in Figs.~\ref{fig:GTD-SMD1}--\ref{fig:GTD-SMD3}). }
\label{tab:opticalstates}
\renewcommand{\arraystretch}{1.4}
\begin{tabular}{c c c c c}
\hline
\textbf{Black hole}
& \textbf{State}
& \textbf{Identifying GTD metric}
& \textbf{Thermodynamic condition}
& $R_{\rm sh}/M$
\\
\hline

& Schwarzschild limit
& $g^{III}$
& $Q\rightarrow0$
& $3\sqrt{3}\simeq5.196$
\\

RN
& D-state
& $g^{II}$
& $S_D=3\pi Q^2$
& $\sqrt{9+6\sqrt{3}}\simeq4.404$
\\

& N-state
& $--$
& $--$
& $--$
\\

& Extremal
& $g^{I}$
& $S_{\rm ext}=\pi Q^2$
& $4$
\\
\hline

& Schwarzschild limit
& $g^{III}$
& $J\rightarrow0$
& $3\sqrt{3}\simeq5.196$
\\

Kerr
& D-state
& $g^{II}$
& $S_D = 2\pi J \sqrt{3 + 2\sqrt{3}}\simeq15.975 J$
& $\simeq5.089$
\\

& N-state
& $g^{II}$
& $S_{N}=2\pi J \sqrt{\frac{1}{5}\left(3 + 2\sqrt{6}\right)}\simeq7.897 J$
& $\simeq4.962$
\\

& Extremal
& $--$
& $S_{ext} = 2\pi J \simeq6.283 J$
& $\simeq4.942$
\\
\hline

& Schwarzschild AdS limit
& $--$
& $J\rightarrow0$
& $\simeq5.201$
\\

& D-state
& $g^{II}$
& $S_{D}\simeq15.975 J$
& $\simeq5.085$
\\

Kerr--AdS
& N-state
& $g^{II}$
& $S_{N}\simeq  7.308 J $
& $\simeq4.915$
\\

& N-state
& $g^{III}$
& $S_{N}\simeq  6.471 J $
& $\simeq4.929$
\\

& Extremal
& $--$
& $S_{ext} \simeq6.283 J$
& $\simeq4.906$
\\
\hline

\end{tabular}
\end{table*}
The SM diagrams for Kerr--AdS black holes in the $R_{\rm sh}^{\rm Kerr-AdS}/M$ versus $J/M^2$ plane, together with the corresponding $T/T_m$ versus $S/S_m$ profiles, are shown in Fig.~\ref{fig:GTD-SMD3}. For ${\cal R}_{\rm Kerr-AdS}^{I}$, the system exhibits repulsive interactions, with RI and RS phases for $R_{\rm sh}^{\rm Kerr-AdS}>R_{\rm sh,m}^{\rm Kerr-AdS}$ and $R_{\rm sh}^{\rm Kerr-AdS}<R_{\rm sh,m}^{\rm Kerr-AdS}$, respectively, as shown in Fig.~\ref{fig:GTD-SMD3}(a). For ${\cal R}_{\rm Kerr-AdS}^{II}$, the RI and RS phases are complemented by an AS phase for ${\cal R}_{\rm Kerr-AdS}^{II}<0$ and an NS phase at ${\cal R}_{\rm Kerr-AdS}^{II}=0$, as shown in Fig.~\ref{fig:GTD-SMD3}(c). Finally, ${\cal R}_{\rm Kerr-AdS}^{III}$ exhibits predominantly attractive behavior, with AI and AS phases, while RS and NS phases occur for ${\cal R}_{\rm Kerr-AdS}^{III}>0$ and ${\cal R}_{\rm Kerr-AdS}^{III}=0$, respectively, as shown in Fig.~\ref{fig:GTD-SMD3}(e).

Finally, Table~\ref{tab:opticalstates} summarizes the correspondence between the distinguished thermodynamic states, their identifying GTD metrics, and their optical representations for RN, Kerr, and Kerr--AdS black holes, highlighting the metric dependence of the SM correspondence.

\subsection{General interpretation and limitations}
\label{subsec:SMlimitations}

The SM diagrams discussed above illustrate the distinction between
thermodynamic information and the geometric structures associated with
the different GTD metrics. Optically, we have shown that it is possible to compute shadow observables corresponding to the intrinsic thermodynamic characteristics of each black hole spacetime, see Table~\ref{tab:opticalstates}. For instance, the Davies points for RN black holes are located at a photon sphere radius and a critical shadow radius of
\be
r_{{\rm ph},D}^{\rm RN} \simeq 2.366M \qquad \text{and} \qquad R_{{\rm sh},D}^{\rm RN} \simeq 4.404M,
\ee
respectively. Meanwhile, for Kerr black holes, the Davies point are located at a photon region and a critical shadow radius of
\be
r_{{\rm sph},D}^{\rm Kerr} \in [2.048M, 3.708M] \qquad \text{and} \qquad R_{{\rm sh},D}^{\rm Kerr} \simeq 5.089M.
\ee
For the Kerr-AdS spacetime, accompanying the critical quantities in Eq.~\eqref{eq:KerrAdScriticalcantidades}, the dimensionless critical shadow radius $R_{{\rm sh},c}^{\rm Kerr-AdS}$ depends on the observer's position, as shown in Ref.~\cite{LadinoRomeroQuevedo2025}. We found, for example, that for an observer angle $\theta_O = 30^\circ$, with radial distances $r_O = 100\sqrt{J}$ and $r_O = 4.1 \times 10^{10} \sqrt{J}$, the critical photon region is always $r_{{\rm sph},c}^{\rm Kerr-AdS} \in [5.068\sqrt{J}, 6.199\sqrt{J}]$ (or dimensionlessly, $r_{{\rm sph},c}^{\rm Kerr-AdS}/M_c \in [2.698, 3.300]$), and the thermodynamic critical shadow radii are at
\be
\left.\frac{|J|}{M^2}\right|_{c-state}\simeq0.283,\qquad R_{{\rm sh},c}^{\rm Kerr-AdS} \simeq 1.075 \qquad \text{and} \qquad R_{{\rm sh},c}^{\rm Kerr-AdS} \simeq 1.071,
\label{eq:KerrAdSshadowcritical}
\ee
respectively. These are optical quantities that we have not yet seen reported in the literature, but they could prove highly valuable for comparisons with observations or with results derived from other spacetimes.

For the RN black hole, the common-potential
formulation identifies the Davies and extremal states through different
GTD curvature singularities, which are mapped to different optical
configurations. For the Kerr black hole, the Davies singularity is
reproduced by $g^{II}$, while $g^I$ is flat and $g^{III}$ remains
regular throughout the rotating physical domain. These results
complement the SM construction of
Ref.~\cite{LadinoRomeroQuevedo2026}, where different
potential--metric pairs were considered. The common-potential
formulation adopted here instead permits the three GTD metrics to be
compared within the same thermodynamic representation.

The most robust content of an SM diagram is the optical image of a
thermodynamic state whose physical meaning has been established
independently. A Davies instability, an extremal state, a critical
point, or a global phase transition can therefore be assigned an
optical configuration once the corresponding thermodynamic state is
mapped into shadow space. GTD curvature signs, regular zeros,
singularities, and metric degeneracies provide additional geometric
information about the same state. These structures have different
thermodynamic significance, however, and their interpretation depends
on the selected GTD metric. In particular, a metric degeneracy does
not necessarily imply a curvature singularity, and neither a regular
curvature zero nor a change of curvature sign should by itself be
identified with a thermodynamic transition.

The statement that the shadow probes thermodynamic microstructure must
therefore be understood conditionally. A shadow observable constrains
the macroscopic black-hole state, while the selected GTD metric assigns
a geometric structure to the corresponding point of the equilibrium
space. The optical observation does not directly measure a GTD
curvature or a microscopic interaction. Moreover, the shadow alone
does not select among $g^I$, $g^{II}$, and $g^{III}$ when these
metrics describe the same macroscopic black-hole state. The complete
inference depends on the spacetime model, the thermodynamic ensemble,
the observer prescription, the optical map, and the GTD metric under
consideration.

This conditional structure is precisely what makes SM diagrams useful
for the inverse problem. An observationally allowed region in shadow
space can be mapped, whenever the optical map contains sufficient
independent information, into an allowed region of black-hole parameter
space and hence into a corresponding region of the equilibrium space.
The three GTD metrics can then be evaluated on these states to determine
which response-function sectors, curvature signs, regular zeros,
curvature singularities, and metric degeneracies are compatible with
the observationally allowed region. In this way, an SM diagram provides
an optical representation of GTD-classified equilibrium states without
identifying optical and thermodynamic-geometric quantities. This
inverse use will be considered in Sec.~\ref{sec:observations}.

\section{Observational implications}
\label{sec:observations}

The observational relevance of the GTD--shadow correspondence is necessarily indirect. Horizon-scale interferometry does not measure a thermodynamic curvature, a heat capacity, or a point of the equilibrium space. Instead, it constrains an image or ring structure. Relating that structure to the mathematical shadow requires a spacetime model, an emission model, radiative transfer, a viewing geometry, and an inference procedure. Only after these ingredients have been specified can the corresponding black-hole parameters be mapped to thermodynamic states and evaluated with a GTD metric.

This distinction is particularly important for the Event Horizon Telescope observations of M87* and Sagittarius A* \cite{EHT2019,EHT2022,EHT2022IV,EHT2022VI}. The critical shadow curve discussed in Sec.~\ref{sec:BHshadows} is determined by null geodesics, whereas the bright emission ring reconstructed from the data depends additionally on the plasma distribution, variability, scattering, calibration, and image reconstruction \cite{PerlickTsupko2022}. The observational problem is therefore an inverse problem with several intermediate modeling steps.

For a specified black-hole model, interferometric data first constrain an image or ring model. Under the assumptions of that model, one infers an allowed region for theoretical shadow observables. These observables constrain the black-hole parameters and hence an allowed region of the equilibrium space. The GTD metrics are then evaluated on that thermodynamic region. This logical sequence is the inverse counterpart of the theoretical construction developed in Secs.~\ref{sec:GTDshadows} and \ref{sec:SMdiagrams}.

\subsection{Sagittarius A* and the inverse problem}
\label{subsec:SgrAobs}

Observational comparisons are most useful when the optical quantities are normalized by the gravitational scale $GM/c^2$. For a spherical black hole, one size observable can be sufficient once the mass and observer configuration are fixed. For a rotating black hole, however, the shadow also contains displacement, asymmetry, horizontal and vertical extent, and other shape information. The inverse problem should therefore be formulated in terms of the multi-observable set ${\cal O}_{\rm sh}^{i}$ introduced in Sec.~\ref{subsec:shadowobservables}. The local criterion of Eq.~\eqref{eq:observabilitycondition} then determines whether the selected optical observables contain enough independent information to reconstruct the relevant thermodynamic parameters after nuisance variables have been fixed or marginalized.

Observational constraints should be interpreted as allowed regions rather than exact shadow values. Mass, distance, inclination, plasma properties, scattering, and image reconstruction all carry uncertainties. A thermodynamic or GTD conclusion should therefore be attached to the corresponding allowed region of equilibrium space. Exact theoretical values remain useful as markers for D-states, extremality, critical endpoints, Hawking--Page transitions, or N-states, but an observational statement is a compatibility test between those theoretical markers and an inferred optical region.

Sagittarius A* provides the first explicit observational application of the GTD--shadow program. Ref.~\cite{LadinoRomeroQuevedo2026} used the asymptotically flat RN and Kerr geometries as charged and rotating reference models and confronted their theoretical shadow sizes with two observational determinations: the EHT-Images bounds obtained from imaging reconstructions and the mG-Rings bounds obtained from analytic model fitting. Using the average of the Keck and VLTI mass--distance determinations, the adopted $1\sigma$ intervals were
\be
4.54M\leq R_{\rm sh}\leq5.22M
\qquad
\hbox{(EHT-Images)},
\label{eq:ObsEHTImages}
\ee
and
\be
4.03M\leq R_{\rm sh}\leq4.82M
\qquad
\hbox{(mG-Rings)}.
\label{eq:ObsmGRings}
\ee
These intervals are observationally inferred shadow constraints within the analysis of Ref.~\cite{LadinoRomeroQuevedo2026}; they should not be confused with a direct measurement of the mathematical critical curve.

For the RN reference model, the theoretical shadow radius decreases monotonically from the Schwarzschild value $3\sqrt{3}M$, corresponding to the RL phase for scalars ${\cal R}_{\rm RN}^{I}$ and ${\cal R}_{\rm RN}^{II}$, and the AL phase for ${\cal R}_{\rm RN}^{III}$, to the extremal value $4M$, corresponding to the RS phase for ${\cal R}_{\rm RN}^{I}$ and ${\cal R}_{\rm RN}^{II}$, and the AS phase for ${\cal R}_{\rm RN}^{III}$. The fixed-charge D-state derived in Secs.~\ref{sec:GTD} and \ref{sec:GTDshadows}, which marks the boundary separating these Large and Small phases, is located at

\be
\left.\frac{|Q|}{M}\right|_{D-state}
=\frac{\sqrt{3}}{2}
\simeq0.866,
\qquad
R_{{\rm sh},D}^{\rm RN}
\simeq4.404M.
\label{eq:ObsRNDavies}
\ee
This optical marker agrees with the value used in Ref.~\cite{LadinoRomeroQuevedo2026}. It lies inside the mG-Rings interval in Eq.~\eqref{eq:ObsmGRings} but below the lower EHT-Images bound in Eq.~\eqref{eq:ObsEHTImages}. By contrast, the Schwarzschild value $3\sqrt{3}M\simeq5.196M$ (RL for ${\cal R}_{\rm RN}^{I, II}$; AL for ${\cal R}_{\rm RN}^{III}$) lies inside the EHT-Images interval, while the extremal value $R_{{\rm sh},ext}^{\rm RN}=4M$ (RS for ${\cal R}_{\rm RN}^{I, II}$; AS for ${\cal R}_{\rm RN}^{III}$) lies just below the adopted mG-Rings lower bound. Thus the observational intervals select different portions of the same RN family, each mapping to distinct microscopic thermodynamic phases.

The common-potential GTD formulation adopted in this review provides an alternative and complementary classification to the potential--metric organization of Ref.~\cite{LadinoRomeroQuevedo2026}. In the common-potential construction adopted in this review, the Davies state is a curvature singularity of $g^{II}[M]$, extremality is a curvature singularity of $g^I[M]$, and the Schwarzschild limit is the degeneracy locus of $g^{III}[M]$. The observational constraint on $\vert{}Q\vert{}/M$ is independent of this reorganization because the spacetime shadow is unchanged. What changes is the GTD classification assigned to the allowed states.

For the Kerr case, the optical inference is more complicated because the critical curve depends on both the spin and the observer inclination. We fix the inclination at $\theta_O=30^\circ$ and reduce the shadow profile to its characteristic size observable. The D-state, acting as the boundary between the Large and Small phases, occurs at
\be
\left.\frac{|J|}{M^2}\right|_{D-state}
=\sqrt{2\sqrt{3}-3}
\simeq0.681,
\qquad
R_{{\rm sh},D}^{\rm Kerr}
\simeq5.089M.
\label{eq:ObsKerrDavies}
\ee
This value lies inside the EHT-Images interval and outside the mG-Rings interval. The same calculation gives an N-state of $g^{II}$ at
\be
\left.\frac{|J|}{M^2}\right|_{N-state}\simeq0.974,
\qquad
R_{{\rm sh},N}^{\rm Kerr}
\simeq4.962M.
\label{eq:ObsKerrZero}
\ee
This state is also inside the EHT-Images interval but slightly above the mG-Rings upper bound. The extremal Kerr shadow in the same fixed-inclination convention is $R_{{\rm sh},ext}^{\rm Kerr} \simeq4.942M$, corresponding to the NS phase for ${\cal R}_{\rm Kerr}^{I}$ and the AS  for both ${\cal R}_{\rm Kerr}^{II}$ and ${\cal R}_{\rm Kerr}^{III}$, and again compatible with the EHT-Images interval but not with the adopted mG-Rings interval \cite{LadinoRomeroQuevedo2026}. For the Kerr-AdS black hole in the asymptotically flat regime (characterized by a large AdS radius $l = 1 \times 10^{12}M$), the shadow analysis reveals a complementary thermodynamic structure. Keeping the observer parameters consistent with Sgr A* ($\theta_O = 30^\circ$, $r_O = 4.1 \times 10^{10}M$), the non-rotating limit yields a shadow size of $R_{\rm sh, J=0}^{\rm Kerr-AdS} \simeq 5.201M$, which fits comfortably within the EHT-Images interval. The critical state characterized by a divergence in the specific heat $C_{lJ} \to \infty$, marking the transition boundary between the Intermediate and Small phases, is located at a shadow radius of
\be
R_{{\rm sh},m}^{\rm Kerr-AdS} \simeq 5.085M.
\label{eq:ObsKerrAdSDavies}
\ee
This critical thermodynamic point is completely consistent with the EHT-Images constraints, although it lies outside the adopted mG-Rings interval. Similarly, the extremal Kerr-AdS black hole gives $R_{{\rm sh},ext}^{\rm Kerr-AdS} \simeq 4.906M$, which also remains compatible with the EHT-Images bounds but slightly above the mG-Rings upper limit. In the common-potential GTD formulation for Kerr-AdS, the thermodynamic curvatures provide distinct phase classifications and non-interacting markers. The N-states predicted by $g^{II}$ and $g^{III}$ occur, respectively, at
\be
R_{{\rm sh},N}^{\rm Kerr-AdS} \simeq 4.915M \quad \text{and} \quad R_{{\rm sh},N}^{\rm Kerr-AdS} \simeq 4.929M.
\label{eq:ObsKerrAdSZero}
\ee
Both of these non-interacting optical markers lie well within the EHT-Images allowed region. The boundary between attractive and repulsive small black hole states depends on the chosen metric, with ${\cal R}^{II}_{\rm Kerr-AdS}$ and ${\cal R}^{III}_{\rm Kerr-AdS}$ changing sign at these specific shadow radii, offering an observational threshold for microscopic interactions within the EHT-Images confidence interval.

The observational application of Ref.~\cite{LadinoRomeroQuevedo2026} used its GTD potential--metric pairs to classify microscopic thermodynamic phases and to derive allowed charge and spin ranges from the EHT-Images and mG-Rings intervals. Those numerical optical constraints remain results of that published construction. The common-potential three-metric formulation considered in this review provides an alternative and complementary classification of the same observationally allowed black-hole states using the curvatures obtained in Secs.~\ref{sec:GTD} and \ref{sec:SMdiagrams}.

The Kerr and Kerr-AdS examples also show why observer variables must be treated as nuisance parameters in a general inverse problem. A size observable alone can be partially degenerate between spin and inclination. Shape observables can reduce this degeneracy. Consequently, a future GTD analysis of real data should use the full optical observable set whenever the observations permit it.

The interpretation of horizon-scale images requires three distinct notions. The critical shadow curve is defined by the unstable null-geodesic structure of the spacetime. The photon ring is generated by highly lensed trajectories that remain close to the critical photon region. The bright emission ring is an astrophysical image feature whose morphology also depends on the plasma distribution, emissivity, radiative transfer, variability, scattering, calibration, and reconstruction method \cite{EHT2022IV,EHT2022VI,PerlickTsupko2022}.

Accordingly, an observational GTD analysis should not equate a measured ring diameter directly with a theoretical shadow diameter. The data determine an allowed image region; a specified inference model maps that region to theoretical shadow observables; those observables constrain the black-hole parameters; and only then are thermodynamic and GTD quantities evaluated.

The same caution applies at the final step. The EHT does not observe a GTD curvature. A shadow-compatible parameter region determines a region of the equilibrium space, and GTD supplies a thermodynamic-geometric classification of that region. That classification can include local stability, proximity to a response-function singularity, metric degeneracy, and the sign or zero of a selected curvature. These are theoretical inferences conditioned on the adopted gravitational, thermodynamic, and optical models.

\subsection{AdS black holes and observational interpretation}
\label{subsec:AdSobservations}

The RN--AdS and Kerr--AdS calculations are essential for establishing the structure of the GTD--shadow correspondence, but their direct astrophysical interpretation differs from that of asymptotically flat RN and Kerr. The AdS curvature radius $l$ is part of the spacetime and thermodynamic description; it is not an environmental pressure surrounding Sagittarius A* or M87*. 

The RN--AdS study of Ref.~\cite{LadinoRomeroQuevedo2024} showed explicitly that the shadow radius can be written in terms of the horizon, photon-sphere, observer, charge, and AdS scales, and that the thermodynamic phase structure depends strongly on $l$. It also imposed quasihomogeneity as a consistency condition of the GTD description. These results establish RN--AdS primarily as a structural example of how thermodynamic curvature and shadow variables can be related.  For Kerr--AdS, the common-potential analysis of Sec.~\ref{Kerr-AdS sec} provides the GTD thermodynamic structure used in this review. The equilibrium space is described by $(S,J,l)$ before an ensemble is selected. Condition~$II$, Eq.~\eqref{eq:KerrAdSII}, contains the physically relevant local singular structure in the domain considered: its thermal branch $M_{,SS}=0$ reproduces the divergence of $C_{lJ}$ and the canonical spinodal structure, whereas the second factor probes the coupled $(J,l)$ response sector. Conditions~$I$, Eq.~\eqref{eq:KerrAdSI}, and $III$, Eqs.~\eqref{eq:KerrAdSIIIa}-\eqref{eq:KerrAdSIIId}, do not generate additional physical singular loci in the regular domain analyzed in Sec.~\ref{sec:GTD}. The optical construction and the ensemble-dependent phase structure are supported by Ref.~\cite{LadinoRomeroQuevedo2025}, which extends the correspondence to rotation and an asymmetric shadow. In particular, the canonical ensemble contains small/large black-hole critical structure, whereas the grand-canonical ensemble contains a distinct phase structure including the Hawking--Page transition.  Thus the AdS examples demonstrate that genuine equilibrium transitions, critical endpoints, response-function singularities, and effective curvature sectors can all be assigned optical configurations while retaining their different thermodynamic meanings. They should not, however, be interpreted as direct models of the astrophysical environment of EHT sources. 

Instead, to connect these structural findings with a more realistic observational scenario, it is highly relevant to analyze the large AdS radius limit ($l \to \infty$), which corresponds to the asymptotically flat spacetime limit ($\Lambda \to 0$). In the study of black hole thermodynamics via shadow images, this limit is crucial because the spacetime between the black hole and a distant observer, particularly in the observer's local vicinity, is expected to be effectively flat. Specifically, for an observer at a finite radial distance $r_0$ where $l \gg r_0$, the AdS background scale does not appreciably alter the curvature in their local frame. 

In this context, the GTD scalars for AdS black holes in $n=3$ GTD-dimensions (the number of thermodynamic degrees of freedom) can be expressed as their asymptotically flat counterparts, supplemented by a perturbative expansion in powers of $1/l^2$. In general, for an AdS black hole in $n=3$ characterized by the extensive variables $E^1$, $E^2$, and $E^3=l$, this is given by
\begin{equation}
    \mathcal{R}^{I/II/III}_{\rm AdS}(n, l, E^1, E^2) = \mathcal{R}^{I/II/III}_{\rm Flat}(E^1, E^2) + (n-2) \sum_{k=0}^{\infty} \frac{\Delta\mathcal{R}^{I/II/III}_{k}(E^1, E^2)}{l^{2k}}.
    \label{eq:AdS_expansion}
\end{equation}
Defining the zeroth-order term as $(n-2)\Delta\mathcal{R}_0 \equiv \lim_{l \to \infty} \mathcal{R}_{\rm AdS} - \mathcal{R}_{\rm Flat}$ guarantees the validity of the expansion, as all $k \ge 1$ terms vanish in the limit $l \to \infty$. If this limit preserved the base geometric structure, the perturbation series would start at $\mathcal{O}(1/l^2)$, yielding $\Delta\mathcal{R}_0 = 0$. However, we generally find $\Delta\mathcal{R}_0 \neq 0$. This zeroth-order residue indicates that the 3D and 2D thermodynamic manifolds do not continuously reduce to one another when the cosmological constant is removed. 

While the asymptotic limit preserves the functional dependence on the variables $E^1$ and $E^2$, the extra degree of freedom $l$ induces a non-trivial geometric shift, fundamentally altering the base curvature of the phase space. The overall $(n-2)$ factor coupled to all correction terms shows that the curvature discrepancy between AdS and flat black holes depends not only on the $l \to \infty$ limit, but on the thermodynamic degrees of freedom ($n$) of the phase space. Thus, even in the asymptotically flat regime, a small negative cosmological constant can modify the fundamental thermodynamic structure and microstructure of the spacetime background. For instance, treating the Kerr-AdS black hole as a Kerr metric plus AdS background corrections, with $E^1=S$ and $E^2=J$, yields
\begin{equation}
    \mathcal{R}^{I/II/III}_{\rm Kerr\text{-}AdS}(n, l, S, J) = \mathcal{R}^{I/II/III}_{\rm Kerr}(S, J) + (n-2) \sum_{k=0}^{\infty} \frac{\Delta\mathcal{R}^{I/II/III}_{k}(S, J)}{l^{2k}},
    \label{eq:KerrAdS_expansion}
\end{equation}
where both  $\mathcal{R}_{\rm Kerr}$ and each perturbative coefficient $\Delta\mathcal{R}_{k}$ are strictly functions of $S$ and $J$. Without loss of generality, to visualize this behavior we consider the slow-rotation approximation ($J \ll S$). Expanding up to $\mathcal{O}(J^2)$ Eq.~\eqref{eq:KerrAdS_expansion} yields the GTD scalars for $n \in \{2,3\}$ as
\begin{align}
    \mathcal{R}^{I}_{\rm Kerr\text{-}AdS} &= \frac{8(n-2)}{3S} \left( 2\pi - \frac{13S}{l^2} \right) + \frac{8(n-2)\pi^2 J^2}{3S^3} \left( -17\pi + \frac{168S}{l^2} \right) + \mathcal{O}(J^4), \\
    \mathcal{R}^{II}_{\rm Kerr\text{-}AdS} &= \frac{8}{3} \left[ \frac{32(n-2)}{l^2} - \frac{(n-17)\pi}{S} \right] + \frac{8\pi^2 J^2}{3S^3} \left[ (349n + 190)\pi + \frac{6360(n-2)S}{l^2} \right] + \mathcal{O}(J^4), \\
    \mathcal{R}^{III}_{\rm Kerr\text{-}AdS} &= \frac{32}{1153977} \left[ \frac{5803987(n-2)}{l^2} + \frac{44(9797n - 43552)\pi}{S} \right] \nonumber \\
    & + \frac{64\pi^2 J^2}{71931233 l^2 S^3} \left[ -11l^2(32189537n + 19697534)\pi + 9646071201(n-2)S \right] + \mathcal{O}(J^4).
\end{align}

For $n=3$, we readily recover the expected Kerr-AdS GTD scalars
\begin{align}
    \mathcal{R}^{I}_{\rm Kerr\text{-}AdS} &= \frac{8}{3S^3} \left[ S^2 \left( 2\pi - \frac{13S}{l^2} \right) + \pi^2 J^2 \left( -17\pi + \frac{168S}{l^2} \right) \right] + \mathcal{O}(J^4), \\
    \mathcal{R}^{II}_{\rm Kerr\text{-}AdS} &= \frac{8}{3} \left[ \frac{32}{l^2} + \frac{14\pi}{S} + \frac{\pi^2 J^2}{S^3} \left( 1237\pi + \frac{6360S}{l^2} \right) \right] + \mathcal{O}(J^4), \\
    \mathcal{R}^{III}_{\rm Kerr\text{-}AdS} &= \frac{32}{43923 l^2 S^3} \left[ 77S^2 \left( 2869S - 308\pi l^2 \right) + 6\pi^2 J^2 \left( 1963377S - 260315\pi l^2 \right) \right] + \mathcal{O}(J^4).
\end{align}

Conversely, for $n=2$, AdS background curvature corrections identically vanish, exactly reproducing the asymptotically flat Kerr GTD scalars
\begin{align}
    \mathcal{R}^{I}_{\rm Kerr} &= 0, \\
    \mathcal{R}^{II}_{\rm Kerr}&= \frac{40\pi}{S} + \frac{2368\pi^3 J^2}{S^3} + \mathcal{O}(J^4), \\
    \mathcal{R}^{III}_{\rm Kerr} &= -\frac{8448\pi}{289S} - \frac{4042752\pi^3 J^2}{4913S^3} + \mathcal{O}(J^4).
\end{align}

Furthermore, taking the non-rotating limit ($J \to 0$) consistently recovers the exact Schwarzschild and Schwarzschild-AdS scalar curvatures, corroborating the validity of the perturbative expansion. 

Depending on $n$, AdS corrections can be separated order by order from the thermodynamic degrees of freedom of the asymptotically flat black hole. Then, in GTD, where scalar curvature encodes effective thermodynamic interactions, the additional AdS parameter acts in two distinct ways: it alters the base phase space geometry through the shift $\Delta\mathcal{R}_0$ and induces scale-dependent corrections via $1/l^{2k}$ terms. Specifically for Kerr-AdS black holes, the residue $\Delta\mathcal{R}_0$ introduces no additional divergences. Consequently, for the majority of the phase space and across the GTD scalars, the AdS background essentially acts merely to rescale the overall phase space curvature, preserving the primary functional dependence on $S$ and $J$. The singular, notable exception to this perturbative scaling behavior is the emergence of novel zeros in the third GTD scalar, $\mathcal{R}^{III}$. This is evidenced in Sec.~\ref{sec:GTDshadows} (see Figs.~\ref{fig:GTD3}), where the three GTD curvature scalars are mapped across the shadow profiles of Kerr--AdS black holes to illustrate these AdS correction effects. Furthermore, we apply this analysis to Sagittarius A* in Sec.~\ref{sec:SMdiagrams}; as shown in Figs.~\ref{fig:GTD-SMD3}, the SM diagrams depict the corresponding constraints on the parameter space. Thus, while the AdS radius $l$ operates predominantly as a small rescaling parameter, it still induces a structural shift evidenced by these new vanishing points.

\subsection{Metric discrimination and observational prospects}
\label{subsec:falsifiability}

The principal observational limitation of the current GTD--shadow program follows from the common spacetime geometry. For a fixed black-hole state, the shadow is determined by the spacetime metric and the observer. The three GTD metrics $g^I$, $g^{II}$, and $g^{III}$ are different geometric descriptions of the same thermodynamic equilibrium state and do not generate different spacetime shadows for that state. Consequently, shadow data can restrict the region of equilibrium space on which the three GTD metrics are evaluated, but the shadow alone cannot determine which GTD metric provides the physically preferred description of the thermodynamic interaction.

Ref.~\cite{LadinoRomeroQuevedo2026} showed that different potential--metric choices can assign different effective microscopic interactions to observationally allowed RN and Kerr states and discussed the possibility that improved observations could help constrain these descriptions. The common-potential analysis developed in this review provides a complementary perspective on this question. Increasing the accuracy of a shadow observable can exclude black-hole parameter regions and therefore exclude thermodynamic states, but it cannot distinguish among $g^I$, $g^{II}$, and $g^{III}$ at a state that remains observationally allowed because all three metrics correspond to the same spacetime shadow.

Since the three GTD metrics correspond to the same spacetime state, the shadow alone cannot distinguish among them. Such a distinction would require an additional observable prediction that depends explicitly on the selected GTD metric. Identifying such a prediction constitutes an open problem within GTD and represents a necessary step toward an observational discrimination among its different metric structures.

No such GTD-specific observable has yet been established in the black-hole shadow literature reviewed here. The relevant future problem is therefore not to return to thermodynamic geometry, whose established results have already served as a point of comparison in the preceding sections, but to determine whether GTD itself leads to an additional physical consequence that can distinguish the predictions associated with $g^I$, $g^{II}$, and $g^{III}$.

The present framework suggests several improvements. Observational SM diagrams should use credible regions rather than exact lines. Rotating black holes should be analyzed with several independent shadow observables rather than only one characteristic size. Uncertainties in mass, distance, inclination, scattering, plasma parameters, and image reconstruction should be propagated consistently into the thermodynamic parameter space. The observer dependence of the shadow should also be kept explicit.

Improved horizon-scale observations can nevertheless sharpen the macroscopic part of the correspondence. Better measurements of size and shape can reduce the allowed regions of charge, spin, and observer geometry. For RN, this can determine more sharply whether the Schwarzschild, Davies, or extremal configurations are compatible with the data. For Kerr, it can test the Davies state, the $g^{II}$ curvature-zero state, the Schwarzschild limit, and extremality within the adopted optical model. These are physically distinct configurations and should not be grouped under a single transition label.

The observational status can therefore be summarized in two parts. The established result is that shadow constraints can be propagated into black-hole parameter space and then into a GTD classification of the corresponding equilibrium states, with Sagittarius A* providing the first explicit example \cite{LadinoRomeroQuevedo2026}. The unresolved problem is to identify, within GTD itself, a measurable consequence that distinguishes the metric structures $g^I$, $g^{II}$, and $g^{III}$ beyond the common spacetime shadow. Future high-resolution imaging can substantially improve the macroscopic parameter constraints even before this specifically GTD problem is solved.

\section{Conclusions and perspectives}
\label{sec:conclusions}

This review has examined black-hole shadows from the perspective of geometrothermodynamics, with the purpose of determining how optical information can represent and eventually constrain the GTD structure of black-hole equilibrium states. The starting point is a clear separation between the geometries involved. A black-hole shadow is determined by null geodesics in spacetime and by the observer configuration, whereas the GTD metrics are defined on the equilibrium space and are induced from Legendre-invariant metrics on the thermodynamic phase space. These geometries are not identified. Their connection arises because the same macroscopic black-hole parameters determine the thermodynamic fundamental equation and the spacetime from which the shadow is constructed.

The thermodynamic analysis provides the physical framework needed to interpret the GTD results. Davies instabilities, extremality, first-order transitions, critical points, and Hawking--Page transitions are distinct thermodynamic structures and must remain distinguishable when represented in optical variables. In particular, the Davies loci of asymptotically flat RN and Kerr are spinodal instabilities associated with divergences of $C_Q$ and $C_J$, respectively \cite{LadinoRomeroQuevedo2026}. The three GTD metrics provide complementary information through different combinations of second derivatives of the fundamental equation, but their physical interpretation requires checking the admissible black-hole domain, possible cancellations in the curvature, and the corresponding response functions. In particular, a metric degeneracy does not necessarily imply a curvature singularity, while a regular curvature zero has a different thermodynamic significance from a curvature singularity.

The RN example illustrates this separation particularly clearly. When the three GTD metrics are constructed from the same fundamental equation $M(S,Q)$, condition~$I$ selects the extremal state $S=\pi Q^2$, condition~$II$ selects the Davies state $S=3\pi Q^2$, and condition~$III$ gives the Schwarzschild limit $Q=0$. Accordingly, ${\cal R}_{\rm RN}^{I}$ is singular at extremality, ${\cal R}_{\rm RN}^{II}$ is singular at the Davies instability, and $g_{\rm RN}^{III}$ becomes degenerate in the Schwarzschild limit. These distinguished states are mapped to characteristic shadow radii and, depending on the selected GTD metric, to different microscopic thermodynamic phases. The Schwarzschild limit corresponds to the Large phase, the Davies state marks the boundary between the Large and Small phases, and extremality corresponds to the Small phase, with the attractive, repulsive, and non-interacting character determined by the curvature sign of the selected metric. Thus, the RN example illustrates how a common spacetime shadow can be combined with different GTD metrics to provide distinct thermodynamic classifications.

Kerr reveals a different distribution of information among the three metrics. With the common fundamental equation $M(S,J)$, the first GTD curvature vanishes identically. The second metric reproduces the fixed-$J$ Davies instability and also possesses an additional regular curvature zero, while the third metric is curved, remains regular at the D-state, and becomes degenerate in the Schwarzschild limit. For the observer inclination used in the Sagittarius A* application, the Davies state acts as the boundary between the Large and Small phases, while the zero-curvature state defines the NS phase for ${\cal R}_{\rm Kerr}^{I}$ and ${\cal R}_{\rm Kerr}^{II}$ \cite{LadinoRomeroQuevedo2026}. In contrast to RN, none of the three Kerr metrics is sensitive to extremality, as extremality is not identified by either a curvature zero or a divergence of a thermodynamic response function. The extremal Kerr state corresponds to the NS phase for ${\cal R}_{\rm Kerr}^{I}$ and the AS phase for both ${\cal R}_{\rm Kerr}^{II}$ and ${\cal R}_{\rm Kerr}^{III}$. The N-state is not a thermodynamic transition, but provides additional information from GTD by separating sectors of opposite curvature sign for $g^{II}$.

Kerr--AdS demonstrates that the construction extends to a multidimensional equilibrium space and to an asymmetric shadow. In the formulation considered here, the equilibrium space is described by $(S,J,l)$, and quasihomogeneity fixes the corresponding quasihomogeneous coefficients. The physically relevant local singular structure in the domain considered is contained in condition~$II$. Its thermal branch, $M_{,SS}=0$, is associated with the divergence of $C_{lJ}$ and the canonical spinodal structure, marking the transition boundary between the Intermediate and Small phases. The non-rotating limit represents the RI phase for scalars ${\cal R}_{\rm Kerr-AdS}^{I}$ and ${\cal R}_{\rm Kerr-AdS}^{II}$, and the AI phase for ${\cal R}_{\rm Kerr-AdS}^{III}$. The extremal Kerr-AdS state corresponds to the RS phase for ${\cal R}_{\rm Kerr-AdS}^{I}$ and ${\cal R}_{\rm Kerr-AdS}^{III}$, and the AS phase for ${\cal R}_{\rm Kerr-AdS}^{II}$. Curvature-zero states  define the NS phase for $g^{II}$ and $g^{III}$. In addition, we determined novel thermodynamic critical shadow radii for Kerr-AdS black holes for the observer configurations considered. Conditions~$I$ and $III$ do not generate additional physical singular loci in the regular domain analyzed in this review. The canonical and grand-canonical ensembles must remain distinct because Legendre invariance does not imply ensemble independence \cite{LadinoRomeroQuevedo2025}.

Furthermore, the optical construction provides the bridge between spacetime and the equilibrium space without identifying them. A shadow observable is not introduced as a thermodynamic variable; rather, it provides an optical representation of a thermodynamic state. For a one-dimensional branch, a regular shadow observable can locally replace the branch parameter. For a multidimensional family, a set of optical observables ${\cal O}_{\rm sh}^{i}$ can locally reconstruct the thermodynamic state only if the corresponding Jacobian has sufficient rank. This local observability criterion provides a useful formulation of the inverse problem, although maximal local rank does not guarantee global injectivity. Disconnected thermodynamic states may still produce the same reduced optical observables.

The SM diagrams provide a compact representation of this GTD--shadow correspondence. Ref.~\cite{LadinoRomeroQuevedo2026} introduced the SM construction for RN and Kerr using GTD metrics associated with different thermodynamic potentials. The common-potential formulation adopted in this review provides an alternative and complementary construction in which $g^I$, $g^{II}$, and $g^{III}$ are obtained from the same fundamental equation. The optical markers associated with a given black-hole state remain unchanged because they are determined by the spacetime and observer configuration, while the three GTD metrics provide complementary geometric classifications of that state. An SM diagram can therefore display thermodynamic stability, GTD curvature structure, and optical information without identifying thermodynamically distinct phenomena or the different metric structures.

The curvature sign supplies an additional GTD classification. Positive, negative, and vanishing curvature have been interpreted phenomenologically as effective repulsive, attractive, and non-interacting thermodynamic sectors, respectively \cite{LadinoRomeroQuevedo2024,LadinoRomeroQuevedo2025,LadinoRomeroQuevedo2026}. The examples considered here show that this classification can depend strongly on the selected GTD metric. For Kerr, for example, $g^I$ is flat, $g^{II}$ changes curvature sign, and $g^{III}$ remains curved. This metric dependence identifies a central problem within GTD: to determine the physical information carried by each of the three Legendre-invariant metric structures and whether an additional physical principle can select the metric relevant to a specified thermodynamic phenomenon.

The observational analysis reverses the theoretical correspondence. Horizon-scale observations constrain an image or ring structure; an astrophysical and gravitational model converts those constraints into an allowed region of theoretical shadow observables; the shadow observables constrain the black-hole parameters; and those parameters determine an allowed region of equilibrium space. GTD can then classify the permitted states. The Sagittarius A* analysis provides the first explicit implementation of this program for RN and Kerr reference geometries \cite{LadinoRomeroQuevedo2026}. The inference remains conditional on the spacetime, emission, observer, and image models, and the bright emission ring must not be identified directly with the mathematical critical shadow. Furthermore, black hole shadow characteristics offer a rich landscape for future exploration. As shown in quantum Oppenheimer-Snyder spacetimes, GTD can successfully link shadow observables, like photon rings and gap parameters, to a black hole's thermodynamic stability and microscopic structure \cite{ladino2026nakedsingularitiesblackhole}. Expanding this analysis to other optical signatures and next-generation high-resolution measurements provides a useful tool for examining black hole thermodynamics via direct astrophysical observations.

Improved shadow measurements can sharpen the allowed regions of black-hole parameter space and, consequently, the regions of equilibrium space on which the GTD metrics are evaluated. For RN they can determine more precisely whether the observationally permitted states include the Schwarzschild limit, the Davies instability, or extremality. For Kerr they can test the compatibility of the D-state, the $g^{II}$ N-state, the Schwarzschild limit, and extremality with a specified optical model. Shadow observations alone, however, cannot distinguish among $g^I$, $g^{II}$, and $g^{III}$ when these metrics are evaluated at the same black-hole state, because the three GTD metrics correspond to the same spacetime and therefore to the same shadow.

This limitation defines one of the principal open problems of the GTD--shadow program. An observational selection among the GTD metrics requires an additional physical prediction whose value depends explicitly on the selected metric. No such GTD-specific observable has yet been established in the black-hole shadow applications considered here. Finding one requires determining whether the distinct curvature structures of $g^I$, $g^{II}$, and $g^{III}$ imply different physical consequences beyond the common equilibrium state and its spacetime shadow. Establishing such a consequence would transform the metric dependence of GTD from a classification issue into a testable prediction.

Several related questions remain open within GTD. The physical significance of regular curvature zeros, additional curvature singularities, and metric degeneracies should be established systematically across broader families of black holes. The phenomenological interpretation of curvature signs as effective microscopic interaction sectors requires a microscopic statistical foundation capable of explaining what information is encoded separately by $g^I$, $g^{II}$, and $g^{III}$. The local Jacobian-rank criterion should be extended to the global observability problem in order to identify optical degeneracies and the minimal set of shadow observables needed for rotating and multidimensional families. It is also natural to investigate whether GTD curvature critical exponents possess universal properties and how such critical behavior is represented in optical variables. These problems can be studied by extending the three-metric analysis systematically while preserving the thermodynamic and physical-domain controls developed in this review.

Finally, the equilibrium character of the present framework should remain explicit. Astrophysical black holes are dynamical systems, whereas the GTD construction used here describes equilibrium states. A natural intermediate problem is to determine when a slowly evolving black hole can be represented by an adiabatic sequence of GTD equilibrium states and whether its optical evolution can correspondingly be represented by a sequence of SM diagrams. A genuinely nonequilibrium formulation would require an extension of GTD itself rather than the introduction of an arbitrary time dependence into equilibrium variables.

The overall picture that emerges is therefore a GTD program. Black-hole shadows provide optical access to families of equilibrium states, while GTD supplies the Legendre-invariant geometric structures used to classify those states. The three GTD metrics reveal complementary curvature singularities, regular structures, degeneracies, and curvature-sign information. The next stage is to understand this internal GTD structure more deeply: to determine the physical roles of $g^I$, $g^{II}$, and $g^{III}$, to establish the significance of their additional geometric structures, to formulate a microscopic interpretation within GTD, and to identify an observable prediction capable of discriminating among them. Progress on these questions would turn black-hole shadow geometrothermodynamics from a geometric classification of optically constrained equilibrium states into a framework with genuinely distinctive physical predictions.

\section*{Acknowledgments}
CRF and JML acknowledge support from SECIHTI-Mexico, grants No. 4003366 and No. 402076.  This work was supported by UNAM-DGAPA-PAPIIT, grant No. 108225 and Conahcyt, grant No. CBF-2025-I-243.

\nocite{apsrev42Control}

\bibliographystyle{apsrev4-2}
\bibliography{references}

\end{document}